\documentclass[
  journal=pasa,
  manuscript=research-paper, 
  year=202X,
  volume=YY,
]{cup-journal}
\usepackage{aas-macros}
\usepackage{xcolor} 
\usepackage{hyperref}
\hypersetup{
    colorlinks=true,
    linkcolor=blue,
    citecolor=violet,
    urlcolor=teal
}
\usepackage{smartdiagram}
\usesmartdiagramlibrary{additions}
\usepackage{longtable}
\usepackage{amsmath}
\usepackage{cprotect}
\usepackage{float}
\usepackage{graphicx}
\usepackage{overpic}
\usepackage{tabularx}
\usepackage{makecell}

\usepackage{subcaption}
\usepackage{amssymb,microtype,siunitx,booktabs}
\def\arcsec{\hbox{$^{\prime\prime}$}}

\newcommand\arcmin{\hbox{$\,\!\!^{\prime}$}}
\def\Msun{\ifmmode{{\rm M}_\odot}\else${\rm M}_\odot$\fi}

\title{The ASKAP Variables and Slow Transients (VAST) Extragalactic Survey -- Data Release 1}

\author{Iris de Ruiter}
\affiliation{Sydney Institute for Astronomy, School of Physics, The University of Sydney, NSW 2006, Australia}
\alsoaffiliation{ARC Centre of Excellence for Gravitational Wave Discovery (OzGrav), Australia}
\email[Iris de Ruiter]{\url{iris.deruiter@sydney.edu.au}}

\author{Dougal Dobie}
\affiliation{Sydney Institute for Astronomy, School of Physics, The University of Sydney, NSW 2006, Australia}
\alsoaffiliation{ARC Centre of Excellence for Gravitational Wave Discovery (OzGrav), Australia}

\author{Tara~Murphy}
\affiliation{Sydney Institute for Astronomy, School of Physics, The University of Sydney, NSW 2006, Australia}
\alsoaffiliation{ARC Centre of Excellence for Gravitational Wave Discovery (OzGrav), Australia}

\author{David L. Kaplan}
\affiliation{Department of Physics \& Astronomy, University of Wisconsin-Milwaukee, P.O. Box 413, Milwaukee, WI 53201, USA}
\alsoaffiliation{ARC Centre of Excellence for Gravitational Wave Discovery (OzGrav), Australia}

\author{Emil Lenc}
\affiliation{Australia Telescope National Facility, CSIRO Space and Astronomy, PO Box 76, Epping, NSW 1710, Australia}

\author{Akash Anumarlapudi}
\affiliation{Department of Physics and Astronomy, University of North Carolina at Chapel Hill, 120 E. Cameron Ave, Chapel Hill, NC, 27599, USA}

\author{Laura~N. Driessen}
\affiliation{Sydney Institute for Astronomy, School of Physics, The University of Sydney, NSW 2006, Australia}

\author{Ashna Gulati}
\affiliation{Sydney Institute for Astronomy, School of Physics, The University of Sydney, NSW 2006, Australia}
\alsoaffiliation{ARC Centre of Excellence for Gravitational Wave Discovery (OzGrav), Australia}
\alsoaffiliation{Australia Telescope National Facility, CSIRO Space and Astronomy, PO Box 76, Epping, NSW 1710, Australia}

\author{Assaf Horesh}
\affiliation{Racah Institute of Physics, The Hebrew University of Jerusalem, Jerusalem 91904, Israel}

\author{James K. Leung}
\affiliation{David A. Dunlap Department of Astronomy and Astrophysics, University of Toronto, 50 St. George Street, Toronto, ON M5S 3H4, Canada}
\alsoaffiliation{Dunlap Institute for Astronomy and Astrophysics, University of Toronto, 50 St. George Street, Toronto, ON M5S 3H4, Canada}
\alsoaffiliation{Racah Institute of Physics, The Hebrew University of Jerusalem, Jerusalem 91904, Israel}

\author{Joshua Pritchard}
\affiliation{Australia Telescope National Facility, CSIRO Space and Astronomy, PO Box 76, Epping, NSW 1710, Australia}

\author{Kovi Rose}
\affiliation{Sydney Institute for Astronomy, School of Physics, The University of Sydney, NSW 2006, Australia}
\alsoaffiliation{Australia Telescope National Facility, CSIRO Space and Astronomy, PO Box 76, Epping, NSW 1710, Australia}

\author{Elaine M. Sadler}
\affiliation{Sydney Institute for Astronomy, School of Physics, The University of Sydney, NSW 2006, Australia}
\alsoaffiliation{ARC Centre of Excellence for Gravitational Wave Discovery (OzGrav), Australia}
\alsoaffiliation{Australia Telescope National Facility, CSIRO Space and Astronomy, PO Box 76, Epping, NSW 1710, Australia}

\author{Gregory Sivakoff}
\affiliation{Department of Physics, University of Alberta, CCIS 4-181, Edmonton AB T6G 2E1, Canada}

\author{Yuanming Wang}
\affiliation{Centre for Astrophysics and Supercomputing, Swinburne University of Technology, Hawthorn, Victoria, 3122, Australia}
\alsoaffiliation{ARC Centre of Excellence for Gravitational Wave Discovery (OzGrav), Australia}

\author{Ziteng Wang}
\affiliation{International Centre for Radio Astronomy Research, Curtin University, Bentley, WA 6102, Australia}

\received {dd Mmm YYYY}
\revised  {dd Mmm YYYY}
\accepted {dd Mmm YYYY}
\published{22 September 202X}


\begin{document}

\begin{abstract}
The Variables and Slow Transients (VAST) Survey on the Australian SKA Pathfinder (ASKAP) is designed to systematically explore the dynamic radio sky, detecting sources that vary on timescales from minutes to several years. In this paper, we present Data Release 1 of the VAST Extragalactic Survey, which targets slowly evolving synchrotron transients in the southern sky. The observations were carried out between June 2023 and May 2025, comprising 2945 images of 276 fields spanning $\sim 12300\ \mathrm{deg}^2$, observed at 888 MHz with a typical rms sensitivity of 0.24 mJy $\rm{beam}^{-1}$ and 12–-20 arcsec resolution. Each field was revisited approximately every two months, yielding 10 or 11 observations per field. The VAST pipeline extracts the light curves for all the observed sources, and additional filters are implemented to improve the reliability of the resulting light curve database. The light curve database contains 0.5 million sources and 6.4 million individual measurements, publicly available through the CSIRO data access portal.
An untargeted variability search yields 117 astrophysical variables, including 27 pulsars, 40 radio stars (10 newly detected at radio wavelengths), 44 active galactic nuclei, two optically identified supernovae, one supernova candidate, one brown dwarf, and two sources without multi-wavelength counterparts that are yet to be identified. This data release provides the first large-scale, high-cadence, uniform view of long-term radio variability in the extragalactic sky and lays the groundwork for future population studies of radio transients with ASKAP.

\end{abstract}

\section{Introduction}

The transient radio sky provides a window to some of the most extreme events in the Universe. The astrophysical processes that give rise to transient radio emission originate from powerful, dynamic phenomena such as the deaths or mergers of stars or highly energetic processes around compact objects, providing insights into physical conditions that cannot be reproduced in terrestrial laboratories. Radio observations contribute a unique perspective on astrophysical transients. For example, radio synchrotron afterglow studies are critical for constraining the energetics and geometry of relativistic outflows \citep{sari1999jets, rhoads1999dynamics, frail2000450}. In addition, certain transients may be heavily obscured by dust in the optical and infrared, making radio wavelengths the only viable window to study their evolution and environments \citep[see e.g.][]{kochanek2011dusty,jencson2018spirits, schroeder2022radio, leung2026first}. Finally, some phenomena are detected primarily in the radio band; for example, pulsars were first discovered, and are still primarily detected, in the radio band \citep{lorimer2005handbook}.\\

Radio variability may be intrinsic to the source or caused by propagation effects that imprint variability on an otherwise steady signal. Most radio transients are caused by either coherent emission processes -- such as those powering fast radio bursts and stellar flares \citep{melrose2017coherent} -- or incoherent synchrotron emission in transients like radio supernovae and gamma-ray burst afterglows \citep{weiler2002radio, chandra2012radio, bietenholz2021radio} or tidal disruption events \citep{alexander2020radio}. Both coherent and synchrotron emission require a reservoir of relativistic particles and strong magnetic fields. For synchrotron emission, shocks between outflows and the surrounding medium are often required to produce a locally amplified magnetic field. For a detailed discussion of radio transients (including stellar flares, scintillating pulsars, intrinsically variable or scintillating active galactic nuclei (AGN), and radio supernovae), we refer the reader to the review by \cite{murphy2026dawes}.
External causes of variability are refractive and diffractive scintillation, originating from density fluctuations of the interplanetary, interstellar and intergalactic media along the line of sight to the source (see e.g. \citealt{jokipii1973turbulence, rickett1990radio, zhou2014fast}). Radio signatures of transients and highly variable sources enable studies of the source energetics, surrounding medium properties, event rates, and population statistics \citep[see e.g.][]{schulze2011circumburst, aksulu2022exploring}.\\

In the discussion below we focus on image-based transient searches. Transient surveys that search for much faster timescale emission, targeting FRBs and pulsars, are not discussed. In addition, we focus on image domain transient searches in the gigahertz regime.
Early untargeted transient searches often relied on comparing archival surveys at similar frequencies taken years apart (see, e.g., \citealt{levinson2002orphan, gal2006radio, law2018discovery, nyland2020quasars,chen2025searching}). Many of these works are based on large-area surveys conducted with the Very Large Array \citep[VLA;][]{napier1983very}, including the NRAO VLA Sky Survey \citep[NVSS;][]{condon1998nrao}, the Faint Images of the Radio Sky at Twenty-cm survey \citep[FIRST;][]{white1997catalog}, and the VLA Sky Survey \citep[VLASS;][]{lacy2020karl}. These transient studies were complicated by the fact that the resulting light curves consisted of only two epochs and, in some cases, by cross-telescope systematics. Such sparse sampling reflects a limited observing time, leading to a low expected transient detection rate. This implies that the light curve can only contribute in a limited way to identifying the transient. \\

In recent years, a more specialised approach has emerged, where dedicated transient surveys have been developed. By increasing the number of repeat observations of a single part of the sky, one expects to detect more flaring radio transients, such as radio stars (see e.g. \citealt{driessen2024sydney}). This approach was adopted by the ThunderKAT collaboration \citep{fender2017thunderkat}, which ran on MeerKAT \citep{jonas2016meerkat}. This survey was designed to monitor fields with X-ray binaries, but led to serendipitous transient discoveries \citep{andersson2022serendipitous, driessen2024frb}, and untargeted transients and variables searches have been done for these fields \citep{rowlinson2022search,driessen202221}. \cite{chastain2023commensal} describe another example of such a commensal transient search on MeerKAT data. These approaches yield rich datasets for a limited number of sources, due to relatively small surveyed sky areas.\\

To characterise extragalactic synchrotron transients, wide-field surveys that revisit the same regions of sky on weeks to months timescales are essential. This requires a telescope with a large field of view, with a dedicated project that allows for a large number of repeat observations over time.
An example of a transient search effort that specifically targets slowly evolving synchrotron transients uses the individual VLASS epochs. The comparison of individual high-sensitivity observations has shown great potential to uncover a large number of transient sources \citep{dong2023shocks, sharma2025fast}. However, since VLASS consists of only three epochs, additional observations are required to characterise the full light curve (which is often difficult for archival discoveries), and transient searches are limited to a variability timescale of approximately a year, implying that faster evolving transients will be missed. \\

The Variables And Slow Transients \citep[VAST\footnote{\url{https://vast-survey.org.}};][]{murphy2013vast} survey is designed to search the Southern sky for transients and variables with both a large sky coverage and a high number of repeat observations. VAST is one of the key Survey Science Projects running on the Australian SKA Pathfinder \citep[ASKAP;][]{johnston2008science, hotan2021australian}. ASKAP is a radio interferometer comprised of 36 12-metre prime-focus antennas located on Inyarrimanha Ilgari Bundara, the CSIRO Murchison Radio-astronomy Observatory in Western Australia. ASKAP is an ideal instrument for wide-field, time-domain radio astronomy in the Southern hemisphere, combining a large instantaneous field of view and good sensitivity. VAST was designed to detect phenomena that vary on timescales of tens of seconds to several years. The survey has two components: (1) An extragalactic component with repeated observations every two months; and (2) a Galactic survey with repeated observations on much shorter timescales (days to weeks). In this paper we present the extragalactic data release.\footnote{Note that the final approved survey strategy presented here is different to the planned surveys discussed in previous papers \citep{murphy2013vast,murphy2021askap}. In particular, VAST-wide and VAST-deep evolved into the current VAST extragalactic survey.}

Prior to the full survey, several pilot surveys were conducted \citep{bhandari2018pilot, murphy2021askap}. An untargeted transient search in the first VAST pilot survey yielded 28 transient or highly variable sources \citep{murphy2021askap}. The VAST pilot data have also been used to characterise radio stars \citep{pritchard2024multi}. ASKAP has also detected a diverse range of extragalactic synchrotron phenomena, including novae \citep{gulati2023classical}, tidal disruption events \citep{anumarlapudi2024radio, dykaar2024untargeted}, supernovae \citep{rose2024late}, and gamma-ray burst afterglows \citep{leung2021search}. Many of these detections used early VAST data. The VAST pilot surveys and studies laid the foundation for the full VAST science program; in this work we present a first data release of the VAST Extragalactic Survey.

In Section~\ref{sec:obs_and_strategy} we discuss the ASKAP observations and VAST Extragalactic Survey strategy, including the survey footprint and observing cadence. In Section~\ref{sec:pp_and_qc} we describe the additional post-processing steps that we apply to the observations, and the data quality in terms of image noise, astrometry and flux density scale. In Section~\ref{sec:light_curve_database} we discuss how we build a light curve database and how we construct a sample of reliable light curves. In Section~\ref{sec:untargeted_variability_search}, we present an untargeted variability search,  highlighting the most variable sources in this data set. Finally, in Section~\ref{sec:discussion} we discuss how we imagine VAST Extragalactic data products will be used, compare our results to the pilot survey and discuss future improvements.\\

\section{Observations and survey strategy} \label{sec:obs_and_strategy}
\begin{figure*}
    \includegraphics[width=\textwidth]{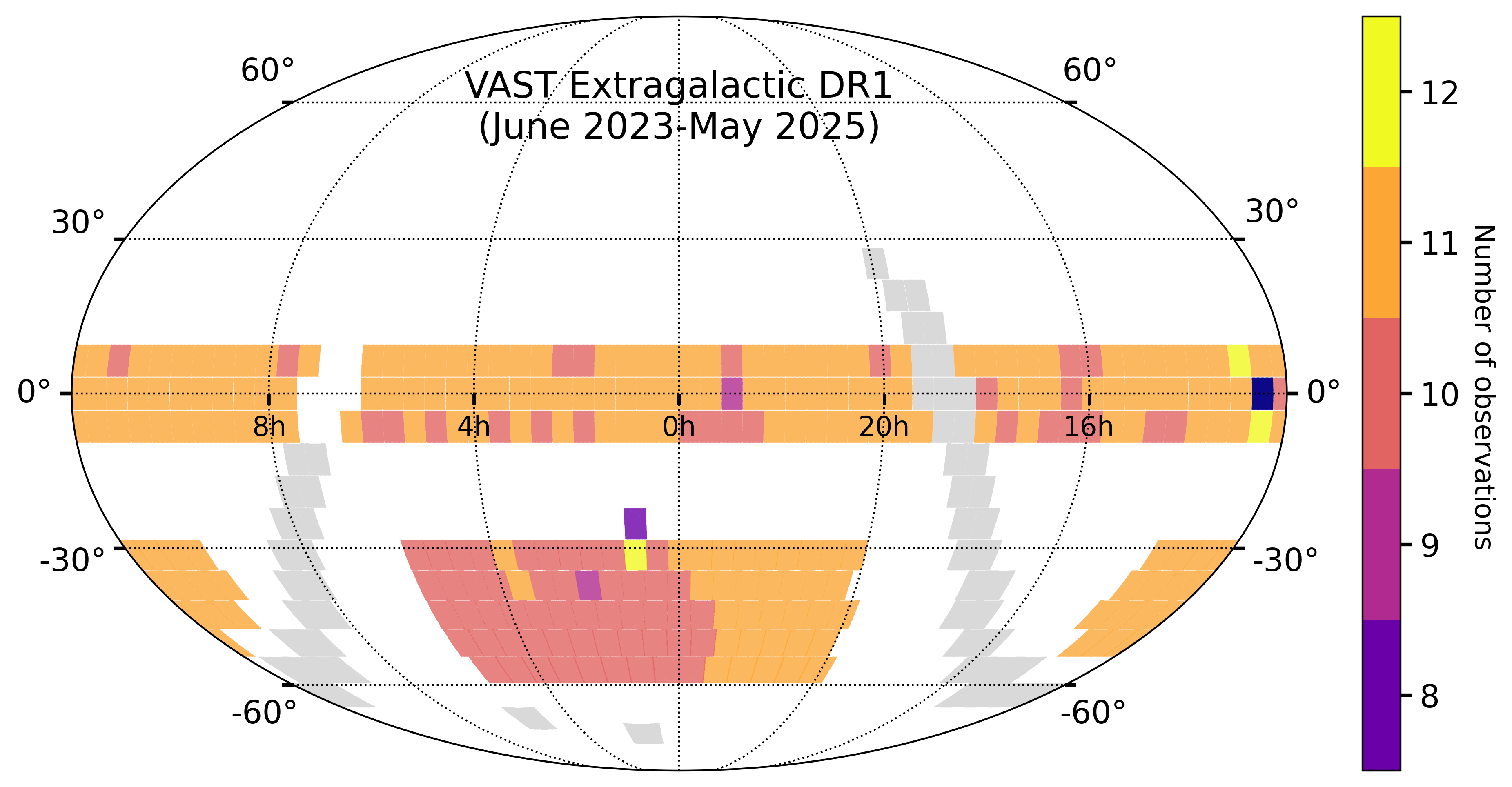}
    \cprotect\caption{The VAST Extragalactic Survey footprint, showing the number of observations of each field. The sky map is plotted with J2000 equatorial coordinates in the Mollweide projection. The VAST Galactic survey is plotted in grey for reference. Typically, each field has been observed 10-11 times to date. There is a single field \verb|VAST 1237+00| (navy) that has only been observed three times, before it was removed from the survey footprint, as it contains 3C273 (64 Jy), which causes poor image quality.}
    \label{fig:nof_obs_per_field}
\end{figure*}

In this paper we present the first data release for the extragalactic component of the VAST survey. Due to the different scientific goals and technical specifications for the surveys, we will publish the Galactic component separately.

The VAST survey uses the lowest ASKAP frequency band, centred at 888 MHz with 288 MHz of bandwidth. All observations use the \texttt{square\_6x6} beam footprint for the arrangement of the phased-array feed (PAF) beams \citep[see][]{hotan2021australian}.

The full VAST survey commenced in November 2022, and is scheduled to run for 4 years. The first VAST observations mainly covered the Galactic part of the survey, and extragalactic observations started in June 2023. In this work, we publish the roughly first half (first two years) of the extragalactic part of the ongoing VAST survey. We refer to the data released in this work as the VAST Extragalactic Data Release 1 (DR1).

Throughout this work, we refer to an observation of a single VAST field as a \textit{field} or an \textit{observation}. The tiling pattern that makes up the VAST Extragalactic Survey footprint is detailed in Figure \ref{fig:nof_obs_per_field} and in Section \ref{sec:obs_survey_footprint}. These fields are observed in groups at a time, each group making up an observing \textit{epoch}. The observing cadence is detailed in Figure~\ref{fig:fields_per_epoch} and in Section~\ref{sec:obs_survey_strategy}. We adopt the equatorial coordinate system referenced to the J2000 reference frame throughout.

\subsection{Rapid ASKAP Continuum Survey (RACS)}
The Rapid ASKAP Continuum Survey \citep[RACS\footnote{\url{https://research.csiro.au/racs/}};][]{mcconnell2020rapid} is a collection of large-area ASKAP continuum surveys conducted in three frequency bands centred on 888, 1296 and 1656 MHz, corresponding to RACS-low, RACS-mid and RACS-high. Multiple epochs exist for each band, and the most relevant RACS versions for this work are RACS-low1 \citep{mcconnell2020rapid,hale2021rapid} and RACS-low2 (Duchesne et al. in prep), observed at the same frequency and with the same PAF footprint as VAST. Improvements in the calibration and imaging strategy as described in Duchesne et al. (in prep), lead to higher resolution, lower rms, and thus more detected sources in RACS-low2 compared to RACS-low1. Table \ref{tab:racs_low_versions} summarises all RACS surveys referenced in this paper, including their observing setup, sky coverage, and the specific roles they play in the VAST Extragalactic DR1 analysis. Note that throughout, when we refer to these RACS surveys, we refer to the curated all-sky catalogues as published and presented in the papers. Similar to the VAST survey (see Section \ref{sec:obs_validation}), the raw data products for RACS are available on CSIRO's ASKAP Science Data Archive \citep[CASDA;][]{huynh2020csiro} immediately after validation.

\begin{table}
\centering
\caption{Summary of RACS survey versions referenced in this work. For each RACS epoch, the central observing frequency, PAF footprint \citep[see][]{hotan2021australian} and sky coverage are listed, along with their key uses in this paper. [1]: \cite{mcconnell2020rapid} [2]: \cite{hale2021rapid} [3]: Duchesne et al. in prep, [4]:\cite{duchesne2023rapid}, [5]: \cite{duchesne2024rapid}.}
\label{tab:racs_low_versions}
\begin{tabularx}{\columnwidth}{l l l X}
\hline
Epoch & \makecell[l]{Freq. (MHz) and \\ footprint} & Coverage & \makecell[l]{Key uses in VAST \\ Extragalactic DR1} \\
\hline
\hline
\makecell[l]{RACS-low1 \\ $[1,2]$} & \makecell{888 MHz \\ \texttt{square\_6x6}} & $<41^\circ$ decl. &
Investigating artificial variability from PAF response (\ref{app:data_obs_issues}) \\ \hline

\makecell[l]{RACS-low2  \\ $[3]$} & \makecell{888 MHz \\ \texttt{square\_6x6}} & $<51^\circ$ decl. &
Post-processing (Sec. \ref{sec:vast_pp}) and
absolute astrometry (Sec. \ref{sec:absolute_astrometry}) \\ \hline

\makecell[l]{RACS-mid1  \\ $[4,5]$} & \makecell{1296 MHz  \\ \texttt{closepack36}} & $<49^\circ$ decl. &
Absolute astrometry (Section \ref{sec:absolute_astrometry}) \\ \hline

\makecell[l]{RACS-low3  \\ (in prep.)} & \makecell{944 MHz \\ \texttt{closepack36}} & $<49^\circ$ decl. &
\textit{Improved astrometry in future VAST releases} \\ 
\hline
\end{tabularx}
\end{table}

 
\subsection{Survey footprint} \label{sec:obs_survey_footprint}
Similar to the VAST pilot survey \citep{murphy2021askap}, the VAST survey adopts the RACS-low tiling pattern \citep{mcconnell2020rapid}. The VAST Extragalactic Survey comprises 276 fields, with sky coverage shown in Figure \ref{fig:nof_obs_per_field}; the VAST Galactic footprint is included in grey for comparison.

The extragalactic fields were selected to maximise overlap with existing multi-wavelength surveys. The equatorial regions (centred at $\mathrm{Dec}=0^\circ$) provide strong northern-hemisphere multi-wavelength coverage and can be observed by both northern and southern facilities, aiding follow-up of transient events. Fields between $-30^\circ < \mathrm{Dec} < -60^\circ$ and RA = 20–6 h were chosen for their overlap with the Dark Energy Survey \citep[DES;][]{dark2016dark}. Additional southern fields extend the overall coverage and distribute the footprint across a wide range of right ascensions to facilitate scheduling. The equatorial fields overlap with FIRST, NVSS and VLASS, while fields below $\mathrm{Dec}=-30^\circ$ overlap with the Sydney University Molonglo Sky Survey \citep[SUMSS;][]{bock1999sumss, mauch2003sumss}.

The RACS tiling pattern introduces some field overlap; the right ascension spacing varies from $6.2^\circ$ at $\mathrm{Dec}=0^\circ$ to $10.6^\circ$, corresponding to a projected separation of $\cos(56.3^\circ)\times10.6^\circ = 5.9^\circ$ at $\mathrm{Dec}=-56.3^\circ$. As described in Section \ref{sec:vast_pp}, VAST light curves use the central $6.67^\circ \times 6.67^\circ$ of each ASKAP field. Some sources are therefore observed as part of multiple fields and receive correspondingly more light curve points. Assuming a $6.67^\circ \times 6.67^\circ$ effective area for all 276 fields gives a total survey footprint of $\sim 12300\ \mathrm{deg}^2$.

\subsection{Survey strategy} \label{sec:obs_survey_strategy}
The primary goal of the VAST Extragalactic Survey is to detect extragalactic synchrotron transients, which typically evolve on month-long timescales \citep[see e.g.][]{pietka2015variability}. Such slow evolution means that relatively sparse repeat observations are sufficient to characterise them. The survey is also expected to detect radio stars, which are isotropically distributed at local distances \citep{driessen2024sydney}.

The VAST Extragalactic Survey observes the full footprint (as described in Section \ref{sec:obs_survey_footprint}) every two months. Each VAST observation has an integration time of 10 to 12 minutes, yielding a median image rms noise of $0.24$ mJy $\rm{beam}^{-1}$. This implies that all VAST Extragalactic fields can be observed within 2 to 3 days of observing time (not considering overheads and the schedulability of other ASKAP surveys).

\begin{figure}[t]
    \centering
    \includegraphics[width=\linewidth]{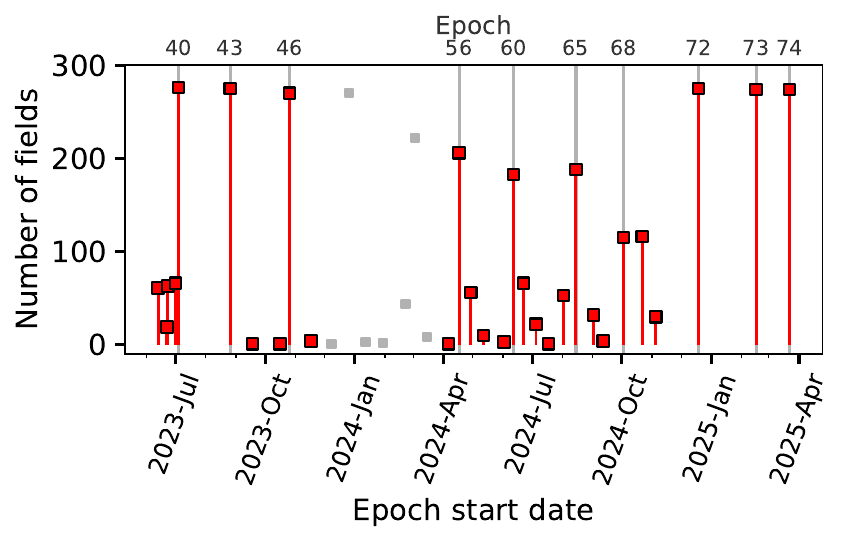}
    \caption{Summary of VAST Extragalactic observations, giving the number of fields in each epoch. From 2024 the strict 2-month cadence was removed, meaning that a full reobservation of all 276 fields was often spread over multiple epochs. The greyed out epochs are not included in the VAST Extragalactic DR1 (see \ref{app:data_obs_issues} for details).}
    \label{fig:fields_per_epoch}
\end{figure}

Figure \ref{fig:fields_per_epoch} summarises the number of fields observed over time and per epoch. Extragalactic observations began in epoch 36; earlier epochs contained only Galactic or pilot observations. Epochs vary in length—from 2 to 52 days—and their definition has evolved throughout the survey. Early in the full VAST survey, all 276 extragalactic fields were typically observed within a few days, producing short epochs. By late 2023, relaxed scheduling and the removal of the strict two-month cadence meant that full-field coverage was often distributed across multiple epochs. Several epochs (40, 43, 46, 72, 73, 74) captured nearly all extragalactic fields, with durations of 8, 16, 8, 20, 26, and 52 days, respectively. The increased length of epoch 74 can be attributed to a combination of scheduling difficulties (considering higher priority observations and avoiding Solar interference) and software changes that delayed processing.
As of epoch 72, each extragalactic epoch includes one observation of every extragalactic field. Figure \ref{fig:fields_per_epoch} also shows greyed out observations from December 2023 to April 2024 corresponding to epochs 48–55, which are excluded from VAST Extragalactic DR1 due to incorrect holography calibrations causing flux density inconsistencies (see \ref{app:data_obs_issues}). After excluding these epochs, VAST Extragalactic DR1 includes 2945 observations from June 2023 to May 2025.

\subsection{Main changes with respect to the VAST pilot surveys}

Below we list the main changes in survey footprint and strategy between the full survey and the pilot survey \citep{murphy2021askap}.

\begin{itemize}
    \item The full VAST Extragalactic Survey consists of 276 fields. This survey footprint has changed substantially compared to the pilot survey. In the pilot survey, only 113 Galactic and extragalactic fields were included.
    \item The full VAST Extragalactic survey only observes in ASKAP low-band, with a central frequency of 888 MHz and 288 MHz of bandwidth. In the pilot surveys, part of the VAST observations were made using the mid band (1296 MHz). After the pilot surveys, we decided to remove the mid-frequency part of the VAST survey since this would allow for direct comparison of all VAST observations, without having to account for spectral effects. Furthermore, the radio frequency interference (RFI) conditions are worse in the mid-band (see e.g. Section 2.1 in \citet{mcconnell2020rapid}) and the low band has a larger field of view per observation. 
    \item In the full VAST survey we treat every observation and image independently. In the pilot surveys, images would be combined to a larger mosaic image. In the pilot, back-to-back observations allowed combining fields within an epoch for uniform sensitivity. In the full survey, scheduling constraints create gaps of up to weeks between adjacent fields, and therefore, mosaicking adjacent observations no longer makes physical sense.
\end{itemize}
 
We exclude the pilot survey observations (AS107 in CASDA) from this VAST Extragalactic data release for three main reasons:
\begin{itemize}
    \item There is a large temporal gap between the pilot observations and the start of the full survey, with no data taken between mid-2021 and mid-2023.
    \item The full survey footprint is more than twice as large, so incorporating the pilot data -- available for only a subset of sources -- would lead to uneven light-curve sampling.
    \item Significant improvements in ASKAP processing have introduced different systematic effects between the pilot and full surveys (see Section \ref{sec:processing_changes}).
\end{itemize}

\subsection{Data processing} \label{sec:data_processing}

\begin{figure*}
    \centering
    \includegraphics[width=0.8\textwidth]{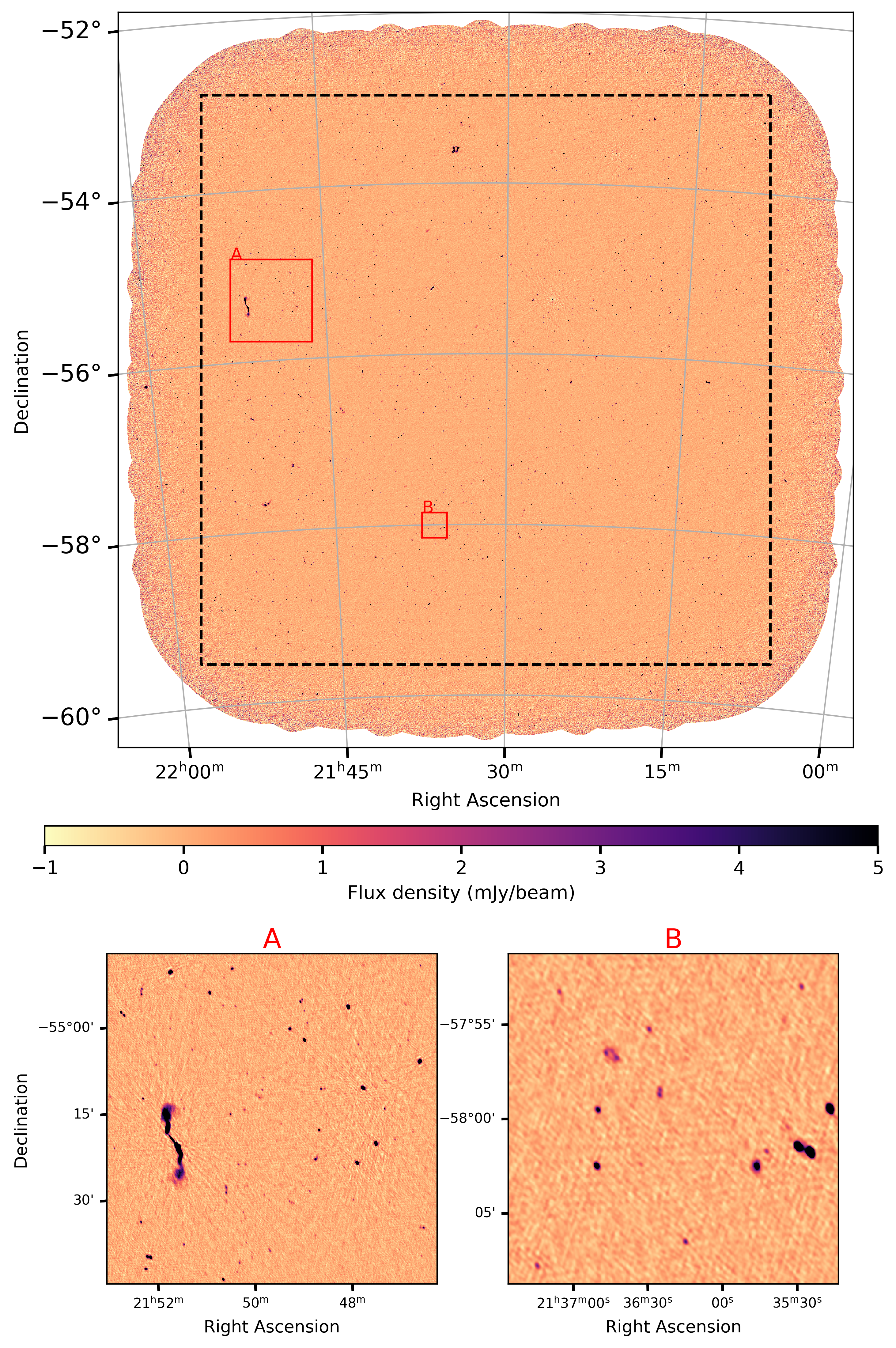}
    \cprotect\caption{A VAST image (\verb|VAST 2131-56|) from epoch 74 with two cutouts. Cutout A: a 1 deg image centred on (J2000) (12:49:26.5, -55:17:52.87) containing several bright sources, including the large radio galaxy 2MASX J21512991-5520124. Cutout B: a 0.3 deg image centred on (J2000) (21:36:18.9, -58:00:12.68) containing a range of source morphologies. The central part of the full image, shown by the black dashed line, indicates the coverage included in the post-processed data products.}
    \label{fig:example_img_cutouts}
\end{figure*}

The VAST data are processed using the same calibration and imaging strategy as used in the VAST pilot survey \citep{murphy2021askap} with two improvements introduced as a result of the experience gained with RACS-low1 \citep{mcconnell2020rapid} and RACS-mid1 \citep{duchesne2023rapid,duchesne2024rapid}. The first is to convolve the point spread functions (PSF) of individual PAF beams to a common PSF to maintain integrated flux density integrity across the observed field. The second is the use of appropriate holography to correct for primary beam attenuation; the holography is updated each time a new set of PAF beams is formed, typically once every four months.\\

We use \textsc{ASKAPSoft} \citep{cornwell2011askap} to create images and source catalogues for the VAST observations. When imaging, we removed short ($<100$ m) baselines in order to (i) minimise solar interference; and (ii) as there is insufficient $(u,v)$-coverage to reliably recover extended structure in the short VAST observations. Figure \ref{fig:example_img_cutouts} shows a typical VAST Extragalactic DR1 image. To allow for a comparison with the pilot survey data, we have chosen the same representative image as shown in \cite{murphy2021askap}. The region indicated with the black dashed line indicates the coverage included in the post-processed data products (Section \ref{sec:pp_and_qc}). In addition to the images, the \textsc{Selavy} source finding software \citep{whiting2012duchamp}, incorporated into the \textsc{ASKAPSoft} pipeline, produces source catalogues. We used the default parameter settings in \textsc{ASKAPSoft}, where sources are considered `detected' if they exceed 5 times the local rms noise, which corresponds to $5\cdot 0.24\approx1.2$ $\rm{mJy\, beam}^{-1}$ in the VAST Extragalactic Survey (see Figure \ref{fig:image_rms_values}).

\subsubsection{Processing changes during and after DR1} \label{sec:processing_changes}
This section describes the changes that were made in processing during VAST Extragalactic DR1, and the processing changes that will be implemented for future VAST data releases. From epoch 63 onward, the observatory processing shifted to using a signal-to-noise based cut-off for deconvolution to prevent deconvolving too deeply in some instances. Efficiency improvements were also implemented in the handling of the $w$-term to allow a larger number of $w$-planes and a higher degree of oversampling to be used within the same memory constraints. Increasing the $w$-planes and oversampling results in a reduction of $w$-term-related wide-field imaging artefacts.\\

From epoch 74 onward, visibility data were initially phase-self-calibrated against the RACS-low2 source catalogue (Duchesne et al. in prep). This effectively introduced phase-referencing to correct for variations that occur as a function of time and direction with respect to the original bandpass corrections applied. These corrections aid subsequent deconvolution and also tie all beams to the common astrometry of RACS-low2 (to reduce image smearing effects that otherwise may occur when beams each have their own independent astrometry errors). It should be noted that RACS-low2 astrometry is not tied to a reference frame. As such, VAST observations will inherit any astrometric errors present in RACS-low2. With the phase-referencing to a consistent catalogue, these errors should be consistent from epoch to epoch for a given field. In the future, this will be further addressed by phase referencing to a new catalogue based on RACS-low3 data with astrometric calibration. The absolute astrometry in VAST Extragalactic DR1 is discussed in more detail in Section \ref{sec:absolute_astrometry}.\\

In addition to the improved phase-referencing (leading to improved astrometry) in future VAST data releases, we anticipate upgrades to \textsc{ASKAPSoft} regarding RFI removal, reduced phase delay errors by using a reference field, reduced PSF side-lobe contamination from bright sources outside of the field and reduced wide-field artefacts from antenna pointing errors.

\subsection{Data validation} \label{sec:obs_validation}

ASKAP Survey Science Projects have no proprietary period, and are released on CASDA after validation. VAST observations are grouped in epochs, and all observations in a single epoch are released to CASDA simultaneously. Before the data are released on CASDA, a few quality control checks are performed on the astrometric accuracy and flux density scale relative to RACS-low1 \citep{hale2021rapid} and RACS-low2 (Duchesne et al. in prep). Systematic offsets in astrometry and or flux density are corrected in post-processing (see Section \ref{sec:pp_and_qc}), but observations with large structures in astrometric offsets or large spreads in flux density offset are flagged as `uncertain' or `rejected'. Additionally, each image is visually inspected to check the overall image quality. If the image quality is poor in a small part of the image, the observation is still included in VAST Extragalactic. However, if the overall image quality is severely reduced, for example, by artefacts across most of the image, it is rejected. This occurs when there is RFI or solar interference during the observation, or when a bad antenna is not flagged. Within the VAST Extragalactic DR1, only 4 observations were initially rejected and reobserved, compared to a total of 2945 observations.

\subsection{Data flow}
\begin{figure*}[h]
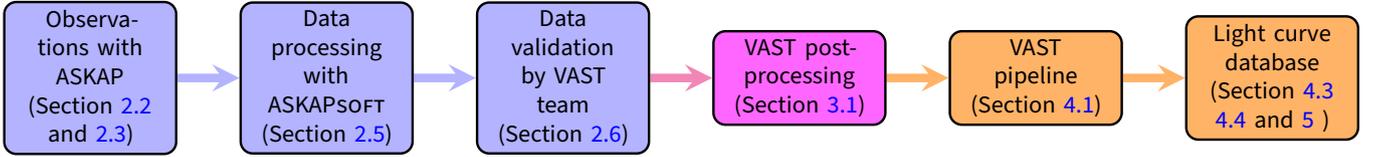

\centering
\tikzset{
  every shadow/.style={
    fill=none,
    shadow xshift=0pt,
    shadow yshift=0pt}
}
\tikzset{module/.append style={top color=\col,bottom color=\col}}
\resizebox{\textwidth}{!}{%
    \smartdiagramset{border color=black,
    set color list={blue!30,blue!30,blue!30,magenta!60,orange!60,orange!60},
    font=\footnotesize,
    back arrow disabled=true,
    }
    \smartdiagram[flow diagram:horizontal]{
    Observations with ASKAP (Section \ref{sec:obs_survey_footprint} and \ref{sec:obs_survey_strategy}),
    Data processing with \textsc{ASKAPsoft} (Section \ref{sec:data_processing}),
    Data validation by VAST team (Section \ref{sec:obs_validation}),
    VAST post-processing (Section \ref{sec:vast_pp}), 
    VAST pipeline (Section \ref{sec:VAST_pipeline}), 
    Light curve database (Section \ref{sec:selection_criteria} \ref{sec:dataproducts} and \ref{sec:untargeted_variability_search} )}
}
\caption{A flow chart describing the high-level data flow of VAST observations. Each block refers to a section describing that step in more detail. The purple/blue steps result in the data products as available on CASDA quickly after an observation. The pink and orange steps are additional steps as introduced in this work.} \label{fig:flow_chart}
\end{figure*}

Figure \ref{fig:flow_chart} presents a high-level overview of the data flow of VAST observations. In Sections \ref{sec:obs_survey_footprint}, \ref{sec:obs_survey_strategy}, \ref{sec:data_processing} and \ref{sec:obs_validation} we highlighted the steps that produce the raw images and source catalogues (available through CASDA). These steps are highlighted in purple in Figure \ref{fig:flow_chart}. Section \ref{sec:vast_pp} below details the post-processing step before data is ingested into the VAST transient detection pipeline (Section \ref{sec:VAST_pipeline}), highlighted in purple and orange in Figure \ref{fig:flow_chart} respectively.

\section{Post-processing and data quality control} \label{sec:pp_and_qc}

\subsection{VAST post-processing} \label{sec:vast_pp}

The VAST post-processing\footnote{\url{https://github.com/askap-vast/vast-post-processing}} \citep{vpp_software} applies corrections to the catalogues and images so we can extract reliable light curves. Below we discuss each of the steps in the post-processing pipeline.

\subsubsection{Calibrator catalogue} \label{sec:vast_pp_cal_cat}

The corrections to the astrometry and integrated flux density of a single observation are calculated using a sample of a few hundred compact, bright and isolated sources per field. These are selected by finding sources in the VAST field with an integrated flux density to peak flux density ratio of less than 1.5, a signal-to-noise ratio of 20, and a distance to their nearest neighbour of at least $1 \arcmin$. This source sample is cross-matched to RACS-low2 (Duchesne et al. in prep). We will refer to these sources as the RACS-calibrator sample. The astrometry and flux density correction are described in more detail in the following two subsections.

\subsubsection{Astrometric corrections}\label{sec:pp_astrometry_corrections}
Per VAST field, we construct a RACS-calibrator sample. For all cross-matched sources, we calculate the right ascension and declination offsets between RACS and VAST. We define the astrometric offset as the median offset of all cross-matched sources in a single VAST field. We subsequently use this median offset as the astrometric correction to the VAST field, which is applied to all the VAST sources in the \textsc{Selavy} source catalogues and the FITS image.

\begin{figure}[h]
    \centering
    \includegraphics[width=\linewidth]{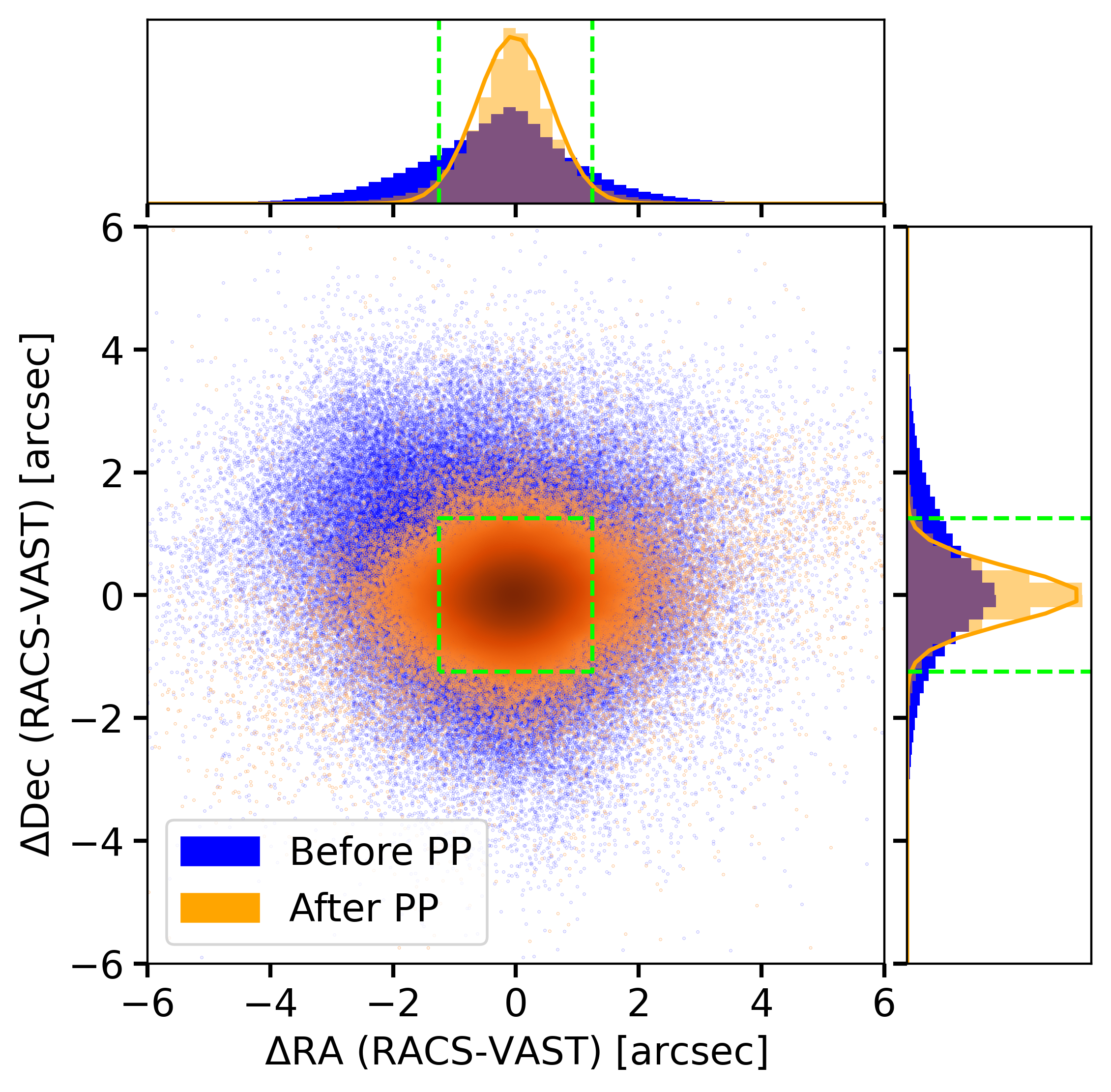}
    \caption{Astrometric offset between RACS-low2 and VAST before post-processing (in blue) and after post-processing (in orange). Each dot represents a single observation of a VAST source cross-matched to the RACS-calibrator sample (see text). The gradient in the orange datapoints represents the point density. The green dashed line shows the typical $2.5 \times 2.5 \arcsec$ pixel size in VAST images.}
    \label{fig:PP_astrometry}
\end{figure}

Figure \ref{fig:PP_astrometry} shows how the astrometric offset between RACS-low2 and VAST Extragalactic improves from the post-processing corrections. The blue dots represent the offset of RACS-calibrator sources in all VAST Extragalactic observations, as calculated using the source catalogues before post-processing. The orange dots represent the offsets of RACS-calibrator sources after we apply a single astrometric shift per observation. After post-processing, the offsets in both right ascension and declination are closer to zero, and the width of the offset distribution has decreased. The green dashed line shows the typical $2.5 \times 2.5 \arcsec$ pixel size in VAST images. The VAST post-processing reduces the average offset in right ascension from $-0.23 \pm 1.18 \arcsec$ to $-0.04 \pm 0.60 \arcsec$, and in declination from $0.10 \pm 0.98 \arcsec$ to $0.00 \pm 0.44 \arcsec$. \\

\begin{figure}[ht!]
    \centering
    \includegraphics[width=\linewidth]{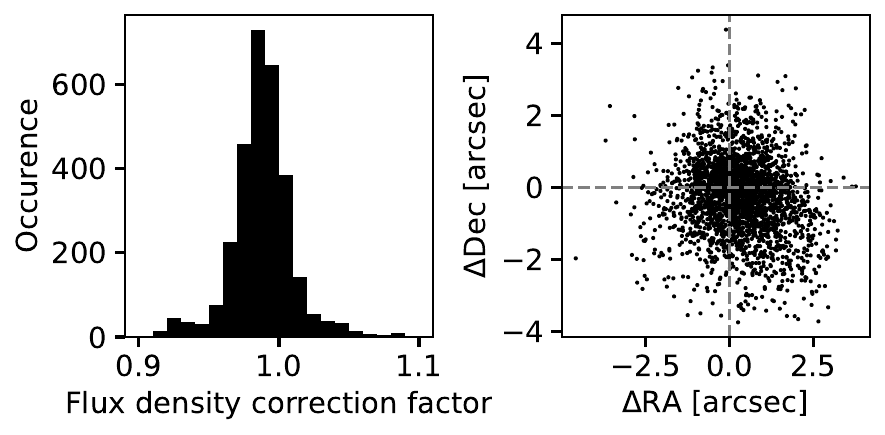}
    \caption{Post-processing corrections for each of the 2945 VAST Extragalactic DR1 observations. Left: flux density correction factor. Right: right ascension and declination corrections.}
    \label{fig:PP_factors}
\end{figure}

The astrometry correction in the post-processing effectively ties the astrometry of all VAST observations to the RACS-low2 astrometry. This is important for the VAST pipeline (Section \ref{sec:VAST_pipeline}), where we build light curves by associating new measurements to sources detected in previous epochs, based on their position. The post-processing reduces the position scatter of individual measurements of a single source, decreasing the number of false associations. The right side of Figure \ref{fig:PP_factors} shows the astrometry corrections that are applied to all 2945 observations in the VAST Extragalactic Survey. Some observations have astrometry corrections of over $3\arcsec$, showing why these corrections are important to ensure proper associations of measurements in the VAST pipeline.

\subsubsection{Flux density scale corrections} \label{sec:pp_flux_corrections}
We perform a flux density correction in the post-processing to account for any flux density offsets that might be present in an individual VAST observation (for example due to observing/ionospheric conditions or calibration solutions). Outlier observations would introduce artificial variability to all sources in a field if unaccounted for. We use the RACS-calibrator sample to scale the VAST integrated flux density to the RACS integrated flux density. 

We use the integrated flux density rather than the peak flux density to account for possible beam shape differences between the VAST and RACS observations. We calculate a single scaling factor per observation that is applied to the integrated and peak flux density of the sources detected in that observation.  The left side of Figure \ref{fig:PP_factors} shows the flux density correction factors that are applied per observation in VAST Extragalactic DR1. 
The flux scaling factor is calculated by fitting the VAST integrated flux density as a function RACS-low2 integrated flux density for the RACS-calibrator sources. We use a Huber regressor to fit the data, as this statistic is more robust against different levels of variance and outliers, compared to the more commonly used orthogonal distance regressor \citep{huber1992robust}. We opt against simply using the mean or median of the histogram of the ratio between the VAST and RACS flux density, as the errors on these quantities are ill-defined for skewed distributions.

\begin{figure}[h]
    \centering
    \includegraphics[width=\linewidth]{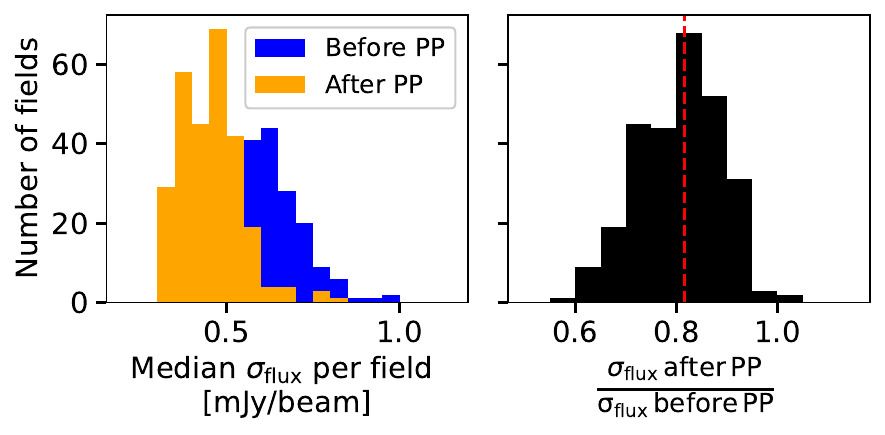}
    \caption{Left: Median root mean square deviation ($\sigma_{\rm{flux}}$) of RACS-calibrator sources in a single field over the duration of VAST Extragalactic. The blue distribution shows the median $\sigma_{\rm{flux}}$ before post-processing, while the orange distribution shows the median $\sigma_{\rm{flux}}$ after post-processing. Right: ratio between the $\sigma_{\rm{flux}}$ after and before post-processing. The red dashed line shows the median of this distribution at $0.82$.}
    \label{fig:PP_flux}
\end{figure}

We quantify how the flux density correction process impacts the light curve variability by calculating the root mean square deviation, $\sigma_{\rm{flux}}$, in the light curve for all sources in the RACS-calibrator sample. In other words, how much do the individual measurements in the light curve deviate from the mean flux density? $\sigma_{\rm{flux}}$ is defined as 
\begin{equation}
    \sigma_{\rm{flux}} = \sqrt{\frac{\sum^{N_T}_{t=1} \left( S_t - \overline{S_t} \right)^2}{N_T}}
\end{equation}
with $N_T$ the number of samples in the light curve, $S_t$ the integrated flux density at a given time $t$ and $\overline{S_t}$ the mean integrated flux.
The left panel of Figure \ref{fig:PP_flux} shows the median root mean square deviation for RACS-calibrator sources in a single observation over the duration of VAST Extragalactic before (blue) and after (orange) post-processing. The right panel of Figure \ref{fig:PP_flux} shows the ratio between the median $\sigma_{\rm{flux}}$ after and before post-processing. The flux density correction in post-processing reduces the median $\sigma_{\rm{flux}}$ to $0.82 \pm 0.08$ of its original value.

This single flux density correction factor removes artificial variability that is present for all sources in a single observation. The left side of Figure \ref{fig:PP_factors} shows that for most VAST Extragalactic DR1 observations, the flux density correction factors are close to 1.0. However, there are a few observations where the flux density correction factor corrects for flux density variations of $5-10\%$. If unaccounted for, these flux density variations could be interpreted as real astrophysical variability. Applying a flux density correction allows us to probe deeper in a search for astrophysical variability.

\subsubsection{Cropping and compression} \label{sec:pp_crop_compress}
In the final post-processing step, the VAST images and catalogues are cropped to include only the inner $6.67 \times 6.67 ^{\circ}$ of each observation. This is visualised by the black dashed line in Figure \ref{fig:example_img_cutouts}. We choose this cut-off as it is the point where the noise, on average, increases by a factor of two compared to the innermost region of the image. As detailed in Section \ref{sec:obs_survey_footprint}, adjacent fields are at most $6.2^\circ$ apart. Thus, the cropping introduced here does not introduce any gaps in the survey footprint; at all declinations there will be overlap between adjacent post-processed VAST fields. After cropping, we compress the images using the \textsc{astropy} CompImageHDU module with a quantisation level of 16. The cropping and compression reduce the image size from 700 MB to 80 MB. \\

The post-processed images and catalogues are available through CASDA under project code AS207 at \url{https://doi.org/10.25919/7597-df49}.

\subsection{Image quality} \label{sec:image_quality}

\begin{figure}[ht!]
    \centering
    \includegraphics[width=\linewidth]{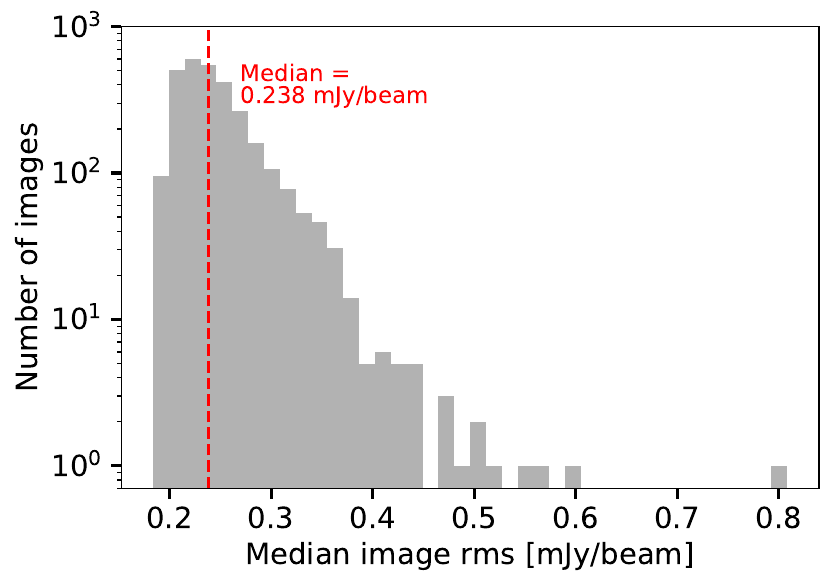}
    \caption{Distribution of median image rms values (computed over the central half of the image), for each observation in VAST Extragalactic DR1.}
    \label{fig:image_rms_values}
\end{figure}

\begin{figure*}[ht!]
    \centering
    \includegraphics[width=\textwidth]{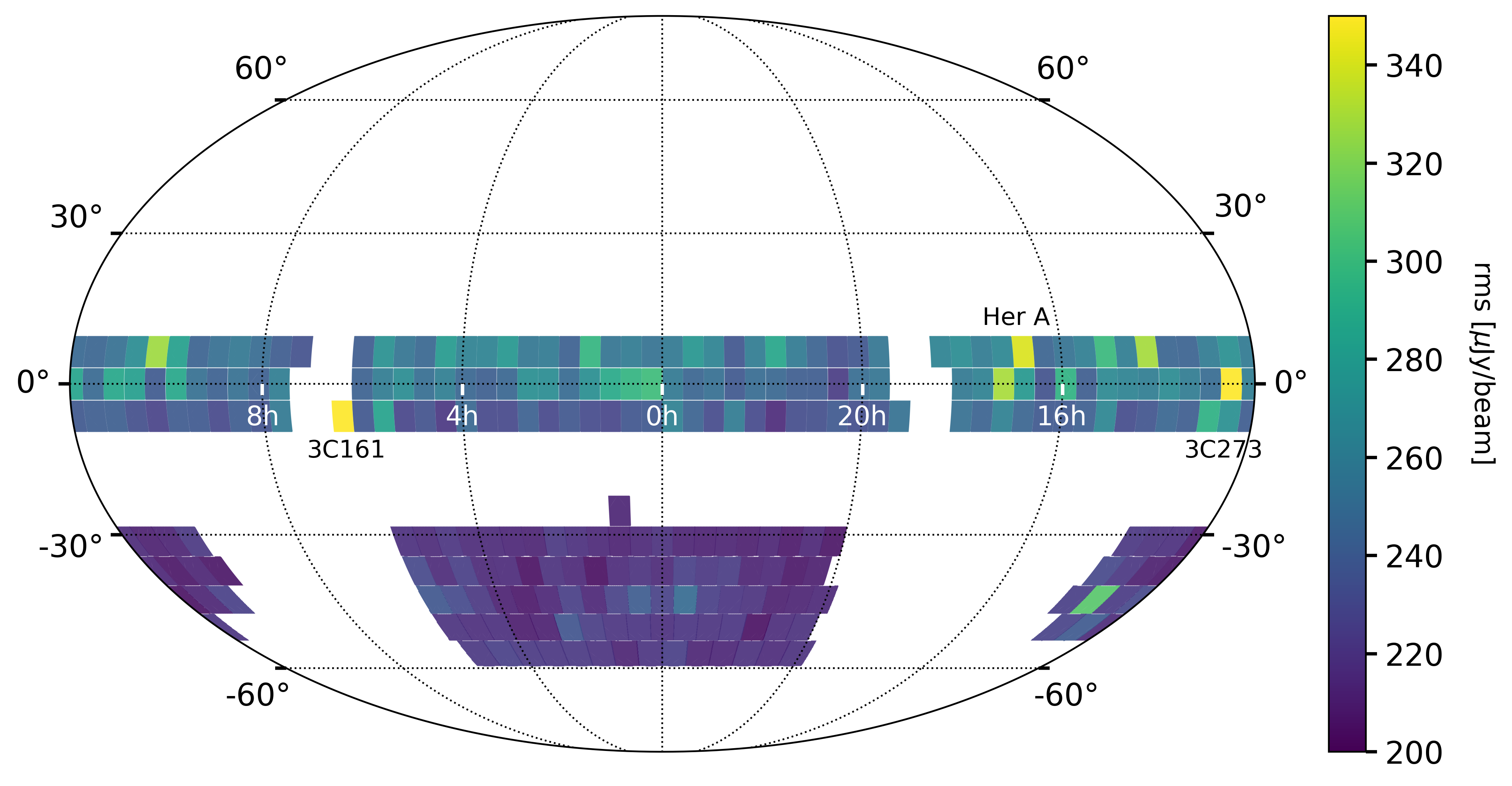}
    \caption{Distribution of image rms values for each field after post-processing, showing the median of all observations for a single field. The yellow fields (with high rms) contain bright radio sources 3C161, Her A and 3C273, respectively.}
    \label{fig:avg_rms_VAST_extragalactic}
\end{figure*}

A histogram of the rms noise in each Stokes I VAST Extragalactic image is shown in Figure \ref{fig:image_rms_values}. The median rms is 0.24 $\rm{mJy\, beam}^{-1}$. This rms is influenced by source confusion, sidelobe confusion, and deconvolution artefacts.  
Figure \ref{fig:avg_rms_VAST_extragalactic} shows the median rms value over all observations for each field. An increased noise at higher declination is consistent with what was observed in RACS-mid \citep{duchesne2023rapid}. There are three fields with significantly higher noise values, which can be attributed to particularly bright sources in the field. For example, \verb|VAST 1237+00|, an equatorial field around RA=12:36:00, shows strong artefacts in the North-West portion of the field, around 3C273. This is a bright quasar with an integrated Stokes I flux density of around 64 Jy, which causes deconvolution artefacts over a large part of the image. The typical resolution of these images is 12–-20$\arcsec$.

\subsection{Absolute flux density scale} \label{sec:absolute_flux_scale}

We compare the overall flux density scale of VAST Extragalactic DR1 to SUMSS \citep{mauch2003sumss}, since the surveys are observed at roughly similar frequencies of 843 MHz and 888 MHz, respectively, and the survey footprints overlap. The SUMSS fluxes have been calibrated against the Molonglo calibrator sample and have been found to be consistent with NVSS \citep{hunstead2000long}. Following the approach outlined by \cite{hale2021rapid}, comparing VAST to SUMSS minimises the effect of a spectral index correction for flux density comparisons. Assuming an overall spectral index of $\alpha =-0.8\pm0.1$, we expect flux density offsets of only $\pm 0.5\%$ when comparing VAST and SUMSS, whereas comparing VAST to a 1.4 GHz survey such as NVSS would impose a $\pm5\%$ flux density error due to the spectral index scaling.

\begin{figure}
    \centering
    \includegraphics[width=\linewidth]{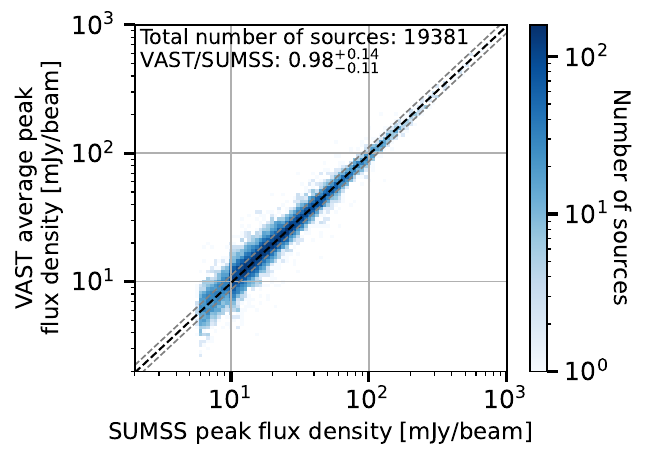}
    \caption{Flux density comparison between the average VAST peak flux density and the SUMSS peak flux density at a frequency of 888 MHz (assuming $\alpha=-0.8$). The black dashed line indicates the median of the flux density ratio distribution, while the grey dashed lines indicate the 16th and 84th percentiles.}
    \label{fig:absolute_flux_SUMSS}
\end{figure}

We assess the absolute flux density scale of VAST sources. The filtering steps we undertake to ensure a reliable sample of sources and light curves are detailed in Section \ref{sec:selection_criteria}. We match these VAST sources to SUMSS, including only VAST sources that are highly compact (integrated flux density over peak flux density ratio below 1.2) and have no forced flux density extractions (i.e. all their flux density measurements are over $5\sigma$) and finally only SUMSS-VAST matches that are within $5\arcsec$. This yields a sample of 19381 SUMSS-VAST source pairs. Figure \ref{fig:absolute_flux_SUMSS} shows the average VAST peak flux density as a function of the SUMSS peak flux density for each of these sources. The SUMSS peak flux density is scaled to VAST observing frequencies, assuming a spectral index of $\alpha=-0.8$. The median flux density ratio is $0.98^{+0.14}_{-0.11}$. The associated errors indicate the 16th and 84th percentile of the flux density ratio distribution. Lines with these slopes are overplotted on the distribution in Figure \ref{fig:absolute_flux_SUMSS}.

\subsection{Absolute astrometry} \label{sec:absolute_astrometry}

The absolute astrometry in VAST depends on the absolute astrometric accuracy of RACS-low2 (Duchesne et al. in prep). There are two main considerations in the RACS astrometry: the astrometric precision and astrometric accuracy. The astrometric precision is determined by the ASKAP PAF beam-to-beam variations, which were found to be $\lesssim 2.5\arcsec$. This was noted in \cite{mcconnell2020rapid}, and detailed in Figure 36 in \cite{duchesne2023rapid}. 
The astrometric precision implies that the overall astrometry scale of RACS (and thereby VAST) cannot be much better than this $2.5\arcsec$ limit. This limitation comes from a lack of phase referencing to a suitable catalogue during processing. A new catalogue based on RACS-low3 data with astrometric calibration will be used for phase referencing in future VAST data releases (see Section \ref{sec:processing_changes}).\\

The astrometric accuracy is determined by cross-matching to external surveys. When cross-matching RACS-low1 to the International Celestial Reference Frame 3 (ICRF3) catalogue \citep{charlot2020third}, it was found that there is a systematic offset between the two; $\Delta \rm{RA} \cdot \rm{cos(Dec)} = -0.6\pm0.6\arcsec$ and $\Delta \rm{Dec} = -0.4\pm0.7\arcsec$ \citep{mcconnell2020rapid}. Similar bulk offsets and scatter are found in RACS-mid \citep{duchesne2023rapid} and RACS-low2 (Duchesne et al. in prep). The cause of this offset is currently not understood.

We evaluate the astrometric accuracy of the VAST sources presented in this data release. The filtering steps we undertake to ensure a reliable sample of sources and light curves are detailed in Section \ref{sec:selection_criteria}. We cross-match these VAST sources with the ICRF3. The ICRF3 consists of 4588 radio sources with accurate positions measured using very long baseline interferometry at frequencies of 8.4 and 2.3 GHz. The median positional uncertainty is about 0.1 mas in right ascension and 0.2 mas in declination. We cross-match VAST to the ICRF3, yielding 76 matched pairs that are separated by less than $3\arcsec$. The number of match pairs is insufficient to be robust against the ASKAP PAF beam-to-beam astrometry variations. To obtain a larger number of matches, we also cross-match the VAST sources against the VLASS epoch 1 Quick Look catalogue \citep{gordon2021quick}. The VLASS catalogue consists of 3.4 million sources north of declination $-40^{\circ}$ observed at 3 GHz. The astrometric accuracy is around $1\arcsec$, increasing to $0.5\arcsec$ above declinations $-20^{\circ}$ (VLASS Quick Look catalogue user guide). We cross-match VLASS to VAST, yielding 183793 matched pairs that are separated by less than $5\arcsec$. We note that this cross-matched sample of source only includes sources with declinations North of $-40^{\circ}$ (VLASS coverage).

\begin{figure}[ht]
    \centering
    \includegraphics[width=\linewidth]{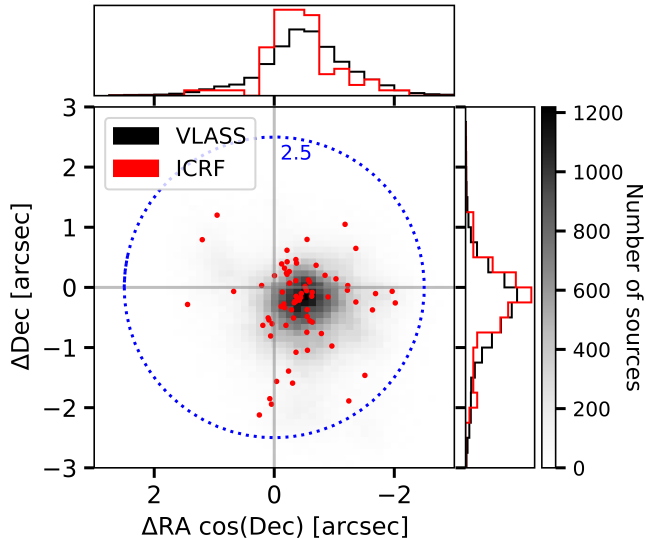}
    \caption{Astrometric offsets for sources in VAST Extragalactic DR1 (selection criteria detailed in Section \ref{sec:selection_criteria}) compared to VLASS (black) and ICRF3 (red). The blue dotted circles show radii of $1.2\arcsec$ and $2.5\arcsec$. }
    \label{fig:absolute_astrometry}
\end{figure}

Figure \ref{fig:absolute_astrometry} shows the astrometric offset between the VAST sources that are cross-matched to ICRF3 (red) and VLASS (black).
For the VAST to ICRF3 cross-matched sources we find $\Delta \rm{RA} \cdot \rm{cos(Dec)}  = -0.4 \pm 0.4$ and $\Delta \rm{Dec} = -0.1 \pm 0.5$. For the VLASS cross-matched sources we find $\Delta \rm{RA} \cdot \rm{cos(Dec)} = -0.5 \pm 0.6$ and $\Delta \rm{Dec} = -0.3 \pm 0.6$. The median systematic offset for both samples is similar to the bulk offset noticed in the RACS surveys (\cite{mcconnell2020rapid, duchesne2023rapid}, Duchesne et al. in prep). In the following, we consider the astrometric offsets of VAST compared to VLASS, as these numbers seem most consistent with the RACS offsets, and the ICRF3 cross-matched sample consists of only 76 sources.


We aim for a conservative and pragmatic astrometry scale to use when cross-matching VAST sources to multi-wavelength counterparts. To this end, we calculate the $3\sigma$ offset in the direction with the largest median offset and scatter. This is the right ascension offset, $-0.527-3\cdot 0.646 \approx 2.5\arcsec$.
Although the actual astrometric accuracy is not uniform, i.e., different systematic offset and variance in different directions, we conservatively use this $2.5\arcsec$ value as a cross-match radius. Additionally, we keep in mind that most beam-to-beam variations (astrometric precision) were contained within $2.5\arcsec$. \textbf{This $2.5\arcsec$ absolute astrometric uncertainty should be added in quadrature to the VAST averaged positional uncertainty ($\sigma_{\rm{pos}}$). One will often find that the absolute astrometric uncertainty dominates.}

We note that for individual sources of interest, it is possible to refine the astrometry by performing local corrections, for example with software like \textsc{fits\_warp} \citep{hurley2018distorting}, which calculates local astrometry corrections to radio images by cross-matching the radio source locations to optical or infra-red surveys.

\section{Light curve database} \label{sec:light_curve_database}

\subsection{Pipeline} \label{sec:VAST_pipeline}
The overall design of the VAST transient pipeline \citep{vast_pipeline} is described in detail in \cite{murphy2021askap} and \cite{pintaldi2021scalable}. For completeness, only a short description of the most important steps and concepts is provided here. 

The main goal of the VAST pipeline is to associate sources from different observing epochs with each other via positional cross-matching. The pipeline takes as input a set of images, rms maps and mean background maps (all as FITS files) as well as the \textsc{Selavy} source finder component list. These data products are all default outputs from the \textsc{ASKAPSoft} continuum imaging process, which means the VAST pipeline can be run immediately after the \textsc{ASKAPSoft} data reduction and VAST post-processing. Once sources have been identified, forced flux density extractions are performed for previously detected sources that are not detected in every subsequent epoch. These forced flux density measurements are obtained by fitting a two‑dimensional Gaussian at the average source position in any observation lacking a $>5\sigma$ detection. The fit solves only for the Gaussian peak amplitude, while the shape and orientation are fixed to match the PSF of the observation. We note that due to the nature of radio images, these fitted flux densities can be negative. This results in a light curve database for every source detected in one or more epochs, and a set of variability metrics.

\subsection{Variability metrics} \label{sec:variability_metrics}
The VAST pipeline calculates two variability metrics ($\eta$ and $V$) for each light curve. These metrics are commonly used to identify highly variable and transient sources \citep{swinbank2015lofar, rowlinson2019identifying}. The flux density coefficient of variation over N measurements is defined as the ratio of the mean flux density to the sample standard deviation $s$, 
\begin{equation} \label{eqn:V}
    V = \frac{s}{\overline{S_t}} = \frac{1}{\overline{S_t}} \sqrt{\frac{N}{N-1} \left( \overline{S_t^2} - \overline{S_t}^2\right)}
\end{equation}
with $S_t$ an individual flux density measurement with uncertainty $\sigma_t$. The significance of the flux density variability, $\eta$, is defined as a reduced-$\Chi^2$ expression in conjunction with the weighted mean flux density $\xi_{S_t} = \frac{\sum^n_{t=1} S_t/\sigma_t^2}{\sum^n_{t=1} 1/\sigma_t^2}$
\begin{equation}
    \eta = \frac{1}{N-1} \sum_{t=1}^N \frac{\left(S_t - \xi_{S_t}\right)^2}{\sigma_t^2}
\end{equation}
Sources that have both a high variability amplitude (V) and a high reduced $\Chi^2$ statistic when compared to the weighted mean flux density ($\eta$), are the most variable sources in the sample.

\subsection{Selection criteria}\label{sec:selection_criteria}
The VAST pipeline run over the 2945 VAST Extragalactic images in this DR1 results in 27.5 million individual measurements for 2.3 million sources. However, we disregard most of these sources ($\sim 76\%$), as they are likely to be spurious or are found at low signal-to-noise ratio. We filter out 
\begin{enumerate}
    \item sources that have fewer than five measurements;
    \item false `sources' that are created based on a single flux density measurement that is due to part of an image having relatively high rms noise;
    \item sources caused by imaging artefacts around bright sources due to incomplete dirty beam subtraction;
    \item extended sources, which can be decomposed into multiple `sources' by the source finder, making the flux density measurements unreliable;
    \item sources found at low signal-to-noise ratios.
\end{enumerate}
Below we detail the consecutive filters and strategies we impose to reduce the spurious sources in our dataset. We suggest future improvements to this filtering approach in Section \ref{sec:disc_improve_lc_database}.

\subsubsection{Filtering out sources with few measurements} \label{sec:filtering_few_measurements}
We filter out sources that have fewer than five measurements, because the field has been observed fewer than five times. As can be seen from Figure \ref{fig:nof_obs_per_field}, this effectively removes the sources in the single field \verb|VAST 1237+00| that was removed from the VAST Extragalactic footprint due to an extremely bright source in the field (3C273). We disregard these light curves, as the sampling for these sources is completely different from the rest of the light curves in the database, which makes it difficult to compare variability statistics.

\subsubsection{Filtering out poor quality measurements} \label{sec:filtering_bad_measurements}
Some VAST images are partially corrupted due to RFI, a bright 3C or MS4 source in the field \citep{edge1959survey, burgess2006a}, or otherwise unfavourable observing conditions. These images are still included in the VAST pipeline run if the majority of the image has low noise and therefore usable flux density measurements. The local noise distribution of the 27.5 million measurements is highly skewed, with a peak around $0.25$ mJy $\rm{beam}^{-1}$, a minimum of $0.14$ mJy $\rm{beam}^{-1}$ and a maximum of $452$ mJy $\rm{beam}^{-1}$. To filter out unreliable measurements, we apply a 99.7 percentile local noise cut ($1.82$ mJy $\rm{beam}^{-1}$), eliminating 82711 measurements from our light curve database. These high-noise measurements often produce spurious flux density outliers and artificial variability. We recalculate the variability metrics for all sources (see Section \ref{sec:variability_metrics}) after excluding these unreliable measurements.


\subsubsection{Filtering out imaging artefacts around bright sources}\label{sec:filtering_artefacts}
\begin{figure}
    \includegraphics[width=\textwidth]{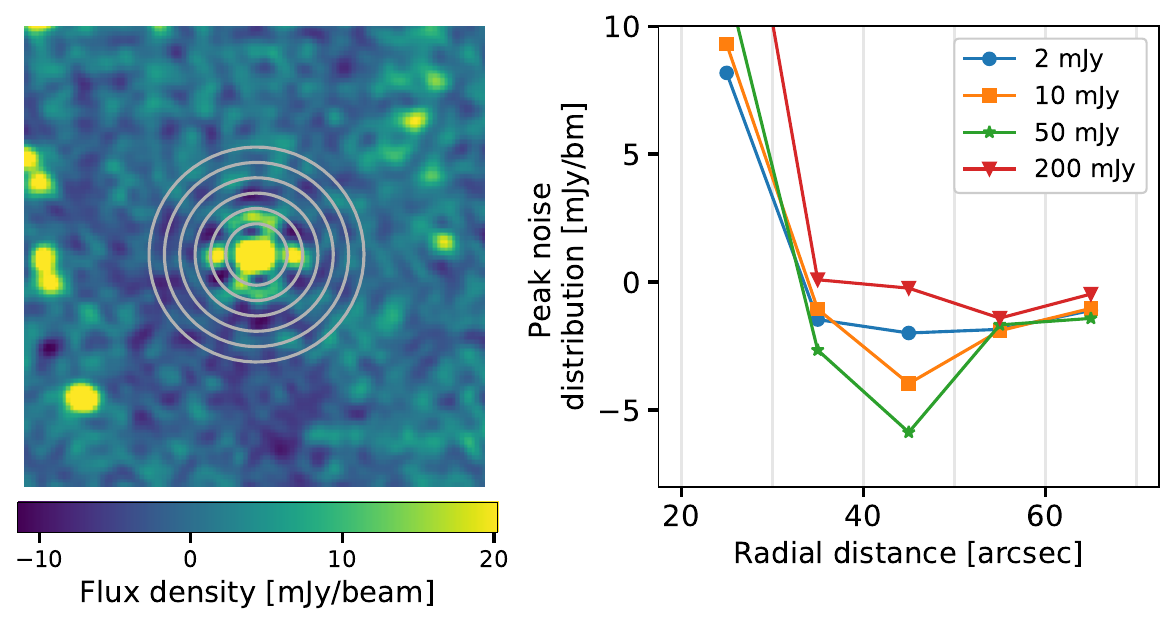}
    \caption{Left: $5\arcmin \times 5 \arcmin$ cutout of a summed stack of 100 2 mJy isolated compact sources. This figure is made for a stack of 10 mJy, 50 mJy and 200 mJy sources as well. Right: peak of the noise distribution in each annulus drawn in the left figure.}\label{fig:noise_annuli}
\end{figure}

To limit contamination from imaging artefacts near bright sources, we require each source to have a nearest neighbour more than $30\arcsec$ away. This is roughly two times the median restoring beam size, which varies between $12\arcsec$ and $20 \arcsec$ and peaks at $15 \arcsec$. We assess this threshold using a sample of truly compact, isolated RACS-low2 sources (Duchesne et al. in prep). This sample of sources has an `S' source code in RACS (meaning they are fit by a single Gaussian), and a nearest neighbour source more than $2\arcmin$ away. We overlay truly compact sources from this sample across integrated flux densities of 2, 10, 50, and 200 mJy. The left panel of Figure \ref{fig:noise_annuli} shows an overlay of a sample of 100 truly compact 2 mJy sources on top of each other. The right panel of Figure \ref{fig:noise_annuli} shows how the peak of the noise distribution changes in radial annuli around the source. Only the stack of 2 mJy sources is shown here as an example. The largest change in noise distribution occurs between the $20-30\arcsec$ and $30-40\arcsec$ annuli for all flux density bins, supporting our choice of a $30\arcsec$ cutoff. A more elaborate filtering strategy could be designed by setting a larger radial filter around brighter sources, but that is beyond the scope of this paper.

We find that $\sim 1$ million sources are filtered out by this distance to nearest neighbour criterion alone (i.e. they satisfy the extended sources and signal-to-noise ratio filter below), effectively reducing the survey size by about 220 $\rm{deg}^2$.

\subsubsection{Filtering out extended sources}\label{sec:filtering_extended sources}
We filter out extended sources using the compactness value - the ratio of integrated to peak flux density - and adopt a relatively high threshold of 2.2 to retain transients in slightly extended radio galaxies. This choice is motivated by the compactness values of giant radio-galaxy cores \citep{koribalski2025askap}, where nuclear transients may occur.

\subsubsection{Filtering out low signal-to-noise ratio sources} \label{sec:filtering_low_snr}
Finally, we only include sources that have been detected with a signal-to-noise ratio larger than 8 in our sample, implying that at least one measurement in the source light curve has a signal-to-noise ratio above 8. This is to decrease the probability of finding noise spikes as `sources' with a single detection over the duration of the VAST Extragalactic Survey. As a last step, we further reduce the number of spurious sources by disregarding sources with a negative average integrated flux density.\\

\textbf{After all filtering steps detailed in this section, we are left with a sample of 0.55 million sources with 6.4 million measurements.}

\subsection{Pipeline products and data availability} \label{sec:dataproducts}
The main output of the transient pipeline is a light curve database that consists of a measurement, source and image table. The measurements table contains detailed information about each \textsc{Selavy} or forced flux density measurement per source and per timestamp. The source table contains the statistics per source averaged/calculated over all measurements, such as average location, maximum and average integrated/peak flux density, and the variability parameters.  Finally, we include a table with detailed information on the images and observations that are part of VAST Extragalactic DR1. These tables represent the light curve database after applying the filtering steps detailed in the previous section, resulting in 0.5 million light curves. \ref{app:light_curve_database} describes the light curve database in detail. The light curve database tables are presented as CSV and \textsc{parquet} files \citep{vohra2016apache} which can be read by data exploration libraries such as \textsc{Pandas} and \textsc{dask}\footnote{\url{https://docs.dask.org/en/stable/dataframe.html}}. To assess the quality and reliability of the light curves, image cutouts are important. For example, it can be the only way to easily identify spurious sources/measurements due to high local noise, or imaging artefacts around bright sources. A FITS cube is available for each source, containing layers of cutouts of the individual detections of the source. These FITS cubes are labelled with the unique source identifier from the light curve database (see \ref{app:light_curve_database}). 

The light curve database and image cutouts for all sources in this data release are available on the CSIRO Data Access Portal (DAP) under the project 'VAST Extragalactic DR1: light curve database and cutouts': \url{https://doi.org/10.25919/nh9d-t846}. 
This collection additionally provides an example notebook demonstrating how to access and work with the light‑curve database and cutouts.\\

We encourage the wider community to make use of the light curves included in this data release. However, we urge the user to consider the following caveats:
\begin{enumerate}
    \item This data release will contain spurious sources and measurements, despite the filters detailed in the previous section. We encourage users to inspect the image cutouts for their sources, to ensure that the light curves only include reliable flux density measurements.
    \item This data release does not contain light curves for all sources in the VAST Extragalactic survey footprint. By filtering out all sources with a nearest neighbour within $30\arcsec$, we exclude both bright sources that are decomposed into multiple components by the source finder, and any otherwise reliable sources that happen to have another source close by. Additionally, we exclude faint and extended sources in this data release.
\end{enumerate}

We remind the reader that the following data products are available through CASDA under project code AS207:
\begin{itemize}
    \item the measurement sets and images per field (Stokes IQUV) \\ \url{http://hdl.handle.net/102.100.100/473995?index=1}
    \item the \textsc{Selavy} source catalogues (Stokes I) \\ \url{http://hdl.handle.net/102.100.100/473996?index=1}
    \item the images and \textsc{Selavy} source catalogues per field after post-processing (Stokes I) \\ \url{https://doi.org/10.25919/7597-df49}
\end{itemize}
We do not release rms maps and background maps (see \citealt{murphy2021askap}), as we expect limited interest in these data products. However, in case these are of use for your science goals, we encourage you to join the VAST collaboration\footnote{\url{https://vast-survey.org/Team/}}.

\section{Untargeted variability search} \label{sec:untargeted_variability_search}
\begin{figure}
    \centering
    \includegraphics[width=\linewidth]{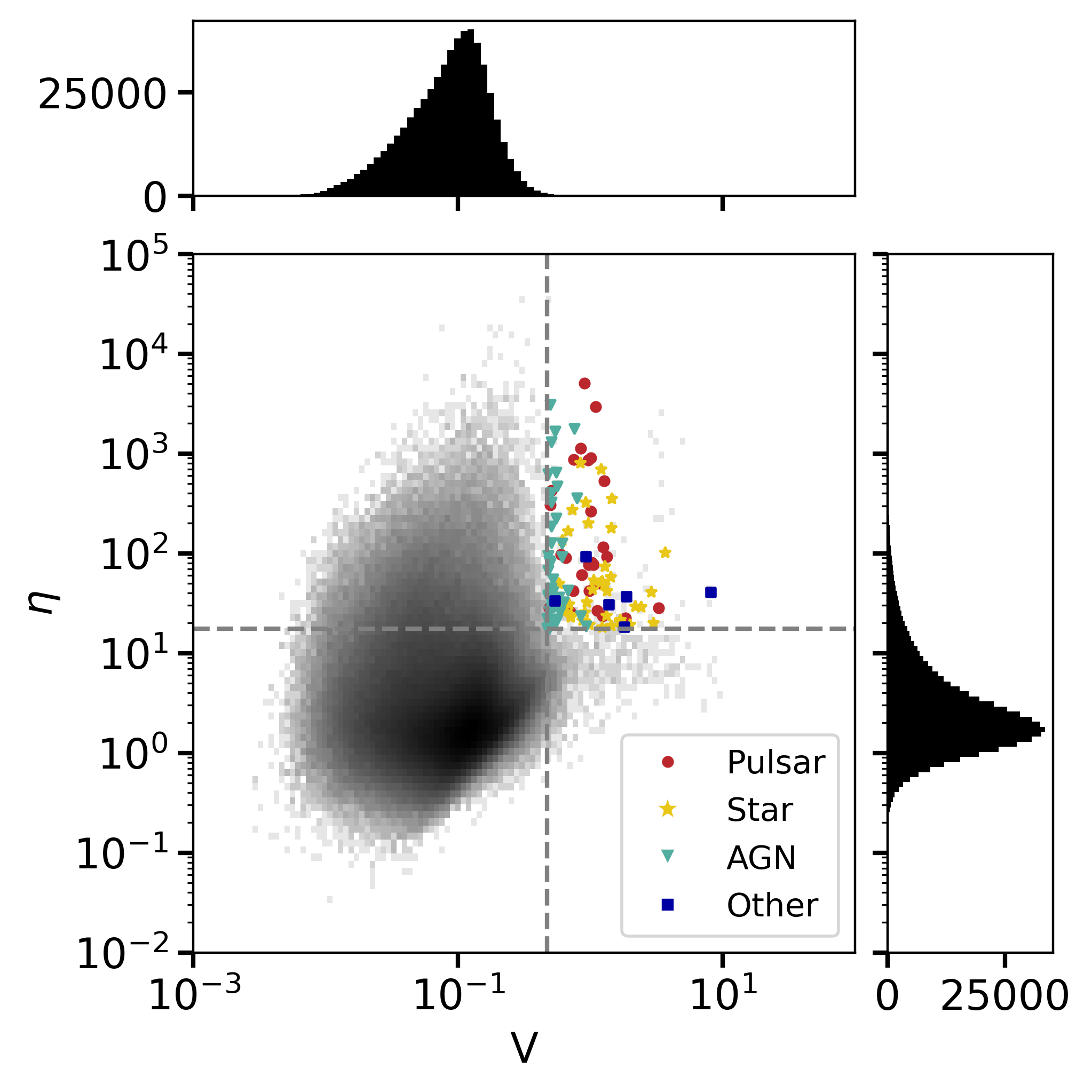}
    \caption{Variability metrics, $\eta$ and V, for the 0.5 million sources in VAST Extragalactic DR1 light curve database. The dashed grey lines indicate the $2.5\sigma$ thresholds on $\eta$ and V calculated by fitting a Gaussian function to the sigma-clipped distributions of each metric. Sources that have been classified as real variables have been marked in various colours.}
    \label{fig:eta_V}
\end{figure}

\begin{figure}
    \centering
    \includegraphics[width=\linewidth]{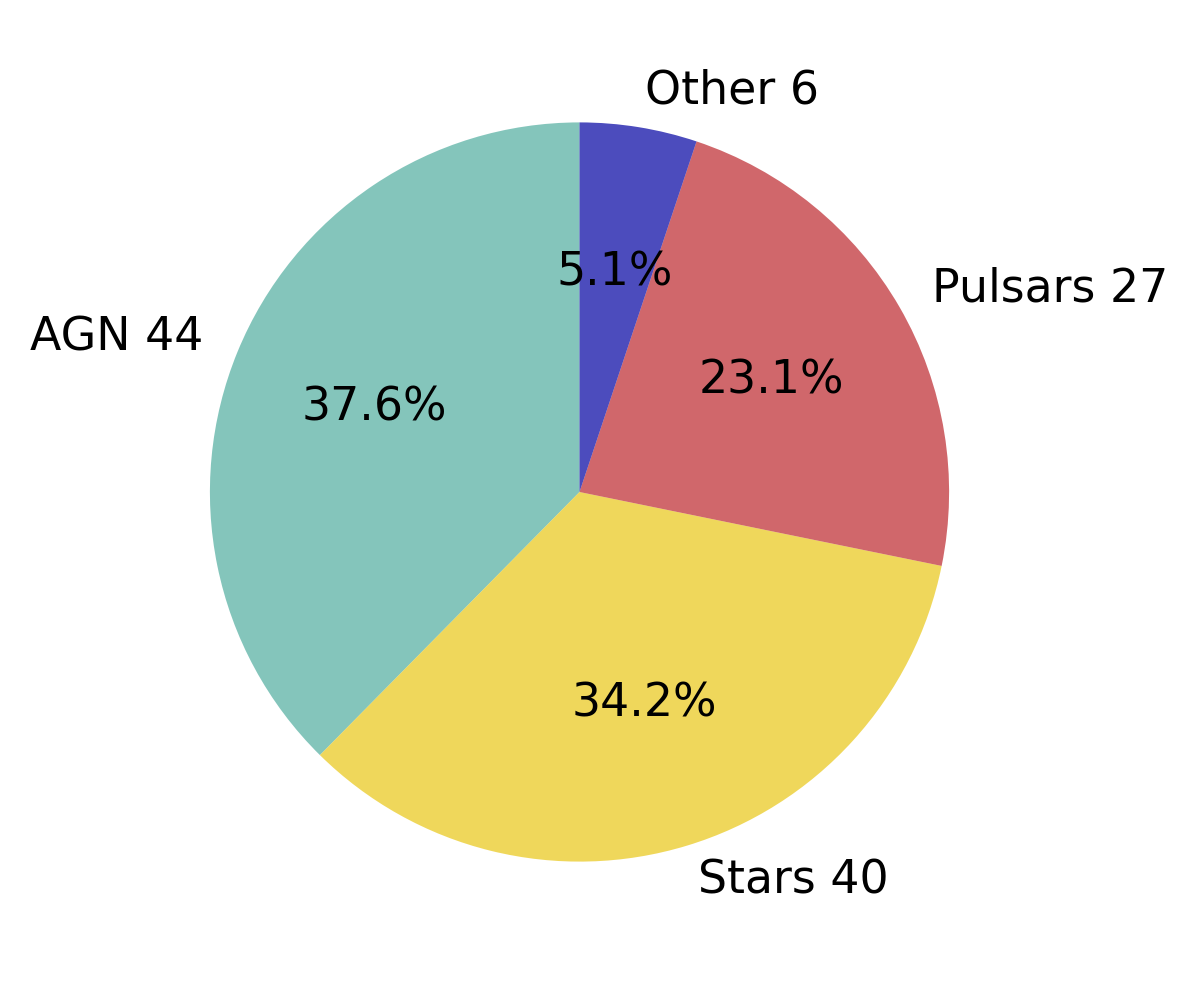}
    \caption{Source class distribution for the 117 transient and variable sources found using the $\eta,V$ search. These are the 126 transients and variables listed in Table \ref{tab:results}, where we exclude the 9 solar system planet detections. The classification schema is detailed in the text.}
    \label{fig:class_pie_chart}
\end{figure}

Figure \ref{fig:eta_V} shows the distribution of the variability metrics $\eta$ and V, as detailed in Section \ref{sec:variability_metrics}, for the 0.5 million sources in VAST Extragalactic DR1 light curve database. Note that we used the peak flux densities to calculate $\eta$ and V. The dashed grey lines in Figure \ref{fig:eta_V} indicate the $2.5\sigma$ thresholds on $\eta$ and V calculated by fitting a Gaussian function to the sigma-clipped distributions of each metric, corresponding to $0.47$ and $17.6$ for V and $\eta$ respectively. There are 170 sources in the top right of the $\eta$,V-phase space, which are the most variable sources in the sample. The $2.5\sigma$ threshold was chosen as it yields a reasonable number of sources for visual inspection.
After inspection of light curves and cutouts, we found that 44 of 170 are spurious; the result of an artefact near a bright source. Table \ref{tab:results} lists the 126 real astrophysical variables in our sample, indicating the source name, position, position uncertainty, $\eta$-value, V-value, number of measurements, number of forced extractions, the maximum integrated flux density, and a column that shows the star/pulsar the VAST source has been matched with, or in the case of the AGN, in which radio survey it was previously detected. A machine-readable version of this Table can be found in the Supplementary Materials. For VAST variables that are identified as stars, the {\it Gaia} \citep{gaia2016, gaia2023} DR3 identifier \citep{vallenari2023gaia, babusiaux2023gaia} is listed in Table \ref{tab:results}. We detect Mercury and Saturn 9 times, as listed in Table \ref{tab:results}, but exclude these detections from Figures \ref{fig:eta_V} and \ref{fig:class_pie_chart}.

Figure \ref{fig:class_pie_chart} shows the source class distribution for the 117 astrophysical variables (excluding the Solar System planet detections). We find that the majority of the transients and variables in our sample are AGN (38\%), radio stars (34\%) and pulsars (23\%). There is a small fraction of `Other' sources (5\%), consisting of two known supernovae, one supernova candidate, a brown dwarf, and two sources without a catalogued multi-wavelength counterpart. In the following sections, we detail how these sources were classified, show example light curves, and highlight interesting sources. 

Figure \ref{fig:examples_lc_cutouts_per_class} shows a gallery of typical light curves and cutouts for each of the main source classes. Figures \ref{fig:examples_lc_cutouts_per_class}, \ref{fig:lc_cutouts_probable_stars}, \ref{fig:lc_cutouts_SN} and \ref{fig:lc_cutouts_unknowns} show a single source per row. 
The left panel shows the VAST Extragalactic DR1 light curve. Black points are integrated flux density measurements from \textsc{Selavy}. White points represent the forced-fitted flux density for images where there was no \textsc{Selavy} detection. The middle panel shows the radio image of the brightest VAST detection. The ellipse in the lower right corner of the radio image shows the full width at half maximum (FWHM) of the restoring beam. The third panel shows composite images from the DESI Legacy Surveys \citep[DESI-LS;][]{dey2019overview} or the Sloan Digital Sky Survey \citep[SDSS;][]{kollmeier2026sloan}, where we overlay the radio contours of the brightest radio detection. The radio contours follow the 50, 70, and 90 per cent peak flux density. We construct an RGB image with red=$i$-band, g=$r$-band and b=$g$-band. In some instances, no $i$-band image is available; in that case, we use a scaled version of the $r$-band for red.

\subsection{Known Pulsars} \label{sec:pulsar_results}
\begin{figure*}[htbp] 
    \centering 
    
    \begin{subfigure}[b]{0.38\textwidth}
        \centering
        \begin{overpic}[width=\textwidth]{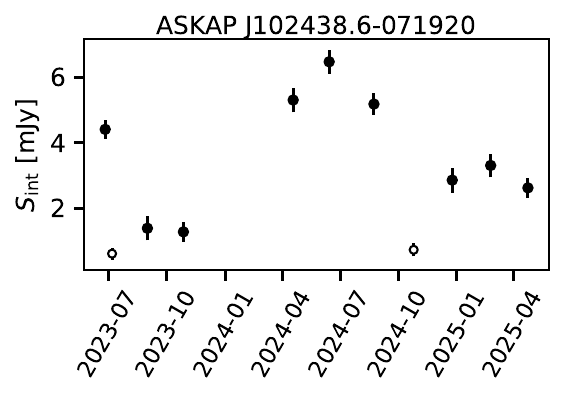}
        \put(0,65){\color{purple}\large Pulsar}
        \end{overpic}
    \end{subfigure}
    \hfill
    \begin{subfigure}[b]{0.6\textwidth}
        \centering
        \includegraphics[width=\textwidth]{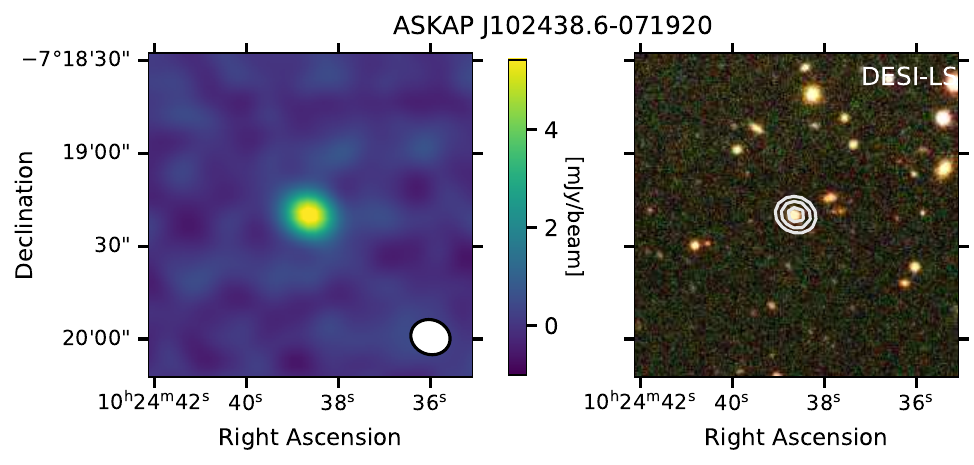} 
    \end{subfigure}

    \begin{subfigure}[b]{0.38\textwidth}
        \centering
        \begin{overpic}[width=\textwidth]{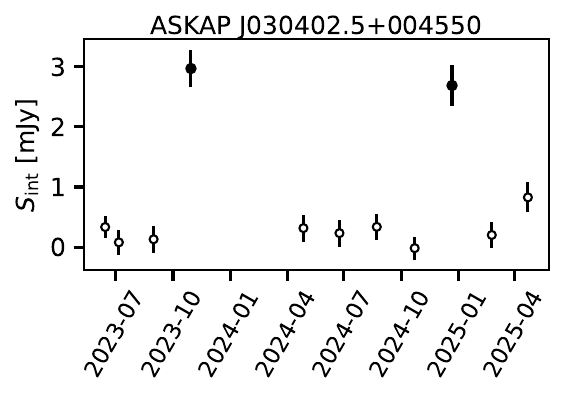}\put(0,65){\color{purple}\large Radio star}
        \end{overpic} 
    \end{subfigure}
    \hfill
    \begin{subfigure}[b]{0.6\textwidth}
        \centering
        \includegraphics[width=\textwidth]{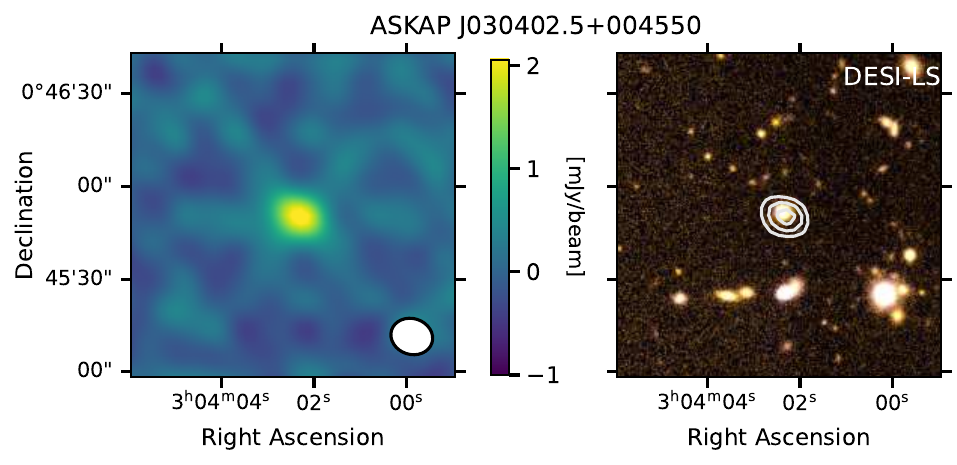} 
    \end{subfigure}

    \begin{subfigure}[b]{0.38\textwidth} 
        \centering
        \begin{overpic}[width=\textwidth]{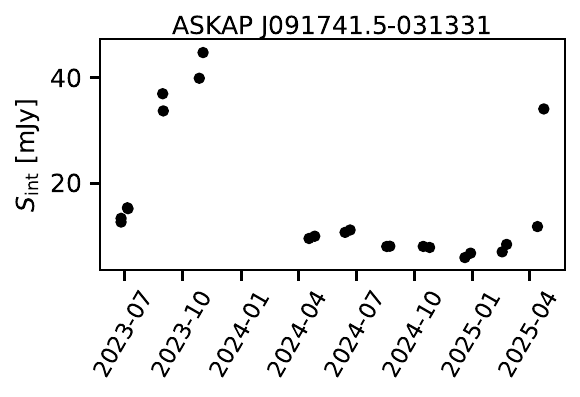} 
        \put(0,65){\color{purple}\large AGN 1}
        \end{overpic}
    \end{subfigure}
    \hfill
    \begin{subfigure}[b]{0.6\textwidth}
        \centering
        \includegraphics[width=\textwidth]{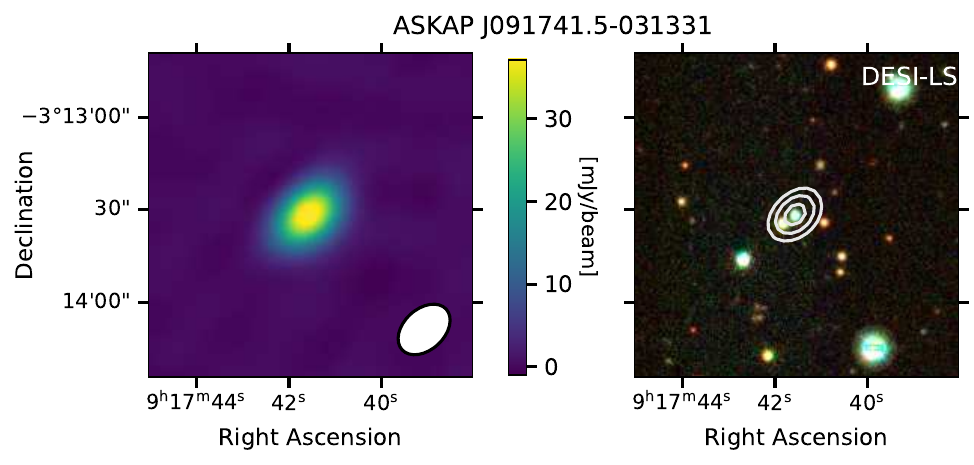} 
    \end{subfigure}
    \vspace{1em} 

    \begin{subfigure}[b]{0.38\textwidth}
        \centering
        \begin{overpic}[width=\textwidth]{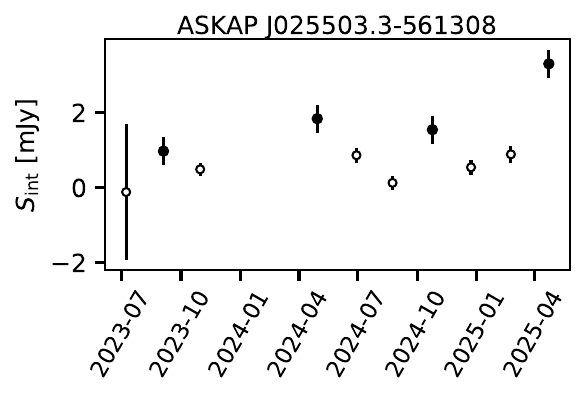}
        \put(0,65){\color{purple}\large AGN 2}
        \end{overpic}
    \end{subfigure}
    \hfill
    \begin{subfigure}[b]{0.6\textwidth}
        \centering
        \includegraphics[width=1.03\textwidth]{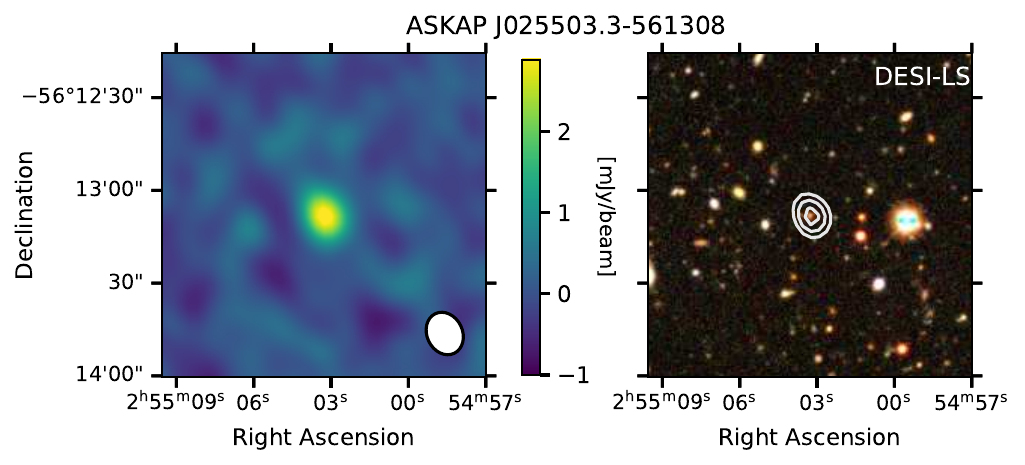} 
    \end{subfigure}

    \caption{Example light curves per source class, each row represents a different source. Left: VAST Extragalactic DR1 light curve. Black points are integrated flux density measurements from \textsc{Selavy}. White points are the forced-fitted flux density for images where there was no \textsc{Selavy} detection. Middle: VAST cutout for the epoch with the maximum flux density, the ellipse in the lower right corner of the radio image shows the FWHM of the restoring beam. Right: RGB image (details in text) of optical data from the Legacy Surveys / D. Lang (Perimeter Institute) with the VAST Stokes I contours at 50\%, 70\% and 90\% of the peak Stokes I flux density overlayed.}
    \label{fig:examples_lc_cutouts_per_class}
\end{figure*}

 All pulsars in the sample of variable sources are identified in an automated manner, by cross-matching the sample to the pulsar survey scraper\footnote{\url{https://pulsar.cgca-hub.org/}} \citep{kaplan2022pss} with a cross-match radius of $10\arcsec$. 27 sources amongst our sample of most variable sources in VAST Extragalactic DR1 are known pulsars. Figure \ref{fig:examples_lc_cutouts_per_class} shows the light curve and cutouts for ASKAP J102438.6-071920, cross-matched to millisecond pulsar PSR J1024-0719, as a representative example of a typical pulsar light curve. For two pulsars, PSR J1556+00 and PSR J0051+0423, corresponding to ASKAP J155540.5+004904 and ASKAP J005129.7+042259, respectively, a larger cross-match radius was required, as the archival pulsar positions have not been determined to sufficient accuracy with timing studies.

Thirteen of the 27 pulsars are confirmed binary millisecond pulsars (MSPs). Most notably, PSR J2222-0137 \citep{boyles2013green} and PSR J2234+0611 \citep{deneva2013goals} are MSPs that have been used for tests of gravity \citep[e.g.][]{guo2021psr, batrakov2024new, stovall2019psr}. 
PSR J2129-0429 \citep{hessels2011350, bellm2016properties} is classified as a redback MSP, while PSR J1227-4853 is a transitional redback \citep{roy2015discovery}. PSR J2051-0827 \citep{stappers1996probing} and PSR J2039-5617 \citep{corongiu2021radio} are eclipsing MSPs. 
In addition, two pulsars are thought to be recycled systems whose binaries were disrupted or transformed. PSR J1320-3512 \citep{belczynski2010double} may have been separated during a second supernova explosion \citep{lorimer2004psr}, while PSR J2124-3358 \citep{reardon2016timing, romani2017asymmetric} may have ablated its companion through spin-down power in an extreme “black widow” phase \citep{ruderman1989accretion}. The remaining MSPs are PSR J1024-0719 \citep{kaplan2016psr, bassa2016millisecond}, PSR J2144-5237 \citep{bhattacharyya2016gmrt, bhattacharyya2019gmrt}, PSR J1337-4441 \citep{spiewak2020survey}, PSR J0034-0534 \citep{bell1995optical}, PSR J2145-0750 \citep{lohmer2004parallax, bell1995optical}, and PSR B0820+02 \citep{kulkarni1986optical}.

\subsection{Radio Stars}

We identify radio stars by cross-matching to within $2.5\arcsec$ of a star in the Sydney Radio Star Catalogue \citep[SRSC;][]{driessen2024sydney}. In addition, we cross-match our highly variable radio sources to the proper-motion-corrected Gaia positions of nearby stars. We classify matches with $2.5\arcsec$ as radio stars, with the exception of ASKAP J014700.3+075128 which has a $3.7\arcsec$ offset, see Section \ref{sec:ASKAPJ014700.3}. We find that 40 of the most variable sources in the VAST Extragalactic DR1 are stars. Figure \ref{fig:examples_lc_cutouts_per_class} shows the light curve and cutouts for ASKAP J030402.5+004550, a highly variable VAST source matched to LP591-256 (Gaia DR3 ID $3266760032372995072$). This light curve is a typical light curve for a radio star. For the cutouts around radio stars (Figures \ref{fig:examples_lc_cutouts_per_class} and \ref{fig:lc_cutouts_probable_stars}), the VAST position of the source has been corrected according to the proper motion of the star, such that the radio detection is corrected to the time of the optical observation.\\

Of the 40 stars in this sample, 30 are known radio stars and 10 are newly detected radio stars. Of the known radio stars, Wolf 359 (aka CN Leo), is not yet in the SRSC but was previously identified as a radio star \citep{1993ApJ...415..236G}. The same holds for HD 34736 \citep{semenko2024hd, das2025vast}. We have counted both of these, and the 28 stars in the SRSC as known radio stars. Some well-known radio stars that are (re)detected in VAST Extragalactic DR 1 are AT Mic and CU Vir. Some of the new radio stars have limited information, but of those with spectral type classifications, they range from M dwarfs to G-type stars. Many of the stars, both new and previously known radio stars, have previously been identified as active with activity indicators such as X-ray emission or Calcium H and K emission. We find that 8 out of the 10 new radio stars have high fractional circular polarisation (between 14 and 83\%). We discuss the two new radio stars with low circular polarisation fractions (ASKAP J014700.3$+$075128 and ASKAP J053954.1-051119) in more detail below. 
\\

\subsubsection{Possible star: ASKAP J014700.3+075128} \label{sec:ASKAPJ014700.3}
\begin{figure*}[htbp]
    \centering 
    
    \begin{subfigure}[b]{0.38\textwidth} 
        \centering
        \includegraphics[width=\textwidth]{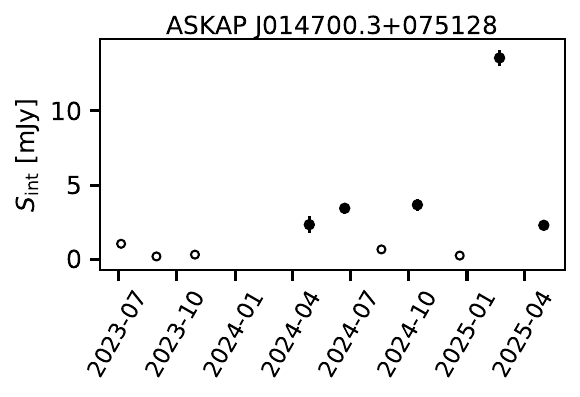} 
    \end{subfigure}
    \hfill
    \begin{subfigure}[b]{0.6\textwidth}
        \centering
        \includegraphics[width=\textwidth]{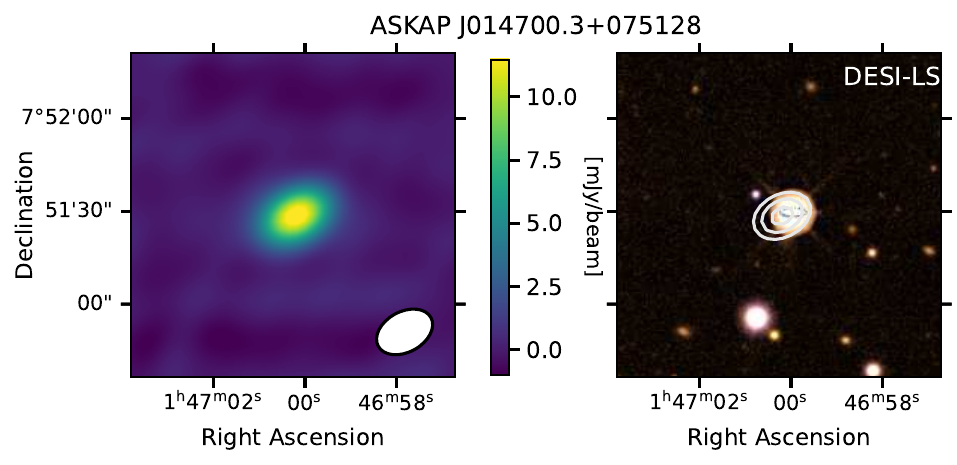} 
    \end{subfigure}
    
    \vspace{1em} 
    \begin{subfigure}[b]{0.38\textwidth}
        \centering
        \includegraphics[width=\textwidth]{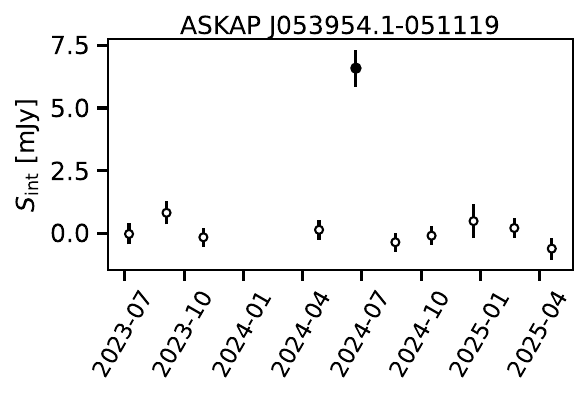}
    \end{subfigure}
    \hfill
    \begin{subfigure}[b]{0.6\textwidth}
        \centering
        \includegraphics[width=\textwidth]{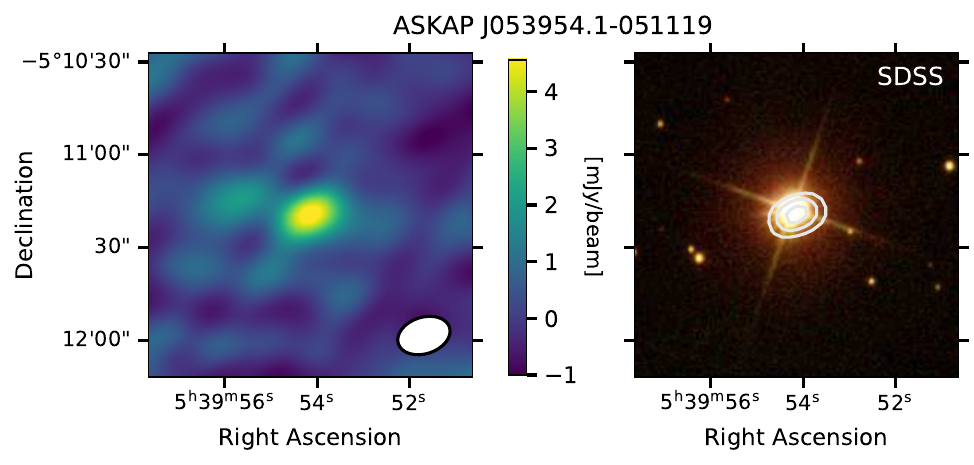} 
    \end{subfigure}
    \caption{Same as Fig. \ref{fig:examples_lc_cutouts_per_class} for the possible radio star ASKAP J014700.3+075128 and the radio star ASKAP J053954.1-051119.}
    \label{fig:lc_cutouts_probable_stars}
\end{figure*}

Figure \ref{fig:lc_cutouts_probable_stars} shows the light curve and cutouts for ASKAP J014700.3$+$075128. Its brightest VAST detection, at 13.7 mJy, lies $3.7\arcsec$ from the Gaia position of a nearby Solar-like star (Gaia DR3 ID $2568706316378318208$; spectral type K3V; $G=12.2$; distance 151 pc; \citealt{gaia2023}), after applying the proper motion correction. This offset exceeds the $2.5\arcsec$ limit from our absolute astrometry (Section \ref{sec:absolute_astrometry}). Inspection of the local astrometry for the VAST observation with the 13.7 mJy detection shows no evidence for significant beam-to-beam systematics (Section \ref{sec:absolute_astrometry}). The offsets between the Gaia position and the remaining VAST detections span $3.5$–$5.3\arcsec$. ASKAP J014700.3$+$075128 is also detected in RACS-mid at 1.05 mJy in May 2022 \citep{duchesne2023rapid}, with a proper motion corrected offset of $2.4\arcsec$. We also produce a dynamic spectrum of the brightest VAST detection using \textsc{DStools}\footnote{\url{https://github.com/askap-vast/dstools}} \citep{Pritchard2025a} (Figure \ref{fig:13805206_DS} in Appendix \ref{app:dynamic_spectra}). The emission is confined to part of the band between 800 and 950 MHz, with no evidence for circularly polarised radiation.

Taken together, these measurements suggest that an association between ASKAP J014700.3$+$075128 and the Gaia star is plausible, particularly given the radio light curve behaviour, which is typical of radio stars. However, the narrow frequency bandwidth of the emission is atypical for a radio star \citep[see e.g.][]{wang2023radio}. In conclusion, improved astrometry and additional monitoring are required to confirm or refute this link, and a scintillating background source remains a viable alternative.

\subsubsection{Probable star: ASKAP J053954.1-051119}
Figure \ref{fig:lc_cutouts_probable_stars} shows the light curve and cutouts for ASKAP J053954.1-051119. The brightest detection of ASKAP J0539 at 6.6 mJy is $2.3 \arcsec$ offset from the Gaia position of a nearby star (Gaia DR3 ID $3023329085698084992$, Gaia magnitude $G=11.0$ at $394$ pc \citep{bailer2021estimating}). The offset between the radio and optical position is quite large, but falls within the $2.5 \arcsec$ limit discussed in Section \ref{sec:absolute_astrometry}. We discuss this star separately because, similar to ASKAP J014700.3$+$075128, we find no evidence for circular polarisation. We make a dynamic spectrum for the flare from ASKAP J053954.1-051119 (see Figure \ref{fig:13829699_DS} in \ref{app:dynamic_spectra}) and do not find any evidence of band-limited behaviour.

\subsection{AGN}
We identify the AGN in our sample by cross-matching to FIRST, SUMSS, NVSS, and the VAST pilot survey. When a source is detected in VAST Extragalactic DR1 and in one of these archival surveys, and there is no stellar or pulsar counterpart, we interpret the variable to most likely be an AGN. This is because a detection in any of the aforementioned archival surveys implies that the source is radio bright over at least 4 years, a variability timescale that is only found for AGN and slow synchrotron transients. 

Using these relatively simple criteria, we classify 44 of our variable sources as AGN. Table \ref{tab:results} lists per AGN candidate in which survey it was detected previously. Note that we list only the first survey we find the source in; we check against archival surveys in the order we list them above. We cross-match sources against the transient name server, and visually inspect each light curve to check for a synchrotron-transient-type evolution. Our final set of AGN light curves all show variability consistent with strong QSO or blazar variability or scintillation. 

Figure \ref{fig:examples_lc_cutouts_per_class} shows two examples of sources classified as AGN. ASKAP~J091741.5-031331 was detected in NVSS with a 1.4\,GHz flux density of 23.6 \,mJy. Its optical counterpart is a known QSO (Gaia DR3 ID $3838208794064660352$) at $z=1.428$ \citep[Gaia QSOC;][]{delchambre2023} and is highly variable in the optical. ASKAP J025503.3-561308 was detected in the VAST pilot survey. An optical/IR  counterpart is visible in both the WISE W1 and W2 images \citep{vizier_wise_2011_cutri, cutri2013explanatory} and the Legacy Survey DR10 \citep{dey2019overview}, with a photometric redshift estimate of $z=0.835\pm0.051$ \citep{zhou2021}. The optical and WISE colours suggest that this is a radio AGN hosted by a massive early-type galaxy. The radio flux density seems to be scintillating around the VAST detection threshold.

\subsection{Other Source Classes}
Six of the highly variable sources do not fall within one of the aforementioned categories. Two of these correspond to the radio counterparts of optically identified supernova, one is a supernova candidate, one is a brown dwarf, and there are two highly variable sources for which we have not been able to determine their origins. We discuss each of these sources briefly below, as most of them are part of (ongoing) follow-up campaigns, and will be described and modelled in detail in separate papers.

\begin{figure*}[htbp] 
    \centering 
    
    \begin{subfigure}[b]{0.38\textwidth} 
        \centering
        \begin{overpic}[width=\textwidth]{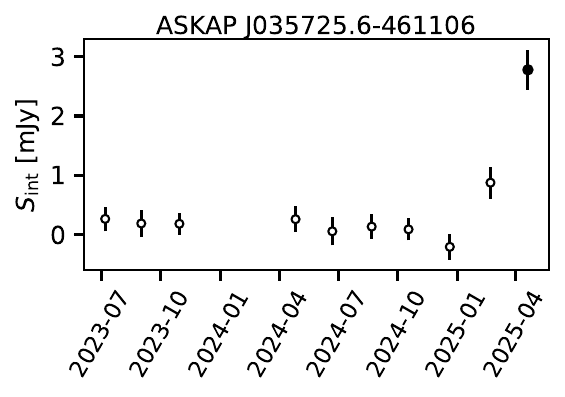} 
        \put(0,70){\color{purple}\large SN 2024abfo}
        \end{overpic}
    \end{subfigure}
    \hfill
    \begin{subfigure}[b]{0.6\textwidth}
        \centering
        \includegraphics[width=\textwidth]{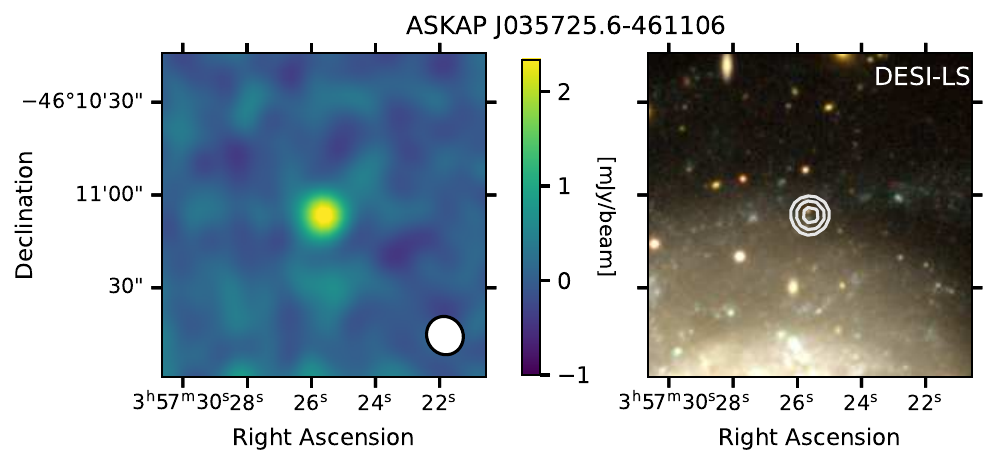} 
    \end{subfigure}
    \vspace{1em} 

    \begin{subfigure}[b]{0.38\textwidth} 
        \centering
        \begin{overpic}[width=\textwidth]{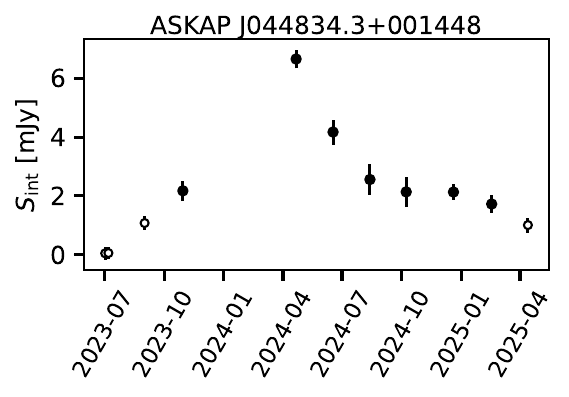} 
        \put(0,70){\color{purple}\large SN 2023mut}
        \end{overpic}
    \end{subfigure}
    \hfill
    \begin{subfigure}[b]{0.6\textwidth}
        \centering
        \includegraphics[width=\textwidth]{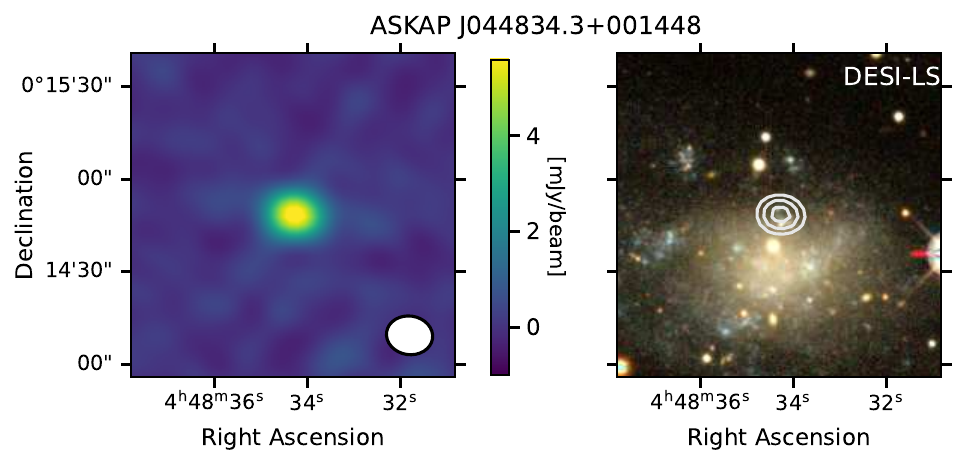} 
    \end{subfigure}
    \vspace{1em} 

    \begin{subfigure}[b]{0.38\textwidth} 
        \centering
        \begin{overpic}[width=\textwidth]{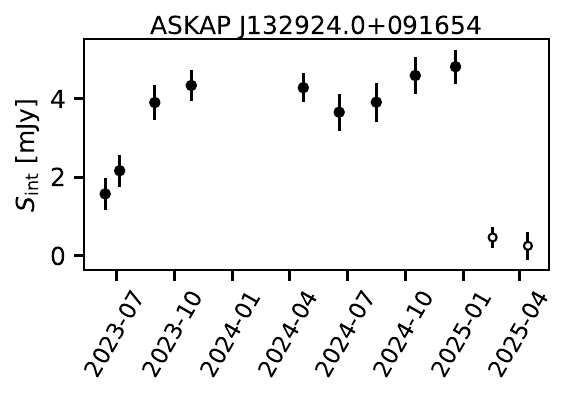} 
        \put(0,70){\color{purple}\large SN candidate}
        \end{overpic}
    \end{subfigure}
    \hfill
    \begin{subfigure}[b]{0.6\textwidth}
        \centering
        \includegraphics[width=\textwidth]{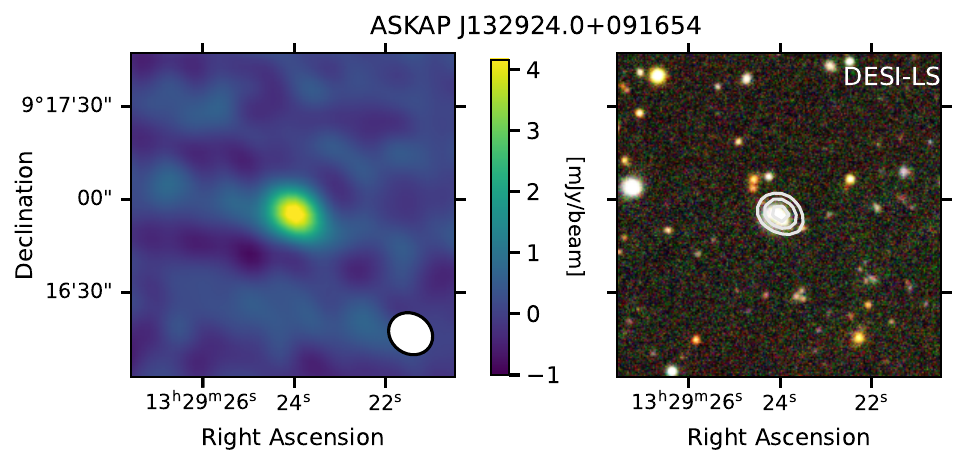} 
    \end{subfigure}
    \vspace{1em} 

    \caption{Same as Fig. \ref{fig:examples_lc_cutouts_per_class} for the supernovae discussed in Sections \ref{sec:other_SN2024_abfo} and \ref{sec:other_SN2023_mut} and the supernova candidate discussed in Section \ref{sec:other_SN_candidate}.  }
    \label{fig:lc_cutouts_SN}
\end{figure*}

\subsubsection{Supernova: ASKAP J035725.6-461106} \label{sec:other_SN2024_abfo}

ASKAP J035725.6-461106 is identified as the radio signature of optically identified supernova SN 2024abfo (TNS: \url{https://www.wis-tns.org/object/2024abfo}). ASKAP upper-limits of SN 2024abfo were reported by \cite{rose_SN2024abfo_ASKAP}, while \cite{rose_SN2024abfo_ATCA}  reported Australia Telescope Compact Array (ATCA) detections at 9 GHz. Figure \ref{fig:lc_cutouts_SN} shows the VAST Extragalactic DR1 light curve of SN 2024abfo. The first indication of radio emission shows at 2025-03-19, while the first ($>5\sigma$) detection is made on 2025-04-20. SN 2024abfo is a partially stripped type II supernova from a yellow supergiant. The host galaxy NGC is a barred spiral galaxy at $z=0.003512$ \citep{reguitti2025sn}. A more comprehensive analysis of a wide range of radio observations of SN 2024abfo will be presented in a separate study (Rose et al. in prep).

\subsubsection{Supernova: ASKAP J044834.3+001448} \label{sec:other_SN2023_mut}
ASKAP J044834.3+001448 is identified as the radio signature of optically identified supernova SN 2023mut (TNS: \url{https://www.wis-tns.org/object/2023mut}). Radio detections were reported previously from the uGMRT \citep{chandra_SN2023mut_GMRT} and AMI-LA \citep{sfaradi_SN2023mut_AMILA}. Figure \ref{fig:lc_cutouts_SN} shows the VAST Extragalactic DR1 light curve of SN 2023mut, where the duration of VAST Extragalactic DR1 seems to fully cover the radio outburst. SN 2023mut is a type IIb supernova at $z=0.0022$. These results will be discussed in more detail as part of ongoing VAST projects, including multi-wavelength (radio) observations and detailed modelling.

\subsubsection{Supernova candidate: ASKAP J132924.0+091654} \label{sec:other_SN_candidate}
ASKAP J132924.0+091654 is cross-matched to AT2023bwp (TNS: \url{https://www.wis-tns.org/object/2023bwp}), which is a supernova candidate classified using ALeRCE's stamp classifier \citep{carrasco2021alert} and the public ZTF stream \citep{2023TNSTR.391....1F}. Figure \ref{fig:lc_cutouts_SN} shows the VAST Extragalactic DR1 light curve, where there seems to be an exceptionally steep flux density decline towards the end of the VAST Extragalactic DR1 coverage. There are no archival ASKAP observations of ASKAP J132924.0+091654, nor are there any radio detections before the first VAST detection in archival radio surveys. The cutouts show that ASKAP J132924.0+091654 appears to be associated with a dwarf galaxy with $z_{phot} = 0.034 \pm 0.015$ \citep{saulder2023target}. If  ASKAP J132924.0+091654 is truly related to this dwarf galaxy, this would imply a peak radio luminosity of $\sim 10^{29} \; \rm{erg}\, \rm{s}^{-1} \, \rm{Hz}^{-1}$, which is one of the highest supernova radio luminosities reported \citep[see e.g. the samples in][]{bietenholz2021radio, mooley2022late}. This source, including planned multi-frequency radio follow-up observations, will be discussed in future work.

\subsubsection{Brown Dwarf: ASKAP J151721.1+052931} \label{sec:other_brown_dwarf}
\begin{figure}[h]
    \centering
    \includegraphics[width=0.8\linewidth]{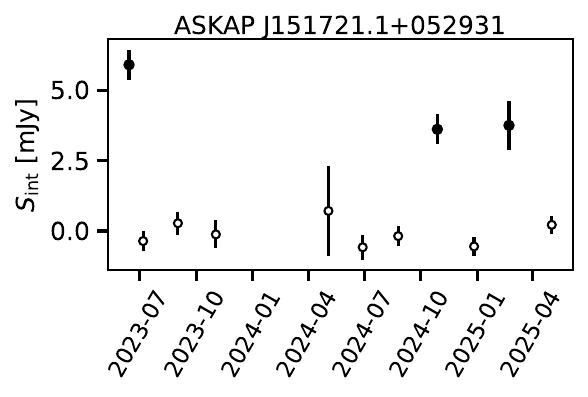}
    \caption{ASKAP J151721.1+052931 VAST Extragalactic DR1 light curve. 
    Black points are integrated flux density measurements from \textsc{Selavy}. White points are the forced-fitted flux density for images where there was no \textsc{Selavy} detection. ASKAP J151721.1+052931 is cross-matched to T8 brown dwarf ULAS J151721.12+052929.0. }
    \label{fig:brown_dwarf_lc}
\end{figure}

ASKAP J151721.1+052931 is cross-matched to ULAS J151721.12+052929.0, using the proper motion corrections from \cite{kirkpatrick2021vizier, best2020vizier}. ULAS J151721.12+052929.0 is spectroscopically confirmed as a T8 brown dwarf \citep{mace2013study, burningham201376}. A detailed discussion of this source will follow in a future study. Figure \ref{fig:brown_dwarf_lc} shows the light curve for ASKAP J151721.1+052931, where three flares are visible. Follow-up observations are underway to characterise the spectral properties of the radio bursts and test for periodicity (e.g. \citealt{rose2023periodic}).

\subsubsection{Unknown: ASKAP J160653.1-085406} \label{sec:other_unknown1}

\begin{figure}
    \centering
    \includegraphics[width=\linewidth]{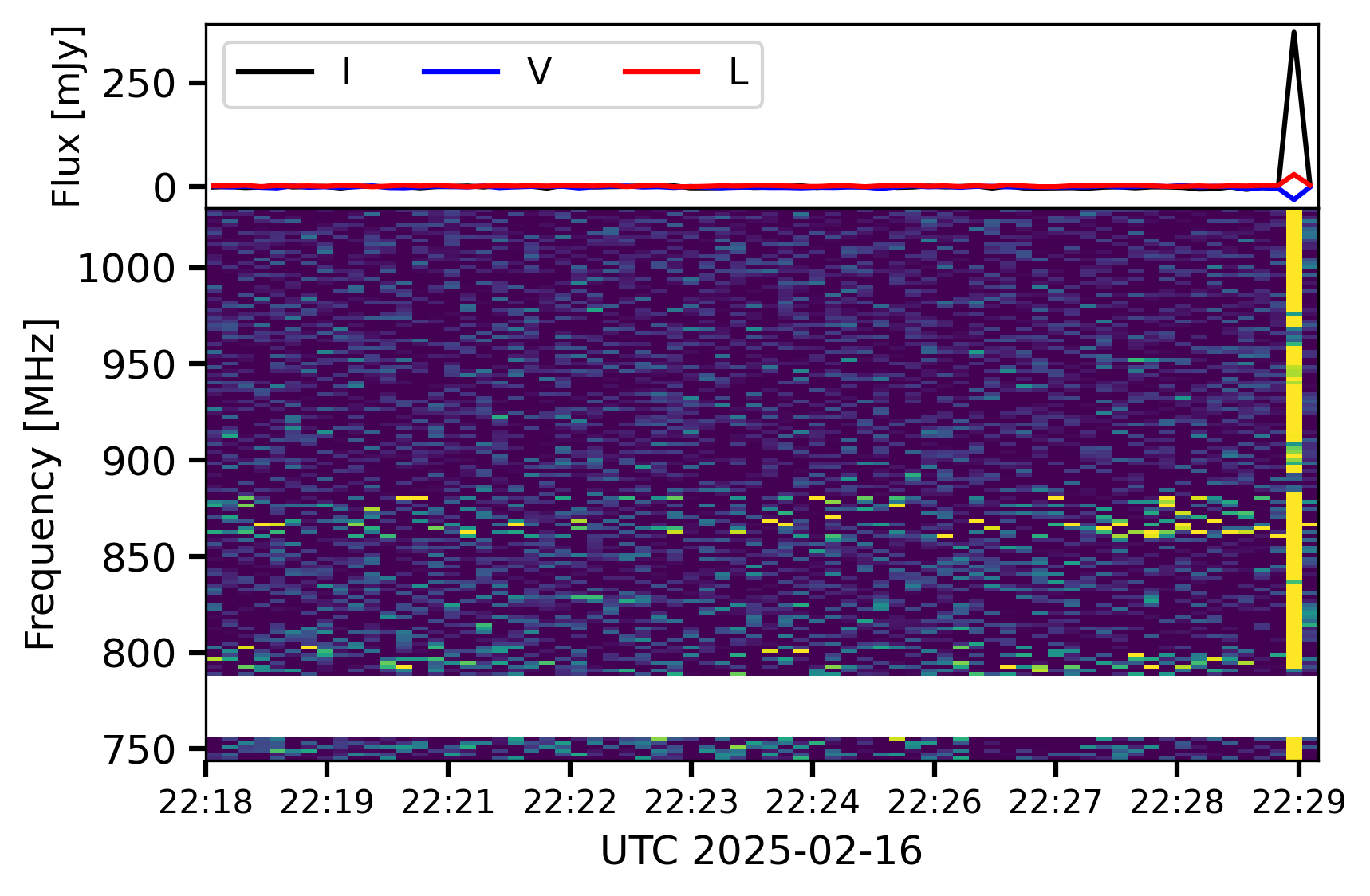}
    \caption{Stokes I dynamic spectrum of the single VAST detection of ASKAP J160653.1-085406 (see Figure \ref{fig:lc_cutouts_unknowns}). The light curves in the top panel show the total (I, black), circular (V, blue), and linear (L, red) intensity, respectively. The linear polarisation is defined as $L=\sqrt{Q^2+U^2}$.}
    \label{fig:single_flare_ds}
\end{figure}

Figure \ref{fig:lc_cutouts_unknowns} shows the light curve and cutouts for ASKAP J160653.1-085406, a source showing a single burst. There are no archival ASKAP observations of ASKAP J160653.1-085406, nor are there any radio detections before the first VAST detection in archival radio surveys.  There is no counterpart in Gaia, DES, WISE, 2MASS, SDSS, the Vista Hemisphere Survey (VHS) or the Skymapper Southern Survey. Figure \ref{fig:single_flare_ds} shows the dynamic spectrum for the single VAST detection. It is evident that the VAST image shows the averaged-out flux density of an extremely bright short-duration burst. The time resolution of the dynamic spectrum is 10 seconds, and the burst is contained well within one time bin, as there are no signs of leakage. The circularly and linearly polarised components are shown in the top panel in blue and red, respectively. The circular polarisation fraction is $|-30|/371\approx 8\%$.
ASKAP~J160653.1–085406 was also independently detected as a single pulse with the Commensal Real-time ASKAP Fast Transients survey COherent upgrade backend \citep[CRACO;][]{2025PASA...42....5W} at the same time. The identification and follow-up observations of this source will be presented in Zhang, Wang, and Gupta et al. (in prep.).

\subsubsection{Unknown: ASKAP J230616.0-395740} \label{sec:other_unkown2}

\begin{figure*}
\centering
\begin{subfigure}[b]{0.38\textwidth}
        \centering
        \begin{overpic}[width=\textwidth]{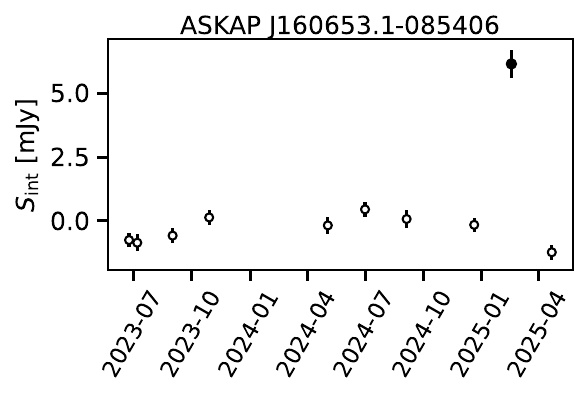} 
        \put(0,65){\color{purple}\large Unknown 1}
        \end{overpic}
    \end{subfigure}
    \hfill
    \begin{subfigure}[b]{0.6\textwidth}
        \centering
        \includegraphics[width=\textwidth]{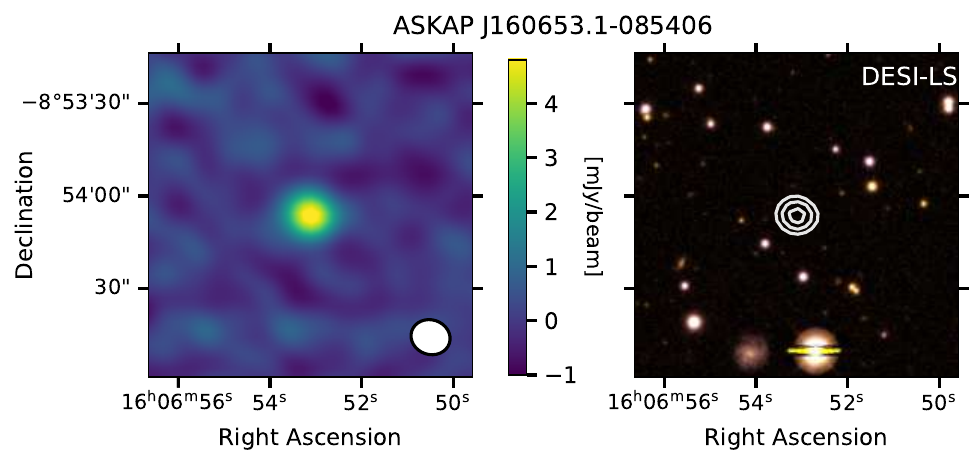} 
    \end{subfigure}
    
    \vspace{1em} 
    
    \begin{subfigure}[b]{0.38\textwidth}
        \centering
        \begin{overpic}[width=\textwidth]{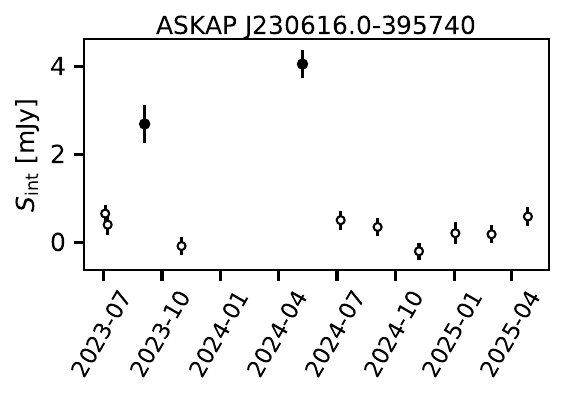} 
        \put(0,70){\color{purple}\large Unknown 2}
        \end{overpic}
    \end{subfigure}
    \hfill
    \begin{subfigure}[b]{0.6\textwidth}
        \centering
        \includegraphics[width=\textwidth]{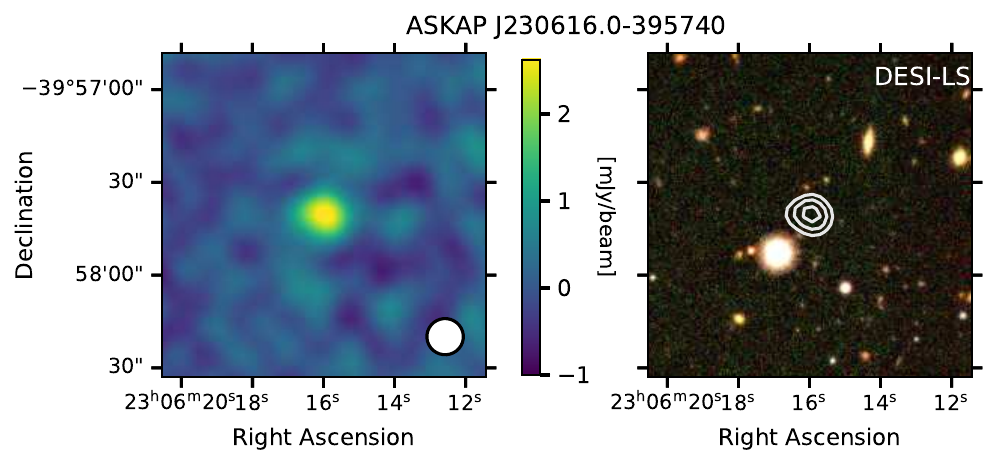} 
    \end{subfigure}
    \caption{Same as Fig. \ref{fig:examples_lc_cutouts_per_class} for the sources with unknown progenitor discussed in Sections  \ref{sec:other_unknown1} and \ref{sec:other_unkown2}. }
    \label{fig:lc_cutouts_unknowns}
\end{figure*}

Figure \ref{fig:lc_cutouts_unknowns} shows the light curve and cutouts for ASKAP J230616.0-395740. This source is detected in two VAST Extragalactic DR1 epochs. There are no previous radio detections with ASKAP or any other public radio survey. There is no counterpart to ASKAP J230616.0-395740 in Gaia, DES, WISE, 2MASS, SDSS, the Vista Hemisphere Survey (VHS) or the Skymapper Southern Survey. The closest catalogued counterpart was found in DES, at $7.2\arcsec$  \citep{abbott2018dark}. The star that is visible in Figure \ref{fig:lc_cutouts_unknowns} is $15.8\arcsec$ offset from ASKAP J230616.0-395740. This difference cannot be accounted for by the proper motion of the star (Gaia DR3 ID $6543420093211391104$) of 18.9 mas/yr. There are no known pulsars within 1 degree. We used the ESO archive to extract deep infrared upper limits from VHS \citep{mcmahon2013first}. J-band (1166-1339 nm) upper limits from VHS with a sensitivity of mag 21.3, and Ks-band (1992-2301 nm) to mag 20.3 (AB magnitudes), and these observations were taken on 2013-07-03.\\

We evaluate whether the radio detections of ASKAP J230616.0-395740 could be explained by refractive scintillation of a background AGN that typically falls below the ASKAP VAST detection limit. To this end, we analyse archival ASKAP observations of ASKAP J230616.0-395740. There are 26 archival  ASKAP observations that cover ASKAP J230616.0-395740 between April 2019 and June 2023 as part of VAST pilot surveys and RACS-low, -mid, and -high \citep{mcconnell2020rapid, duchesne2023rapid, duchesne2025high}. The source is not detected during any of these observations. Additionally, the amplitude of scintillation-induced flux density modulation would be extreme, going from non-detections ($\sim 5\cdot 0.24 =$) $1.2$ mJy to 4.05 mJy, much more than the typical 20--30\% modulations expected from refractive scintillation \citep{walker1998interstellar}. Alternatively, the radio light curve of ASKAP J230616.0-395740 could be explained by extreme scintillation due to local Galactic screens (e.g. \cite{bignall2007observations, de2015intra}). These intra-day/intra-hour variables can have modulation indices up to 50\%. The dynamic spectra for the two epochs in which the source is detected show no evidence of circular or linear polarisation, and the Stokes I flux density shows no bursty behaviour. At the moment it is unclear what type of source produces the radio emission in ASKAP J230616.0-395740. Further monitoring with VAST might shed light on its origin.

\subsection{Solar System Planets}
Throughout VAST Extragalactic DR1, we detect Mercury four times, and Saturn five times (see Table \ref{tab:results}). Solar System planets are detected routinely by ASKAP\footnote{\url{https://research.csiro.au/racs/home/gallery/solar-system-objects/}}. The planets in our Solar System show radio emission due to various processes. At frequencies below 100 MHz, electron cyclotron-maser emission is the dominant mechanism \citep{callingham2024radio}. At higher observing frequencies the emission is a combination of thermal emission \citep{roberts1963radio}, from the heat absorption by the planet atmosphere, and non-thermal emission for the planets with a significant magnetic field. Saturn has a strong magnetic field, and non-thermal emission is generated as highly energetic electrons become trapped in the planetary magnetic field and spiral along its field lines (see e.g. \citealt{hao2020formation}). For Mercury, which does not have a strong magnetic field, the radio emission is dominated by thermal blackbody re-radiation from the solar insolation over a diurnal cycle \citep{burns1987radio, ledlow1992subsurface}. As mentioned before, we do not consider the planet detections in Figure \ref{fig:class_pie_chart}, as the main reason for variability comes from the fact that the Solar System planet positions are not fixed in an equatorial coordinate system.

\section{Discussion} \label{sec:discussion}
We decide not to present our results in the context of a two-epoch transient source surface density, as has been customary in the past, see e.g. \cite{thyagarajan2011variable, dobie2019askap, anderson2020caltech, murphy2021askap}. Traditionally, for surveys with only a few epochs, a transient would be defined as `detected in a single epoch'. However, in VAST Extragalactic DR1 we see sources that would have been classified as transient, that now have multiple detections due to the longer time baseline (see e.g. the star example in Figure \ref{fig:examples_lc_cutouts_per_class}).

In this paper, we do not build a complete sample of transients down to the detection limit of our survey in this work. We only consider the most variable sources for visual inspection and classification. Instead of defining summary statistics for the radio transient sky as a whole, we think that the scientifically relevant questions are related to source classes. For example; what is the typical variability amplitude/timescale of QSOs in the VAST footprint; which radio stars show a flare over the duration of VAST Extragalactic DR1, and do the recurrence times link to stellar properties? Answering these questions requires detailed considerations of selection effects and cross-matching, which is beyond the scope of this work. 

 In Section \ref{sec:disc_pilot_comp} we compare the VAST Extragalactic DR1 results to the pilot survey \citep{murphy2021askap}. We discuss the limitations of the light curve database presented here and the limitations of the untargeted transient search methods adopted in this work in Sections \ref{sec:disc_improve_lc_database} and \ref{sec:disc_improve_search_methods}. Finally, in Section \ref{sec:disc_pulsar_var}, we illustrate how to use the VAST Extragalactic DR1 light curve database to answer more scientifically relevant metrics regarding the sample of highly variable pulsars identified in this work.

\onecolumn

\begin{longtable}{llllllllll}
\hline
name                      & RA     & Dec      & $\sigma_{\rm{pos}}$ & $\eta$     & V    & $N_T$ & $N_f$ & $S_{\rm{max}}$ & Match/previous detection            \\
                      & (J2000)     & (J2000)       & [$\arcsec$] &     &     &  &  & [mJy] &                                      \\ \hline
\textbf{Pulsars}  &  &   &  &  &   &  &  &    \\ \hline
ASKAP J003408.8-072153 & 00:34:08.8 & -07:21:53 & 0.2       & 5069.75 & 0.9  & 11  & 0   & 90.01 & PSR B0031-07                                                                                        \\
ASKAP J003421.8-053437 & 00:34:21.8 & -05:34:37 & 0.4       & 60.69   & 0.86 & 11  & 3   & 6.45  & PSR J0034-0534                                                                                      \\
ASKAP J005129.7+042300 & 00:51:29.7 & +04:23:00 & 0.7       & 23.62   & 1.24 & 11  & 8   & 4.7   & PSR J0051+0423                                                                                      \\
ASKAP J015122.7-063503 & 01:51:22.7 & -06:35:03 & 0.3       & 901.06  & 1.01 & 10  & 0   & 26.61 & PSR B0148-06                                                                                        \\
ASKAP J020601.3-402803 & 02:06:01.3 & -40:28:03 & 0.3       & 96.7    & 0.6  & 9   & 1   & 6.23  & PSR B0203-40                                                                                        \\
ASKAP J034328.0-300025 & 03:43:28.0 & -30:00:25 & 0.4       & 26.09   & 0.7  & 11  & 4   & 3.8   & PSR J0343-3000                                                                                      \\
ASKAP J041803.7-415414 & 04:18:03.7 & -41:54:14 & 0.5       & 26.7    & 1.13 & 10  & 7   & 4.19  & PSR J0418-4154                                                                                      \\
ASKAP J051452.0-440837 & 05:14:52.0 & -44:08:37 & 0.5       & 115.36  & 1.25 & 10  & 5   & 8.88  & PSR J0514-4408                                                                                      \\
ASKAP J082309.7+015913 & 08:23:09.7 & +01:59:13 & 0.2       & 305.11  & 0.5  & 11  & 0   & 18.32 & PSR B0820+02                                                                                        \\
ASKAP J102438.6-071920 & 10:24:38.6 & -07:19:20 & 0.3       & 90.3    & 0.65 & 11  & 2   & 6.48  & PSR J1024-0719 (Fig. \ref{fig:examples_lc_cutouts_per_class})                                       \\
ASKAP J122758.5-485343 & 12:27:58.5 & -48:53:43 & 0.6       & 22.39   & 1.85 & 11  & 8   & 2.6   & PSR J1227-4853                                                                                      \\
ASKAP J132012.6-351227 & 13:20:12.6 & -35:12:27 & 0.2       & 2933.76 & 1.1  & 11  & 0   & 50.66 & PSR J1320-3512                                                                                      \\
ASKAP J133252.4-303219 & 13:32:52.4 & -30:32:19 & 0.3       & 76.72   & 0.97 & 11  & 6   & 4.79  & PSR J1332-3032                                                                                      \\
ASKAP J133727.7-444142 & 13:37:27.7 & -44:41:42 & 0.4       & 42.17   & 0.74 & 11  & 6   & 6.38  & PSR J1337-4441                                                                                      \\
ASKAP J155540.5+004904 & 15:55:40.5 & +00:49:04 & 1.0       & 28.3    & 3.28 & 11  & 10  & 4.82  & PSR J1556+00/RRAT J1555+01                                                                          \\
ASKAP J160712.0-003241 & 16:07:12.0 & -00:32:41 & 0.3       & 867.29  & 0.75 & 11  & 0   & 47.53 & PSR B1604-00                                                                                        \\
ASKAP J200616.4-080701 & 20:06:16.4 & -08:07:01 & 0.3       & 423.82  & 0.51 & 11  & 0   & 17.68 & PSR B2003-08                                                                                        \\
ASKAP J203935.0-561708 & 20:39:35.0 & -56:17:08 & 0.4       & 92.81   & 1.34 & 11  & 6   & 5.69  & PSR J2039-5617                                                                                      \\
ASKAP J205107.5-082738 & 20:51:07.5 & -08:27:38 & 0.3       & 28.39   & 0.49 & 11  & 3   & 4.32  & PSR J2051-0827                                                                                      \\
ASKAP J212443.8-335845 & 21:24:43.8 & -33:58:45 & 0.2       & 857.86  & 0.96 & 22  & 2   & 40.42 & PSR J2124-3358                                                                                      \\
ASKAP J212945.1-042908 & 21:29:45.1 & -04:29:08 & 0.4       & 80.91   & 1.03 & 11  & 5   & 7.03  & PSR J2129-0429                                                                                      \\
ASKAP J214412.0-393358 & 21:44:12.0 & -39:33:58 & 0.4       & 41.93   & 0.98 & 11  & 6   & 7.36  & PSR J2144-3933                                                                                      \\
ASKAP J214435.7-523708 & 21:44:35.7 & -52:37:08 & 0.3       & 263.28  & 1.01 & 11  & 1   & 17.91 & PSR J2144-5237                                                                                      \\
ASKAP J214550.4-075018 & 21:45:50.4 & -07:50:18 & 0.2       & 1128.33 & 0.85 & 11  & 0   & 38.24 & PSR J2145-0750                                                                                      \\
ASKAP J215513.6-311854 & 21:55:13.6 & -31:18:54 & 0.3       & 76.61   & 1.06 & 11  & 2   & 6.99  & PSR B2152-31                                                                                        \\
ASKAP J222206.0-013715 & 22:22:06.0 & -01:37:15 & 0.4       & 530.11  & 1.27 & 11  & 3   & 24.22 & PSR J2222-0137                                                                                      \\
ASKAP J223423.0+061129 & 22:34:23.0 & +06:11:29 & 0.5       & 49.7    & 1.18 & 11  & 7   & 4.31  & PSR J2234+0611                                                                                      \\ \hline
\textbf{Stars}  &  &   &  &  &   &  &  &    \\ \hline
ASKAP J001650.0-071016 & 00:16:50.0 & -07:10:16 & 0.6       & 19.28   & 0.98 & 11  & 8   & 2.6   & SRSC 00031                                                                                                \\
ASKAP J002750.5-323308 & 00:27:50.5 & -32:33:08 & 0.8       & 20.75   & 1.66 & 10  & 9   & 2.69  & GJ2006A/ \footnotesize\textit{2315841869173294080}                                                        \\
ASKAP J010648.9-225124 & 01:06:48.9 & -22:51:24 & 0.5       & 26.04   & 0.62 & 8   & 2   & 4.84  & CS Cet / SRSC 00101                                                                                       \\
ASKAP J012256.8+072514 & 01:22:56.8 & +07:25:14 & 0.4       & 32.62   & 0.6  & 11  & 3   & 4.49  & AR Psc/ SRSC 00287                                                                                        \\
ASKAP J013655.4-064742 & 01:36:55.4 & -06:47:42 & 0.8       & 18.18   & 1.23 & 11  & 9   & 2.98  & G 271-110/ SRSC 00593                                                                                     \\
ASKAP J014345.3-060241 & 01:43:45.3 & -06:02:41 & 0.8       & 29.46   & 2.19 & 10  & 9   & 3.82  & SRSC 00272                                                                                                \\
ASKAP J014700.3+075129 & 01:47:00.3 & +07:51:29 & 0.4       & 180.15  & 1.44 & 10  & 5   & 13.57 & Paenaga 2856/ \footnotesize\textit{2568706316378318208} (Fig. \ref{fig:lc_cutouts_probable_stars})  \\
ASKAP J020202.8+024549 & 02:02:02.8 & +02:45:49 & 0.7       & 32.62   & 0.94 & 11  & 8   & 6.41  & Alf Psc/ SRSC 00192                                                                                       \\
ASKAP J023422.7-434754 & 02:34:22.7 & -43:47:54 & 0.3       & 50.03   & 0.58 & 10  & 1   & 4.64  & CC Eri/ SRSC 00437                                                                                        \\
ASKAP J024012.2-092711 & 02:40:12.2 & -09:27:11 & 0.8       & 21.77   & 1.69 & 9   & 8   & 3.93  & HD16673/ \footnotesize\textit{5173902189571919872}                                                        \\
ASKAP J025040.3-055952 & 02:50:40.3 & -05:59:52 & 0.7       & 25.63   & 0.9  & 10  & 8   & 3.91  & SRSC 00196                                                                                                \\
ASKAP J030402.5+004551 & 03:04:02.5 & +00:45:51 & 0.7       & 23.98   & 1.32 & 11  & 9   & 2.97  & LP591-256/ \footnotesize\textit{3266760032372995072} (Fig. \ref{fig:examples_lc_cutouts_per_class}) \\
ASKAP J031204.7-285856 & 03:12:04.7 & -28:58:56 & 0.2       & 135.8   & 0.61 & 10  & 1   & 8.25  & Alf For B/ SRSC 00119                                                                                     \\
ASKAP J044831.5-494733 & 04:48:31.5 & -49:47:33 & 0.8       & 28.97   & 2.42 & 10  & 9   & 2.95  & HD30850/ \footnotesize\textit{4784730633037533568}                                                        \\
ASKAP J050136.1-444950 & 05:01:36.1 & -44:49:50 & 0.4       & 24.3    & 0.71 & 10  & 4   & 3.14  & HD 32517/ SRSC 00254                                                                                      \\
ASKAP J051921.1-072050 & 05:19:21.1 & -07:20:50 & 0.5       & 23.12   & 0.71 & 11  & 6   & 3.29  & HD 34736/ \footnotesize\textit{3207445915541912832}                                                       \\
ASKAP J052351.8-083409 & 05:23:51.8 & -08:34:09 & 0.8       & 18.94   & 1.44 & 9   & 7   & 3.1   & UCAC4 408-009321/ \footnotesize\textit{3014907303368355712}                                               \\
ASKAP J053954.1-051120 & 05:39:54.1 & -05:11:20 & 0.9       & 20.19   & 3.0  & 10  & 9   & 6.59  & \footnotesize\textit{3023329085698084992} (Fig. \ref{fig:lc_cutouts_probable_stars})                      \\
ASKAP J060003.9+024223 & 06:00:03.9 & +02:42:23 & 0.5       & 74.37   & 1.28 & 11  & 7   & 6.17  & G 99-49/ SRSC 00054                                                                                       \\
ASKAP J061542.2-072346 & 06:15:42.2 & -07:23:46 & 0.6       & 51.58   & 1.2  & 11  & 8   & 5.2   & SRSC 00526                                                                                                \\
ASKAP J074439.5+033258 & 07:44:39.5 & +03:32:58 & 0.6       & 52.36   & 1.23 & 10  & 6   & 6.38  & YZ Cmi/ SRSC 00085                                                                                        \\
ASKAP J074950.6-031720 & 07:49:50.6 & -03:17:20 & 0.4       & 21.44   & 0.87 & 22  & 16  & 6.69  & PM J07498-0317B/ \footnotesize\textit{3080863574242572032}                                                \\
ASKAP J085857.0+082818 & 08:58:57.0 & +08:28:18 & 0.3       & 700.45  & 1.21 & 11  & 0   & 24.65 & G41-14/ SRSC 00206                                                                                        \\
ASKAP J093541.6+002527 & 09:35:41.6 & +00:25:27 & 0.7       & 102.19  & 3.68 & 11  & 10  & 9.01  & \footnotesize\textit{3841045469640297984}                                                                 \\
ASKAP J103600.1+050717 & 10:36:00.1 & +05:07:17 & 0.8       & 48.37   & 1.27 & 11  & 9   & 5.18  & RY Sex/ SRSC 00463                                                                                        \\
ASKAP J105622.5+065948 & 10:56:22.5 & +06:59:48 & 0.9       & 19.47   & 1.99 & 10  & 9   & 3.77  & GJ406/ Wolf 359/ CN Leo/ \footnotesize\textit{3864972938605115520}                                        \\
ASKAP J120805.1-503941 & 12:08:05.1 & -50:39:41 & 0.3       & 32.68   & 0.6  & 11  & 1   & 4.28  & V* V863 Cen/ SRSC 00150                                                                                   \\
ASKAP J121755.4-355714 & 12:17:55.4 & -35:57:14 & 0.7       & 20.62   & 1.66 & 11  & 9   & 2.82  & SCR J1217-3557/ SRSC 00708                                                                                \\
ASKAP J131055.1-484403 & 13:10:55.1 & -48:44:03 & 0.3       & 31.1    & 0.69 & 11  & 5   & 4.72  & V* V1265 Cen/ SRSC 00375                                                                                  \\
ASKAP J131603.0-054008 & 13:16:03.0 & -05:40:08 & 0.2       & 166.35  & 0.68 & 11  & 1   & 8.22  & HD 115247. SRSC 00033                                                                                     \\
ASKAP J141158.3-413224 & 14:11:58.3 & -41:32:24 & 0.4       & 53.61   & 1.06 & 11  & 5   & 7.14  & WT 460/ SRSC 00163                                                                                        \\
ASKAP J141215.7+022432 & 14:12:15.7 & +02:24:32 & 0.3       & 324.6   & 0.92 & 11  & 0   & 22.09 & CU Vir/ SRSC 00146                                                                                        \\
ASKAP J165222.7-063844 & 16:52:22.7 & -06:38:44 & 0.7       & 40.82   & 2.88 & 10  & 9   & 5.67  & PM J16523-0638/ \footnotesize\textit{4340941420508388352}                                                 \\
ASKAP J165527.4-082032 & 16:55:27.4 & -08:20:32 & 0.3       & 352.93  & 1.45 & 10  & 3   & 14.79 & V* V1054 Oph/ SRSC 00391                                                                                  \\
ASKAP J193938.9-060350 & 19:39:38.9 & -06:03:50 & 0.6       & 43.67   & 1.04 & 11  & 8   & 4.75  & V* V1379 Aql/ SRSC 00397                                                                                  \\
ASKAP J195518.8+062426 & 19:55:18.8 & +06:24:26 & 0.3       & 200.93  & 0.96 & 10  & 2   & 8.46  & * bet Aql B/ SRSC 00013                                                                                   \\
ASKAP J200918.2-011348 & 20:09:18.2 & -01:13:48 & 0.8       & 58.05   & 1.43 & 11  & 9   & 5.72  & SCR J2009-0113/ SRSC 00093                                                                                \\
ASKAP J204151.7-322617 & 20:41:51.7 & -32:26:17 & 0.2       & 811.79  & 0.84 & 11  & 0   & 18.06 & AT Mic/ SRSC 00288                                                                                        \\
ASKAP J204510.0-312036 & 20:45:10.0 & -31:20:36 & 0.4       & 42.19   & 1.33 & 11  & 7   & 3.74  & AU Mic/ SRSC 00568                                                                                        \\
ASKAP J231323.8+024032 & 23:13:23.8 & +02:40:32 & 0.2       & 275.28  & 0.73 & 11  & 2   & 17.2  & SZ Psc/ SRSC 00354    \\ \hline
\textbf{AGN}  &  &   &  &  &   &  &  &    \\ \hline
ASKAP J000232.8-585454 & 00:02:32.8 & -58:54:54 & 0.2       & 21.8    & 0.56 & 20  & 6   & 4.56  & VAST pilot                                                                                                \\
ASKAP J015553.4-282944 & 01:55:53.4 & -28:29:44 & 0.3       & 29.14   & 0.51 & 10  & 2   & 4.7   & VAST pilot                                                                                                \\
ASKAP J025503.3-561308 & 02:55:03.3 & -56:13:08 & 0.5       & 18.73   & 0.93 & 10  & 6   & 3.31  & VAST pilot (Fig. \ref{fig:examples_lc_cutouts_per_class})                                                 \\
ASKAP J041424.2-014607 & 04:14:24.2 & -01:46:07 & 0.3       & 323.54  & 0.51 & 10  & 0   & 19.54 & NVSS                                                                                                      \\
ASKAP J041840.2-073256 & 04:18:40.2 & -07:32:56 & 0.4       & 17.75   & 0.49 & 11  & 4   & 3.5   & VAST pilot                                                                                                \\
ASKAP J041950.3-473452 & 04:19:50.3 & -47:34:52 & 0.2       & 36.85   & 0.48 & 10  & 0   & 4.68  & VAST pilot                                                                                                \\
ASKAP J042632.5+022300 & 04:26:32.5 & +02:23:00 & 0.3       & 95.32   & 0.48 & 11  & 0   & 10.12 & NVSS                                                                                                      \\
ASKAP J044600.0-395039 & 04:46:00.0 & -39:50:39 & 0.3       & 29.23   & 0.53 & 10  & 1   & 4.19  & VAST pilot                                                                                                \\
ASKAP J045300.1-492824 & 04:53:00.1 & -49:28:24 & 0.3       & 35.36   & 0.49 & 10  & 1   & 3.85  & VAST pilot                                                                                                \\
ASKAP J052803.1-330348 & 05:28:03.1 & -33:03:48 & 0.2       & 1304.17 & 0.51 & 10  & 0   & 27.62 & NVSS                                                                                                      \\
ASKAP J053736.5-362804 & 05:37:36.5 & -36:28:04 & 0.2       & 404.33  & 0.51 & 10  & 0   & 12.54 & VAST pilot                                                                                                \\
ASKAP J053850.5-345435 & 05:38:50.5 & -34:54:35 & 0.2       & 83.61   & 0.49 & 10  & 0   & 8.17  & VAST pilot                                                                                                \\
ASKAP J054058.5-290059 & 05:40:58.5 & -29:00:59 & 0.3       & 54.93   & 0.52 & 10  & 2   & 5.58  & NVSS                                                                                                      \\
ASKAP J063644.1-053002 & 06:36:44.1 & -05:30:02 & 0.4       & 17.84   & 0.49 & 11  & 3   & 4.7   & VAST pilot                                                                                                \\
ASKAP J085439.5+084326 & 08:54:39.5 & +08:43:26 & 0.3       & 44.88   & 0.51 & 11  & 3   & 4.49  & NVSS                                                                                                      \\
ASKAP J091741.5-031332 & 09:17:41.5 & -03:13:32 & 0.1       & 1783.32 & 0.76 & 22  & 0   & 44.71 & NVSS (Fig. \ref{fig:examples_lc_cutouts_per_class})                                                       \\
ASKAP J091915.9+022041 & 09:19:15.9 & +02:20:41 & 0.2       & 356.94  & 0.79 & 11  & 5   & 13.85 & NVSS                                                                                                      \\
ASKAP J102538.4-290826 & 10:25:38.4 & -29:08:26 & 0.4       & 26.18   & 0.59 & 11  & 4   & 3.66  & NVSS                                                                                                      \\
ASKAP J103752.5-333258 & 10:37:52.5 & -33:32:58 & 0.2       & 650.13  & 0.55 & 11  & 0   & 26.21 & NVSS                                                                                                      \\
ASKAP J104515.9-332549 & 10:45:15.9 & -33:25:49 & 0.3       & 36.41   & 0.49 & 11  & 1   & 5.11  & VAST pilot                                                                                                \\
ASKAP J112205.3-445616 & 11:22:05.3 & -44:56:16 & 0.3       & 17.72   & 0.47 & 11  & 4   & 3.5   & VAST pilot                                                                                                \\
ASKAP J113554.9-421044 & 11:35:54.9 & -42:10:44 & 0.2       & 1679.28 & 0.54 & 11  & 0   & 35.7  & SUMSS                                                                                                     \\
ASKAP J114535.9+014230 & 11:45:35.9 & +01:42:30 & 0.5       & 26.34   & 0.49 & 11  & 1   & 5.47  & FIRST                                                                                                     \\
ASKAP J122046.7-484709 & 12:20:46.7 & -48:47:09 & 0.3       & 32.51   & 0.63 & 11  & 3   & 3.56  & VAST pilot                                                                                                \\
ASKAP J123431.2-345345 & 12:34:31.2 & -34:53:45 & 0.3       & 36.3    & 0.58 & 11  & 3   & 5.6   & VAST pilot                                                                                                \\
ASKAP J123951.7-073814 & 12:39:51.7 & -07:38:14 & 0.4       & 41.01   & 0.52 & 12  & 2   & 5.03  & FIRST                                                                                                     \\
ASKAP J131529.5-411838 & 13:15:29.5 & -41:18:38 & 0.3       & 19.49   & 0.49 & 11  & 2   & 4.11  & VAST pilot                                                                                                \\
ASKAP J135441.4-373333 & 13:54:41.4 & -37:33:33 & 0.2       & 623.79  & 0.48 & 11  & 0   & 14.86 & NVSS                                                                                                      \\
ASKAP J161726.3+063233 & 16:17:26.3 & +06:32:33 & 0.5       & 23.7    & 0.85 & 10  & 5   & 3.78  & VAST pilot                                                                                                \\
ASKAP J165910.9-073710 & 16:59:10.9 & -07:37:10 & 0.2       & 3101.55 & 0.5  & 11  & 0   & 61.42 & NVSS                                                                                                      \\
ASKAP J175921.5+055512 & 17:59:21.5 & +05:55:12 & 0.4       & 18.21   & 0.49 & 11  & 2   & 4.15  & VAST pilot                                                                                                \\
ASKAP J195239.2-581701 & 19:52:39.2 & -58:17:01 & 0.2       & 79.61   & 0.5  & 11  & 0   & 6.33  & VAST pilot                                                                                                \\
ASKAP J202605.4-384549 & 20:26:05.4 & -38:45:49 & 0.2       & 67.27   & 0.48 & 11  & 1   & 6.2   & NVSS                                                                                                      \\
ASKAP J203105.1-362304 & 20:31:05.1 & -36:23:04 & 0.2       & 187.11  & 0.51 & 11  & 0   & 10.64 & VAST pilot                                                                                                \\
ASKAP J203357.7-084725 & 20:33:57.7 & -08:47:25 & 0.4       & 19.64   & 0.51 & 11  & 4   & 3.62  & VAST pilot                                                                                                \\
ASKAP J204402.5-372651 & 20:44:02.5 & -37:26:51 & 0.2       & 468.52  & 0.56 & 11  & 0   & 17.9  & NVSS                                                                                                      \\
ASKAP J204847.1-435933 & 20:48:47.1 & -43:59:33 & 0.2       & 225.42  & 0.55 & 11  & 0   & 9.46  & VAST pilot                                                                                                \\
ASKAP J210046.3+085829 & 21:00:46.3 & +08:58:29 & 0.3       & 125.23  & 0.61 & 11  & 0   & 10.92 & VAST pilot                                                                                                \\
ASKAP J211626.1-561702 & 21:16:26.1 & -56:17:02 & 0.2       & 128.07  & 0.51 & 11  & 0   & 7.68  & VAST pilot                                                                                                \\
ASKAP J212427.5-032842 & 21:24:27.5 & -03:28:42 & 0.3       & 92.32   & 0.61 & 11  & 2   & 9.87  & NVSS                                                                                                      \\
ASKAP J214521.8-314827 & 21:45:21.8 & -31:48:27 & 0.3       & 20.65   & 0.55 & 11  & 3   & 3.16  & VAST pilot                                                                                                \\
ASKAP J223835.7+081504 & 22:38:35.7 & +08:15:04 & 0.4       & 22.11   & 0.47 & 11  & 2   & 4.47  & VAST pilot                                                                                                \\
ASKAP J224712.4-015428 & 22:47:12.4 & -01:54:28 & 0.3       & 38.1    & 0.48 & 9   & 1   & 8.57  & VAST pilot                                                                                                \\
ASKAP J233451.9-585148 & 23:34:51.9 & -58:51:48 & 0.3       & 42.35   & 0.68 & 10  & 2   & 6.49  & VAST pilot                                                                                                \\ \hline
\textbf{Other}  &  &   &  &  &   &  &  &    \\ \hline
ASKAP J035725.6-461107 & 03:57:25.6 & -46:11:07 & 0.6       & 18.13   & 1.82 & 10  & 9   & 2.78  & SN2024abfo (Fig. \ref{fig:lc_cutouts_SN})                                                            \\
ASKAP J044834.3+001448 & 04:48:34.3 & +00:14:48 & 0.4       & 92.07   & 0.93 & 11  & 4   & 6.66  & SN2023mut (Fig. \ref{fig:lc_cutouts_SN})                                                             \\
ASKAP J132924.0+091655 & 13:29:24.0 & +09:16:55 & 0.5       & 33.21   & 0.54 & 11  & 2   & 4.82  & AT2023bwp (Fig. \ref{fig:lc_cutouts_SN})                                                             \\
ASKAP J151721.1+052932 & 15:17:21.1 & +05:29:32 & 0.6       & 36.83   & 1.89 & 11  & 8   & 5.91  & brown dwarf ULAS J151721.12+052929.0 (Fig. \ref{fig:brown_dwarf_lc})                                      \\
ASKAP J160653.1-085406 & 16:06:53.1 & -08:54:06 & 0.8       & 40.77   & 8.2  & 10  & 9   & 6.16  & Unknown 1 (Fig. \ref{fig:lc_cutouts_unknowns})                                                             \\
ASKAP J230616.0-395741 & 23:06:16.0 & -39:57:41 & 0.9       & 30.54   & 1.38 & 11  & 9   & 4.05  & Unknown 2 (Fig. \ref{fig:lc_cutouts_unknowns}) \\ \hline
\textbf{Planets}  &  &   &  &  &   &  &  &    \\ \hline
ASKAP J001340.9-010335 & 00:13:40.9 & -01:03:35 & 0.7       & 255.46  & 4.09 & 11  & 10  & 10.7  & Mercury                                                                                             \\
ASKAP J010058.6+061518 & 01:00:58.6 & +06:15:18 & 0.7       & 207.28  & 3.56 & 11  & 10  & 18.16 & Mercury                                                                                             \\
ASKAP J103955.7+054440 & 10:39:55.7 & +05:44:40 & 0.8       & 240.95  & 3.24 & 11  & 10  & 15.86 & Mercury                                                                                             \\
ASKAP J111436.9-000203 & 11:14:36.9 & -00:02:03 & 1.1       & 515.99  & 3.46 & 11  & 10  & 16.52 & Mercury                                                                                             \\
ASKAP J225825.9-085152 & 22:58:25.9 & -08:51:52 & 0.7       & 1338.02 & 3.15 & 10  & 9   & 35.75 & Saturn                                                                                              \\
ASKAP J230509.2-080028 & 23:05:09.2 & -08:00:28 & 0.5       & 1607.18 & 2.84 & 10  & 9   & 28.79 & Saturn                                                                                              \\
ASKAP J230959.6-074428 & 23:09:59.6 & -07:44:28 & 0.7       & 1365.72 & 5.05 & 20  & 19  & 38.63 & Saturn                                                                                              \\
ASKAP J231741.5-065233 & 23:17:41.5 & -06:52:33 & 0.6       & 2674.24 & 3.59 & 10  & 9   & 37.32 & Saturn                                                                                              \\
ASKAP J232401.7-055718 & 23:24:01.7 & -05:57:18 & 0.6       & 991.19  & 3.39 & 10  & 9   & 25.65 & Saturn                                                                                             \\

\hline
\caption{Highly variable sources identified in VAST Extragalactic DR1. The coordinate of each source is given as the weighted average of all \textsc{Selavy} detections, where the weight is the inverse square of the positional error. $\sigma_{\rm{pos}}$ is the averaged positional uncertainty. $\eta$ and V are the variability parameters as described in Section \ref{sec:variability_metrics}. $N_T$ gives the number of epochs (observations) that cover the source location, which is equal to the number of points in the light curve. $N_f$ gives the number of forced flux extractions. $S_{\rm{max}}$ gives the maximum integrated flux density for this source. The last column shows the source the VAST detection has been matched to, or for the AGN, the survey that previously detected the AGN. The numbers in italics for the stars refer to the Gaia DR3 identifier. A reference to the light curve is included if a source is shown in the text, either as an example of the class (Fig. \ref{fig:examples_lc_cutouts_per_class}) or for a more detailed discussion. A machine-readable version of this Table can be found in the Supplementary Materials.}
\label{tab:results}
\end{longtable}

\twocolumn

\subsection{Comparison to the pilot survey}\label{sec:disc_pilot_comp}
The transient search in the VAST pilot survey \citep{murphy2021askap} covered 1646 square degrees. Observations were conducted between May 2019 and August 2020, obtaining three to eleven epochs per source. In the pilot survey, filters similar to the ones described in Section \ref{sec:selection_criteria} were applied, though with a lower signal-to-noise ratio threshold (7 instead of 8 in this work). Additionally, in the pilot survey, quality cuts were calculated on a per-source basis, whereas in this work, we also remove individual measurements that are of poor quality (see Section \ref{sec:filtering_bad_measurements}). An untargeted $\eta,V$ search using a $2\sigma$ threshold yielded 171 candidates, of which only 28 were genuine variables or transients. VAST Extragalactic DR1 covers $\sim12300$ square degrees, with most sources observed 10–11 times (see Figure \ref{fig:nof_obs_per_field}). An untargeted transient search based on $\eta,V$-statistics, selecting outliers above $2.5\sigma$ produces 170 candidates, 126 of which are real (including Solar System planet detections). This corresponds to a 26\% rejection rate by visual inspection, a major improvement over the pilot survey’s 84\%.\\

It is difficult to directly compare the number of transients identified in this work to the pilot survey, as a higher signal-to-noise threshold was used in addition to different $\eta,V$-thresholds. VAST Extragalactic DR1 contains 802 sources with $\eta,V$-values above $2\sigma$. Assuming a similar fraction of artefacts (26\%), this implies 593 real astrophysical transients. 
This is broadly consistent with the pilot survey when accounting for the larger footprint ($12300/1646=7.5$) and longer duration (factor of $\sim 2$), which would predict $\sim$420 transients in VAST Extragalactic DR1 by extrapolating from the pilot survey. The classification breakdown is also consistent. The pilot survey’s 28 real sources comprised 7 pulsars, 7 stars, 8 AGN/galaxies, and 6 unclassified, similar to the DR1 fractions (38\% AGN/galaxy, 34\% star, 23\% pulsar, 5\% other; Figure \ref{fig:class_pie_chart}). Notably, 27 of 44 AGN classifications in VAST Extragalactic DR1 rely on pilot survey detections (see Table \ref{tab:results}), underscoring the importance of long-term monitoring for distinguishing AGN variability and scintillation.

\subsection{Improvements to the light curve database} \label{sec:disc_improve_lc_database}
Here we discuss future improvements that could be made to the selection criteria we apply in Section \ref{sec:selection_criteria}. There are many ways to improve these filters to get a more complete light curve database while decreasing the spurious measurements due to artefacts.

\subsubsection{Reducing artefacts}
We remove imaging artefacts using two methods: a conservative local rms-threshold to discard poor-quality measurements, and a nearest neighbour distance threshold, since artefacts cluster around bright sources. The rms cut removes only the most severe artefacts, as they occur across a wide range of rms values. The distance cut is overly aggressive, discarding not only artefacts but also genuine sources located near other sources, as it cannot distinguish between the two. This could be improved in future studies by making this distance threshold flux density dependent \citep[e.g.][]{deruiter2024transient}.

Image inspection is often the only way to determine whether a measurement is a valid radio source or an (imaging) artefact \citep[see e.g.][]{andersson2025finding}. This is why we publish the image cutouts for all sources in VAST Extragalactic DR1 (see Section \ref{sec:dataproducts}). In the future, machine learning methods may automate this classification, as they are already used for radio-galaxy morphology classification \citep[e.g.][]{lukic2018radio} and for monitoring calibration-algorithm convergence \citep{de2025scalable}.

\subsubsection{Improving completeness for faint transients}
In the current filtering approach, a source must reach a signal-to-noise ratio of 8 at least once to enter the catalogue. Lowering this threshold to 5 in future releases will include more, fainter sources, but without additional filtering, it will also increase false positives. More robust filtering criteria, such as a multi-parameter compactness test that incorporates local noise \citep[see e.g. Figure 8 in ][]{hale2021rapid}, could mitigate this. Cross-matching marginal ($\sim5\sigma$) detections to archival radio surveys would further help distinguish real sources from noise.

Together with improved artefact filtering, lowering the detection threshold will enhance VAST Extragalactic DR1’s completeness for faint synchrotron transients. These events often lie close to or atop their radio-bright host galaxies \citep[e.g. Figure 3 in][]{leung2021search}, making them vulnerable to rejection by the nearest-neighbour filter, and their intrinsically low radio flux densities \citep[e.g. Figure 2 in][]{leung2021search} mean that reaching the survey limit is essential for detecting more distant and fainter extragalactic synchrotron transients.

\subsection{Improvements to untargeted transient search methods} \label{sec:disc_improve_search_methods}

Untargeted transient searches based on the $\eta$ and $V$ variability parameters (Section \ref{sec:variability_metrics}) are known to be incomplete. Their effectiveness declines as light curves become more densely sampled, since genuine flares are diluted by many non-variable points. Indeed, \cite{andersson2023bursts} show that most volunteer-identified variables lie within the bulk of the $\eta$–$V$ distributions and would be missed by these metrics alone.

Figure \ref{fig:class_pie_chart} shows that the “most variable’’ VAST Extragalactic sources are predominantly Galactic (stars and pulsars), consistent with the fact that most radio AGN exhibit only weak variability (e.g. $<4\%$ vary by $>30\%$ at 3 GHz; \citealt{mooley2016}). In Figure \ref{fig:eta_V}, nearly all extragalactic sources cluster near the vertical cutoff in $V$, implying that many variable AGN fall below the imposed $\eta$ and $V$ thresholds. Lower-level variability can instead be assessed using the Debiased Variability Index \citep[DVI;][]{akritas1996,barvainis2005,sadler2006,hogan2015}, which accounts for flux density uncertainties.

Improving completeness will require incorporating information beyond simple variability statistics. Promising avenues include identifying burst-like events \citep{dobie2024two}, applying matched-filter searches for characteristic temporal signatures \citep{leung2023matched}, using Gaussian-process modelling for more robust variability characterisation \citep{fu2025new}, and exploring richer time-series features for unsupervised anomaly detection \citep{andersson2025finding}.

\subsection{Use case example: pulsar variability in the image domain} \label{sec:disc_pulsar_var} 
Although pulsars are typically studied with beamformed, high–time-resolution data, many show strong image-domain variability, and some are even discovered through their spectral, polarisation, or variability properties in the image domain \citep[see e.g.][]{dai2017prospects, wang2022discovery, wang2024discovery}. This variability may be due to intrinsic reasons, where the pulsar is nulling \citep{backer1970pulsar} or (periodically) intermittent \citep{kramer2006periodically, lyne2010switched}. For binary systems, the pulsar may be eclipsing behind its companion \citep{broderick2016low,zic2024discovery, petrou2025investigating, ghosh2025exploring} or the radio pulses may be severely affected by scattering due to the dense gas in binary systems \citep{polzin2020study, kudale2020study}. Pulsars may also be variable due to interstellar scintillation, which is a propagation effect \citep{rickett1990radio, walker1998interstellar}, see \cite{kumamoto2021flux} for an example. \\

The data presented in VAST Extragalactic DR1 can be used to characterise the variability of pulsars in the image domain. We cross-match the sources in VAST Extragalactic DR1 to the proper motion corrected positions of the pulsars in the ATNF pulsar catalogue\footnote{\url{http://www.atnf.csiro.au/research/pulsar/psrcat}} \citep{manchester2005australia} out to $5\arcsec$. We find that there are 73 pulsars in VAST Extragalactic DR1, 20 of which are binary millisecond pulsars. We found 27 highly variable pulsars (Section \ref{sec:pulsar_results}), 13 of which are MSPs, suggesting that MSPs exhibit stronger image-domain variability. 

We provide this example to showcase the types of follow-up studies that are possible with VAST Extragalactic DR1. In this simplified analysis, many caveats have not been taken into account. For example, to get a more comprehensive sample of the fraction of highly variable pulsars, one should cross-match down to the VAST detection limit, consider the flux density distribution of the pulsar and MSP populations, take into account the effects of scintillation (due to potential galactic latitude differences), and carefully consider false associations and potential imaging artefacts carefully (by visual inspection).

\section{Summary and Outlook}
We have introduced the first data release of the VAST Extragalactic Survey, including observations between June 2023 and May 2025. We outline the survey footprint and observing strategy, where we observe the fields in the VAST Extragalactic sky once every two months. Using the VAST transient detection pipeline, we construct light curves for 0.5 million sources, where most sources have 10 or 11 measurements. These light curves are available to the community on the CSIRO DAP. The VAST survey is still ongoing, and future data releases will include new observations for the extragalactic part of the survey. Additionally, we anticipate the data release of the observations of the Galactic part of the survey. \\

We perform an untargeted variability search on VAST Extragalactic DR1, yielding 117 highly variable and transient sources. These sources include 44 AGN, 40 stars (10 of which were new radio stars), 27 known pulsars, 2 supernovae, 1 supernova candidate, 1 brown dwarf, and 2 unknowns. The number of transient sources is consistent with the extrapolated rates from the VAST pilot survey \citep{murphy2021askap}. The light curves that are produced as part of this data release will include many more transient and variable sources, at lower variability levels. We anticipate the wider community to use the light curves in this data release to cross-match to their targets of interest and thereby uncover the radio variability of specific source classes.

\section{Acknowledgements}
The authors thank Minh Huynh for her support in making this data release available on CSIRO platforms, and Dominic Hogan for his help with the CSIRO DAP API. We thank the anonymous referees for constructive comments that have improved the clarity of the paper, Alex Andersson for feedback on the Jupyter notebook, Stefan Duchesne and Tim Galvin for helpful discussions around RACS-low data products, and Andrew Zic for input on the post-processing statistics.

This research was supported by the Australian Research Council Centre of Excellence for Gravitational Wave Discovery (OzGrav), project number CE230100016. DLK was supported by NSF grants AST-1816492 and AST-2511757. 
JKL acknowledges support from the University of Toronto and Hebrew University of Jerusalem through the University of Toronto - Hebrew University of Jerusalem Research and Training Alliance program. 
The Dunlap Institute is funded through an endowment established by the David Dunlap family and the University of Toronto. AH is grateful for the support by the Israel Science Foundation (ISF grant 1679/23) and by the the United States-Israel Binational Science Foundation (BSF grant 2020203).  GRS is supported by NSERC Discovery Grant RGPIN-2021-04001. 

This research was supported by the Sydney Informatics Hub, a Core Research Facility of the University of Sydney. This research made use of resources provided under the JA3 project on the Pawsey Supercomputing Research Centre (2023; Supercomputing, Data and Visualisation Services; Perth, Western Australia; \url{https://ror.org/04f2f0537}), as well as the Nimbus Research Cloud (Pawsey Supercomputing Research Centre, 2023;  Nimbus Research Cloud; Perth, Western Australia;  \url{https://doi.org/10.48569/v0j3-qd51}), and the Nectar Research Cloud, a collaborative Australian research platform supported by the NCRIS-funded Australian Research Data Commons (ARDC). This research used VAST Tools \citep{vast_tools}, a Python module to interact with results from the VAST Pipeline and the VAST Survey data.

The Australian SKA Pathfinder is part of the Australia Telescope National Facility which is managed by CSIRO. Operation of ASKAP is funded by the Australian Government with support from the National Collaborative Research Infrastructure Strategy. ASKAP uses the resources of the Pawsey Supercomputing Centre. Establishment of ASKAP, the Murchison Radio-astronomy Observatory and the Pawsey Supercomputing Centre are initiatives of the Australian Government, with support from the Government of Western Australia and the Science and Industry Endowment Fund. We acknowledge the Wajarri Yamatji people as the traditional owners of the Observatory site. This paper includes archived data obtained through the CSIRO ASKAP Science Data Archive, CASDA (\url{http://data.csiro.au}).

The Legacy Surveys consist of three individual and complementary projects: the Dark Energy Camera Legacy Survey (DECaLS; Proposal ID \#2014B-0404; PIs: David Schlegel and Arjun Dey), the Beijing-Arizona Sky Survey (BASS; NOAO Prop. ID \#2015A-0801; PIs: Zhou Xu and Xiaohui Fan), and the Mayall z-band Legacy Survey (MzLS; Prop. ID \#2016A-0453; PI: Arjun Dey). DECaLS, BASS and MzLS together include data obtained, respectively, at the Blanco telescope, Cerro Tololo Inter-American Observatory, NSF’s NOIRLab; the Bok telescope, Steward Observatory, University of Arizona; and the Mayall telescope, Kitt Peak National Observatory, NOIRLab. Pipeline processing and analyses of the data were supported by NOIRLab and the Lawrence Berkeley National Laboratory (LBNL). The Legacy Surveys project is honored to be permitted to conduct astronomical research on Iolkam Du’ag (Kitt Peak), a mountain with particular significance to the Tohono O’odham Nation.\\

NOIRLab is operated by the Association of Universities for Research in Astronomy (AURA) under a cooperative agreement with the National Science Foundation. LBNL is managed by the Regents of the University of California under contract to the U.S. Department of Energy.\\

This project used data obtained with the Dark Energy Camera (DECam), which was constructed by the Dark Energy Survey (DES) collaboration. Funding for the DES Projects has been provided by the U.S. Department of Energy, the U.S. National Science Foundation, the Ministry of Science and Education of Spain, the Science and Technology Facilities Council of the United Kingdom, the Higher Education Funding Council for England, the National Center for Supercomputing Applications at the University of Illinois at Urbana-Champaign, the Kavli Institute of Cosmological Physics at the University of Chicago, Center for Cosmology and Astro-Particle Physics at the Ohio State University, the Mitchell Institute for Fundamental Physics and Astronomy at Texas A\&M University, Financiadora de Estudos e Projetos, Fundacao Carlos Chagas Filho de Amparo, Financiadora de Estudos e Projetos, Fundacao Carlos Chagas Filho de Amparo a Pesquisa do Estado do Rio de Janeiro, Conselho Nacional de Desenvolvimento Cientifico e Tecnologico and the Ministerio da Ciencia, Tecnologia e Inovacao, the Deutsche Forschungsgemeinschaft and the Collaborating Institutions in the Dark Energy Survey. The Collaborating Institutions are Argonne National Laboratory, the University of California at Santa Cruz, the University of Cambridge, Centro de Investigaciones Energeticas, Medioambientales y Tecnologicas-Madrid, the University of Chicago, University College London, the DES-Brazil Consortium, the University of Edinburgh, the Eidgenossische Technische Hochschule (ETH) Zurich, Fermi National Accelerator Laboratory, the University of Illinois at Urbana-Champaign, the Institut de Ciencies de l’Espai (IEEC/CSIC), the Institut de Fisica d’Altes Energies, Lawrence Berkeley National Laboratory, the Ludwig Maximilians Universitat Munchen and the associated Excellence Cluster Universe, the University of Michigan, NSF’s NOIRLab, the University of Nottingham, the Ohio State University, the University of Pennsylvania, the University of Portsmouth, SLAC National Accelerator Laboratory, Stanford University, the University of Sussex, and Texas A\&M University.\\

BASS is a key project of the Telescope Access Program (TAP), which has been funded by the National Astronomical Observatories of China, the Chinese Academy of Sciences (the Strategic Priority Research Program “The Emergence of Cosmological Structures” Grant \# XDB09000000), and the Special Fund for Astronomy from the Ministry of Finance. The BASS is also supported by the External Cooperation Program of Chinese Academy of Sciences (Grant \# 114A11KYSB20160057), and Chinese National Natural Science Foundation (Grant \# 12120101003, \# 11433005).

The Legacy Survey team makes use of data products from the Near-Earth Object Wide-field Infrared Survey Explorer (NEOWISE), which is a project of the Jet Propulsion Laboratory/California Institute of Technology. NEOWISE is funded by the National Aeronautics and Space Administration.

The Legacy Surveys imaging of the DESI footprint is supported by the Director, Office of Science, Office of High Energy Physics of the U.S. Department of Energy under Contract No. DE-AC02-05CH1123, by the National Energy Research Scientific Computing Center, a DOE Office of Science User Facility under the same contract; and by the U.S. National Science Foundation, Division of Astronomical Sciences under Contract No. AST-0950945 to NOAO.\\

Funding for the Sloan Digital Sky Survey V has been provided by the Alfred P. Sloan Foundation, the Heising-Simons Foundation, the National Science Foundation, and the Participating Institutions. SDSS acknowledges support and resources from the Center for High-Performance Computing at the University of Utah. SDSS telescopes are located at Apache Point Observatory, funded by the Astrophysical Research Consortium and operated by New Mexico State University, and at Las Campanas Observatory, operated by the Carnegie Institution for Science. The SDSS web site is \url{www.sdss.org}.

SDSS is managed by the Astrophysical Research Consortium for the Participating Institutions of the SDSS Collaboration, including the Carnegie Institution for Science, Chilean National Time Allocation Committee (CNTAC) ratified researchers, Caltech, the Gotham Participation Group, Harvard University, Heidelberg University, The Flatiron Institute, The Johns Hopkins University, L'Ecole polytechnique f\'{e}d\'{e}rale de Lausanne (EPFL), Leibniz-Institut f\"{u}r Astrophysik Potsdam (AIP), Max-Planck-Institut f\"{u}r Astronomie (MPIA Heidelberg), Max-Planck-Institut f\"{u}r Extraterrestrische Physik (MPE), Nanjing University, National Astronomical Observatories of China (NAOC), New Mexico State University, The Ohio State University, Pennsylvania State University, Smithsonian Astrophysical Observatory, Space Telescope Science Institute (STScI), the Stellar Astrophysics Participation Group, Universidad Nacional Aut\'{o}noma de M\'{e}xico, University of Arizona, University of Colorado Boulder, University of Illinois at Urbana-Champaign, University of Toronto, University of Utah, University of Virginia, Yale University, and Yunnan University. 

This work has made use of data from the European Space Agency (ESA) mission
{\it Gaia} (\url{https://www.cosmos.esa.int/gaia}), processed by the {\it Gaia}
Data Processing and Analysis Consortium (DPAC,
\url{https://www.cosmos.esa.int/web/gaia/dpac/consortium}). Funding for the DPAC
has been provided by national institutions, in particular the institutions
participating in the {\it Gaia} Multilateral Agreement.

This publication makes use of data products from the Wide-field Infrared Survey Explorer, which is a joint project of the University of California, Los Angeles, and the Jet Propulsion Laboratory/California Institute of Technology, funded by the National Aeronautics and Space Administration.

This research has made use of NASA’s Astrophysics Data System. This research has made use of the SIMBAD database \citep{wenger2000simbad} and the VizieR catalogue access tool, operated at CDS, Strasbourg, France \citep{10.26093/cds/vizier}. The original description of the VizieR service was published in \cite{vizier_wise_2011_cutri}. This research made use of ds9, a tool for data vizualisation supported by the Chandra X-ray Science Center (CXC) and the High Energy Astrophysics Science Archive Center (HEASARC) with support from the JWST Mission office at the Space Telescope Science Institute for 3D visualization \citep{joye2003new}. This work makes use of the Python programming language\footnote{Python Software Foundation; \url{https://www.python.org/}}. The following python packages were crucial for this work: Astropy:\footnote{\url{http://www.astropy.org}} a community-developed core Python package and an ecosystem of tools and resources for astronomy \citep{astropy:2013, astropy:2018, astropy:2022}, \textsc{numpy} \citep{oliphant2006guide, van2011numpy}, \textsc{scipy} \citep{2020SciPy-NMeth}, and \textsc{matplotlib} \citep{Hunter:2007}. ChatGPT (OpenAI) was used for editorial assistance, specifically improving clarity and phrasing in parts of the manuscript. All scientific content, analysis, and conclusions are the authors' own.

\section{Data availability and supplementary materials}
Section \ref{sec:dataproducts} outlines the data products generated as part of this work and provides links to each.
A machine-readable version of Table \ref{tab:results} is included in the supplementary materials.

\appendix

\section{Artificial variability introduced by incorrect measurement of phased array feed beam response} \label{app:data_obs_issues}
Each ASKAP dish is fitted with a phased array feed (PAF) at its focal point \citep{chippendale2010phased, hotan2021australian}. Instead of a single receiver, these PAFs contain multiple receivers, allowing ASKAP to create 36 separate simultaneous beams on the sky. This technique gives ASKAP its large 30 square degree field of view. In radio telescopes with a single receiver, the primary beam shape is defined by a physical feed structure and is fixed by the manufactured characteristics of that feed. For ASKAP however, the primary beam can be changed by altering the digital weights of the beamforming hardware, which forms a single primary beam by taking the weighted sum of the individual receivers in the PAF \citep{hotan2014australian, hotan2021australian, mcconnell2016australian}. This primary beam response is calibrated by holography measurements with a cadence of a few weeks to a few months, a procedure described in detail in \citep{hotan2016holographic}. Application of mismatched primary beam responses can result in brightness scale errors, especially pronounced at the beam edges, as already noticed by \cite{duchesne2023rapid}.\\

Upon inspection of the VAST Extragalactic data we find a large number of compact, bright ($\sim 100$ mJy), isolated sources that show a significant ($\sim 20\%$) flux density increase in observations taken between December 2023 and March 2024. Upon closer inspection, we find that these sources are not restricted to a single field. This flux density increase is clearly artificially introduced by our observing or processing techniques. Observations in epochs 48 to 55, corresponding to the aforementioned period of December 2023 to March 2024, all use the same holography calibration. This holography calibration turned out to be affected by severe ducted RFI, which was either inadequately flagged or the flagged regions were not robustly interpolated over. 

\begin{figure}
    \centering
    \includegraphics[width=\linewidth]{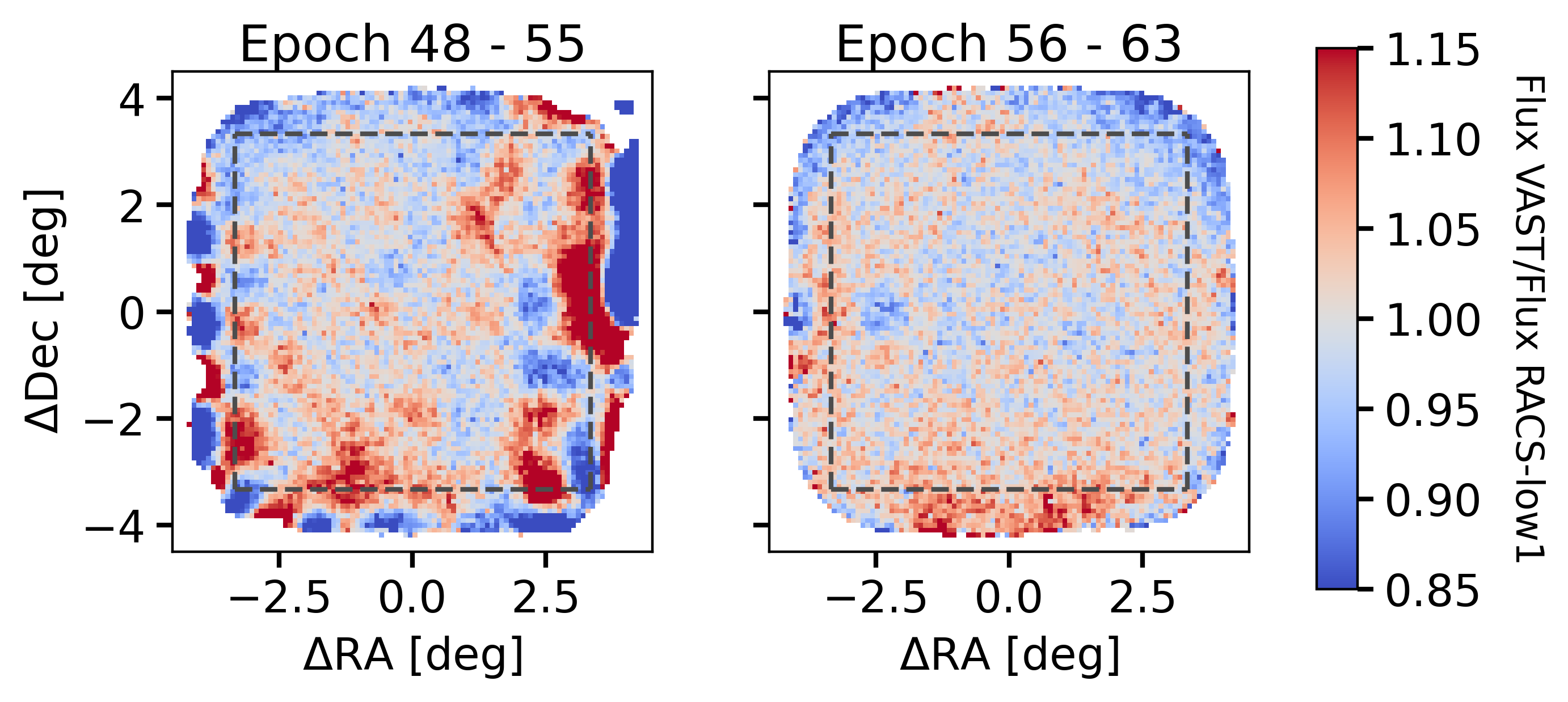}
    \caption{Spatial distribution across a field (after averaging and interpolating over all VAST fields) of the VAST flux density divided by the RACS-low1 flux density. The left panel shows epochs 48-55, which were affected by erroneous holography calibration, where extreme, localised variations in fluxdensity scale are visible. The right panel shows a similar number of unaffected observations for comparison. The dashed line indicates part of the image that is fed into the VAST pipeline.}
    \label{fig:holography}
\end{figure}

Below, we assess the impact of the erroneous holography calibration on the flux density scale in these observations. For a sample of compact, isolated, and bright sources, we calculate the mean flux density from the two observations taken as part of epochs 48-55 and divide that by the RACS-low1 \citep{hale2021rapid} flux density. Figure \ref{fig:holography} shows the distribution of these ratios across the field, after averaging and interpolating all VAST fields. The left panel shows the flux density ratio distribution in the epochs affected by the holography calibration, while the right panel shows a similar number of unaffected observations for comparison. The left panel of Figure \ref{fig:holography} clearly shows extreme, localised variations in the flux density scale across the field, where the flux density is either over- or under-corrected with the erroneous holography, whereas the right panel shows a more uniform pattern. The dashed lines in Figure \ref{fig:holography} indicate the portion of the image that is fed into the VAST pipeline, and the sources that are used to build light curves. Even in this inner part of the image, there are large flux density scale variations, making these observations unusable for a transient survey like VAST. We decide to disregard the epochs affected by this issue, as indicated by the greyed out points in Figure \ref{fig:fields_per_epoch}.

\section{Light curve database} \label{app:light_curve_database}
The light curve database consists of three tables: a measurement, source and image catalogue. The measurements catalogue contains all measurements of each source in VAST Extragalactic DR1. The source catalogue has one entry per source in VAST Extragalactic DR1. Additionally, we publish a table with details per image that is run through the VAST pipeline. As mentioned in Section \ref{sec:dataproducts}, these tables:
\begin{itemize}
    \item \texttt{VAST\_Extragal\_DR1\_measurement\_df.parquet}
    \item \texttt{VAST\_Extragal\_DR1\_source\_df.parquet}
    \item \texttt{VAST\_Extragal\_DR1\_image\_df.csv}
\end{itemize}
and a notebook with example code showing how to read and use these tables is available at \url{https://doi.org/10.25919/nh9d-t846}. The columns of these measurements,source and image tables, including example entries are given in the sections below.

\subsection{Measurements catalogue} \label{app:lcdb_measurement}
The measurement catalogue describes the properties of each individual measurement of each source in VAST Extragalactic DR1. Table \ref{tab:meas_cat} shows the top 11 entries in the measurement catalogue that correspond to the first source: 13297628. The measurements catalogue contains 6449202 measurements (rows), with 36 columns:

\begin{itemize}
    \item \texttt{source} - The unique source identifier, which can be used to match to the source table, and to retrieve cutouts.
    \item \texttt{ra}, \texttt{dec} (and errors) [deg] - The right ascension/declination coordinate of the measurement, peak of the fitted Gaussian.
    \item \texttt{flux\_peak}, \texttt{flux\_int} (and errors) [mJy/bm, mJy] - The measured peak/integrated flux density of the component.
    \item \texttt{bmaj}, \texttt{bmin}, \texttt{bpa} (and errors) [arscec, arcsec, deg] - The major/minor axis size (FWHM) and position angle of the fitted Gaussian.
    \item \texttt{psf\_bmin}, \texttt{psf\_bmaj}, \texttt{psf\_bpa} - The Selavy deconvolved size of the major/minor axis and position angle of the fitted Gaussian.
    \item \texttt{flag\_c4} - Selavy flag denoting whether the component is considered formally bad (does not meet chi-squared criterion).
    \item \texttt{chi\_squared\_fit} - Selavy quality of the fit.
    \item \texttt{spectral\_index} - Selavy spectral index of the component.
    \item \texttt{spectral\_index\_from\_TT} - Selavy flag to denote if the spectral index has been derived from the Taylor-term images (True or False).
    \item \texttt{has\_siblings} - Selavy flag to denote whether the component is one of many fitted to the same island.
    \item \texttt{image\_id} - The unique identifier of the image the measurement is from, which can be used to match to the image table.
    \item \texttt{time} - The date and time of observation the measurement is from (obtained from the image).
    \item \texttt{snr} - The signal-to-noise ratio of the measurement.
    \item \texttt{compactness} - The compactness of the measurement (flux\_int/flux\_peak).
    \item \texttt{forced} - Flag to denote whether the measurement is produced from the forced fitting procedure (True or False).
\end{itemize}

Certain columns are inherited from the \textsc{Selavy} source finder component catalogue, more details can be found at \url{https://www.atnf.csiro.au/computing/software/askapsoft/sdp/docs/current/analysis/postprocessing.html#component-catalogue}.

\begin{table*} 
\begin{minipage}{\textwidth}
\begin{center}
\centering
\small
\begin{tabular}{llllllllll} 
\\ \hline
source & local\_ & ra & ra\_err & dec & dec\_err & flux\_ & flux\_ & flux\_ & flux\_\\
 & rms &  &  &  &  & peak & peak\_err & int & int\_err\\ \hline
 & mJy/bm & deg & deg & deg & deg & mJy/bm & mJy/bm & mJy & mJy\\ \hline
13297628 & 0.258 & 321.489 & 1.08e-04 & -3.982 & 1.89e-04 & 182.19 & 0.27 & 192.72 & 0.45\\
13297628 & 0.258 & 321.489 & 1.50e-04 & -3.982 & 2.08e-04 & 179.87 & 0.31 & 192.68 & 0.48\\
13297628 & 0.300 & 321.489 & 1.17e-04 & -3.983 & 1.53e-04 & 183.45 & 0.33 & 193.24 & 0.54\\
13297628 & 0.273 & 321.489 & 1.97e-04 & -3.982 & 1.78e-04 & 181.58 & 0.30 & 191.78 & 0.49\\
13297628 & 0.953 & 321.489 & 3.69e-04 & -3.982 & 3.97e-04 & 182.17 & 0.95 & 189.63 & 1.64\\
13297628 & 0.336 & 321.489 & 1.33e-04 & -3.982 & 1.89e-04 & 184.64 & 0.35 & 193.40 & 0.59\\
13297628 & 0.261 & 321.489 & 1.28e-04 & -3.982 & 1.92e-04 & 181.38 & 0.27 & 191.59 & 0.46\\
13297628 & 0.252 & 321.489 & 1.58e-04 & -3.982 & 1.94e-04 & 181.02 & 0.27 & 190.86 & 0.44\\
13297628 & 0.268 & 321.489 & 1.44e-04 & -3.982 & 1.83e-04 & 182.68 & 0.29 & 193.62 & 0.48\\
13297628 & 0.263 & 321.489 & 1.36e-04 & -3.983 & 1.89e-04 & 180.65 & 0.30 & 190.61 & 0.48\\
13297628 & 0.314 & 321.489 & 1.14e-04 & -3.982 & 2.00e-04 & 183.90 & 0.32 & 193.75 & 0.55\\\hline 
 \end{tabular}
 \end{center}
\end{minipage}

\begin{minipage}{\textwidth}
\begin{center}
\centering
\small
\begin{tabular}{lllllllllll} 
\\ \hline
bmaj & err\_ & bmin & err\_ & pa & err\_ & psf\_ & psf\_ & psf\_ & flag\_ & chi\_\\
 & bmaj &  & bmin &  & pa & bmaj & bmin & pa & c4 & squared\_\\
 &  &  &  &  &  &  &  &  &  & fit\\ \hline
arcsec & arcsec & arcsec & arcsec & deg & deg & deg & deg & deg &  & \\ \hline
13.53 & 0.02 & 11.66 & 0.02 & 71.98 & 0.01 & 2.96 & 2.89 & 88.76 & False & 270.71\\
13.50 & 0.02 & 11.95 & 0.02 & 74.15 & 0.01 & 3.36 & 3.17 & -64.50 & True & 4186.10\\
13.20 & 0.02 & 11.32 & 0.02 & 75.60 & 0.01 & 2.87 & 2.61 & 46.04 & False & 251.35\\
19.40 & 0.03 & 12.20 & 0.02 & 111.62 & 0.00 & 4.08 & 2.91 & 89.36 & False & 511.01\\
24.10 & 0.12 & 14.10 & 0.07 & 102.12 & 0.01 & 3.72 & 2.21 & 8.36 & False & 654.47\\
17.41 & 0.03 & 11.93 & 0.02 & 117.21 & 0.00 & 3.28 & 2.80 & -59.85 & False & 139.11\\
15.41 & 0.02 & 11.39 & 0.02 & 83.32 & 0.00 & 3.21 & 2.79 & -61.12 & False & 224.41\\
15.48 & 0.02 & 11.80 & 0.02 & 80.30 & 0.00 & 3.24 & 2.73 & -49.47 & False & 138.22\\
13.55 & 0.02 & 11.77 & 0.02 & 72.19 & 0.01 & 3.08 & 2.91 & 80.56 & False & 565.98\\
13.94 & 0.02 & 11.94 & 0.02 & 81.57 & 0.01 & 3.07 & 2.82 & -87.15 & False & 466.39\\
15.43 & 0.02 & 11.55 & 0.02 & 80.01 & 0.00 & 3.16 & 2.82 & 86.05 & False & 140.22\\\hline 
 \end{tabular}
 \end{center}
\end{minipage} 

\begin{minipage}{\textwidth}
\begin{center}
\centering
\small
\begin{tabular}{llllllll} 
 \\ \hline
spectral\_ & spectral\_ & has\_ & image\_ & time & snr & compact- & forced\\
index & index\_from\_TT & siblings & id &  &  & ness & \\ \hline
 &  &  &  &  &  &  & \\ \hline
-1.38 & True & False & 8721 & 2024-06-17 19:50:38.3 & 706.18 & 1.06 & False\\
-1.29 & True & False & 9203 & 2024-10-17 11:40:44.5 & 697.18 & 1.07 & False\\
-1.33 & True & False & 10154 & 2025-04-22 22:54:56 & 611.49 & 1.05 & False\\
-1.13 & True & False & 7694 & 2023-07-05 19:37:23.1 & 665.14 & 1.06 & False\\
-1.37 & True & False & 8871 & 2024-08-14 17:26:28.5 & 191.15 & 1.04 & False\\
-1.21 & True & False & 9881 & 2025-02-21 03:42:11.3 & 549.53 & 1.05 & False\\
-1.20 & True & False & 7969 & 2023-08-30 16:14:00.4 & 694.93 & 1.06 & False\\
-1.33 & True & False & 8469 & 2024-04-21 00:43:27.9 & 718.35 & 1.05 & False\\
-1.29 & True & False & 9606 & 2024-12-20 07:07:33.1 & 681.63 & 1.06 & False\\
-1.16 & True & False & 8241 & 2023-10-27 10:13:10.9 & 686.87 & 1.06 & False\\
-1.15 & True & False & 7310 & 2023-06-13 21:29:09.6 & 585.66 & 1.05 & False\\ \hline 
  \end{tabular} 
\end{center} 
 \caption{The 11 lines in the measurements catalogue that correspond to source 13297628. The columns are described in \ref{app:lcdb_measurement}.}
 \label{tab:meas_cat}
 \end{minipage}
\end{table*} 

\subsection{Source catalogue} \label{app:lcdb_source}
The source catalogue describes the general properties and summary statistics of all sources in VAST Extragalactic DR1. Table \ref{tab:source_cat} shows the first 10 entries of the source catalogue. The source catalogue contains 548113 sources (rows), with 24 columns:
\begin{itemize}
    \item \texttt{source} - The unique source identifier, which can be used to match to the measurements table, and to retrieve cutouts.
    \item \texttt{wavg\_ra}, \texttt{wavg\_dec} [deg] - The source position in right ascension and declination, calculated by the weighted average of \textsc{Selavy} measurement (i.e. not including forced measurements).
    \item \texttt{wavg\_uncertainty\_ew}, \texttt{wavg\_uncertainty\_ns} [deg] - The uncertainty of the weighted average right ascension/declination value (excluding forced measurements).
    \item \texttt{avg\_compactness} - The average compactness (ratio of the integrated and peak flux density) value of the associated measurements (excluding forced measurements).
    \item \texttt{min\_snr},\texttt{max\_snr} - The minimum/maximum signal-to-noise ratio of the associated measurements (including forced measurements).
    \item \texttt{avg\_flux\_int}, \texttt{max\_flux\_int}, \texttt{min\_flux\_int} [mJy] - The average/maximum/minimum integrated flux density value of the measurements associated with the source (including forced measurements).
    \item \texttt{avg\_flux\_peak}, \texttt{max\_flux\_peak}, \texttt{min\_flux\_peak} [mJy $\rm{beam}^{-1}$] - The average/maximum/minimum peak flux density value of the measurements associated to the source (including forced measurements).
    \item \texttt{n\_measurements} - The total number of measurements associated to the source (selavy and forced).
    \item \texttt{n\_selavy} - The total number of selavy measurements associated to the source.
    \item \texttt{n\_forced} - The total number of forced measurements associated to the source.
    \item \texttt{n\_siblings} - The total number of measurements that have a \texttt{has\_sibling} value of \texttt{True}. \footnote{We note that the 'siblings' terminology comes from the \textsc{Selavy} source finder, where pixels that lie above a primary threshold are identified and grouped together into ‘islands’. When such an island is fit with multiple Gaussian components, these components are called siblings.}
    \item \texttt{n\_relations} - 	The total number of relations the source has.
    \item \texttt{v\_int}, \texttt{v\_peak} - The calculated variability $V$
     metric using the integrated/peak flux density values. See Section \ref{sec:variability_metrics}.
     \item \texttt{eta\_int}, \texttt{eta\_peak} - The calculated variability $\eta$
     metric using the integrated/peak flux density values. See Section \ref{sec:variability_metrics}.
     \item \texttt{n\_neighbour\_dist} [deg] - The on-sky separation to the nearest source within VAST Extragalactic DR1. 
\end{itemize}

\begin{table*} 
\begin{minipage}{\textwidth}
\begin{center}
\centering
\small
\begin{tabular}{lllllllllll} 
\\ \hline
source & wavg\_ & wavg\_ & wavg\_ & wavg\_ & avg\_ & min\_ & max\_ & avg\_ & avg\_ & max\_\\
 & ra & dec & uncertainty\_ & uncertainty\_ & compact- & snr & snr & flux\_ & flux\_ & flux\_\\
 &  &  & ew & ns & ness &  &  & peak & int & peak\\ \hline
 & deg & deg & deg & deg &  &  &  & mJy/bm & mJy & mJy/bm\\ \hline
13297621 & 322.580 & -9.460 & 4.59e-05 & 4.59e-05 & 1.07 & 231.32 & 528.52 & 230.85 & 247.43 & 281.83 \\
13297628 & 321.489 & -3.982 & 7.30e-05 & 7.30e-05 & 1.06 & 191.15 & 718.35 & 182.14 & 192.17 & 184.64 \\
13297679 & 320.202 & -3.509 & 7.30e-05 & 7.30e-05 & 1.07 & 134.76 & 478.45 & 105.96 & 113.62 & 109.79 \\
13297696 & 322.242 & -3.063 & 5.35e-05 & 5.35e-05 & 1.11 & 64.87 & 301.78 & 98.46 & 109.04 & 106.57 \\
13297702 & 319.625 & -3.217 & 5.35e-05 & 5.35e-05 & 1.20 & 98.23 & 305.36 & 90.53 & 108.67 & 96.14 \\
13297726 & 322.246 & -5.276 & 7.30e-05 & 7.30e-05 & 1.05 & 90.54 & 385.32 & 82.98 & 87.48 & 90.94 \\
13297727 & 321.892 & -4.747 & 7.30e-05 & 7.30e-05 & 1.06 & 79.63 & 406.16 & 72.50 & 76.65 & 94.64 \\
13297728 & 322.816 & -3.083 & 3.40e-05 & 3.40e-05 & 1.06 & 87.06 & 160.67 & 79.01 & 83.93 & 87.72 \\
13297739 & 318.219 & -4.750 & 7.30e-05 & 7.30e-05 & 1.07 & 94.99 & 332.48 & 72.20 & 77.36 & 74.79 \\
13297741 & 321.417 & -3.155 & 5.37e-05 & 5.37e-05 & 1.43 & 28.54 & 88.83 & 28.36 & 40.45 & 33.59 \\ \hline 
 \end{tabular}
 \end{center}
\end{minipage} 

\begin{minipage}{\textwidth}
\begin{center}
\centering
\small
\begin{tabular}{lllllllllllll} 
 \\ \hline
max\_ & min\_ & min\_ & n\_ & n\_ & n\_ & n\_ & n\_ & v\_ & v\_ & eta\_ & eta\_ & n\_\\
flux\_ & flux\_ & flux\_ & measure- & selavy & forced & siblings & relations & int & peak & int & peak & neighbour\_\\
int & peak & int & ments &  &  &  &  &  &  &  &  & dist\\ \hline
mJy & mJy/bm & mJy &  &  &  &  &  &  &  &  &  & deg\\ \hline
296.57 & 188.45 & 203.31 & 21 & 21 & 0 & 0 & 0 & 0.11 & 0.11 & 659.19 & 1969.06 & 1.12e-01\\
193.75 & 179.87 & 189.63 & 11 & 11 & 0 & 0 & 0 & 0.01 & 0.01 & 5.21 & 20.69 & 1.04e-01\\
117.35 & 99.51 & 107.22 & 11 & 11 & 0 & 0 & 0 & 0.03 & 0.03 & 61.43 & 202.25 & 5.39e-02\\
114.01 & 89.24 & 104.56 & 22 & 22 & 0 & 0 & 0 & 0.03 & 0.05 & 22.36 & 140.51 & 8.18e-02\\
114.08 & 85.06 & 104.65 & 22 & 22 & 0 & 0 & 0 & 0.02 & 0.03 & 18.66 & 76.33 & 1.52e-01\\
95.61 & 77.15 & 81.48 & 11 & 11 & 0 & 0 & 0 & 0.05 & 0.05 & 96.38 & 250.30 & 1.01e-01\\
99.17 & 62.59 & 65.96 & 11 & 11 & 0 & 0 & 0 & 0.11 & 0.12 & 434.71 & 1244.91 & 9.50e-02\\
92.01 & 69.57 & 77.07 & 43 & 43 & 0 & 0 & 0 & 0.04 & 0.05 & 12.63 & 53.78 & 5.07e-02\\
80.22 & 70.49 & 75.98 & 11 & 11 & 0 & 0 & 0 & 0.02 & 0.02 & 9.60 & 23.29 & 5.74e-02\\
46.73 & 25.32 & 35.22 & 22 & 22 & 0 & 22 & 0 & 0.07 & 0.07 & 21.38 & 47.25 & 2.37e-02\\ \hline 
  \end{tabular} 
\end{center} 
 \caption{{The first 10 lines from the source catalogue. The columns are described in \ref{app:lcdb_source}.}}
 \label{tab:source_cat}
 \end{minipage}
\end{table*}

\subsection{Image catalogue} \label{app:lcdb_images}
The image catalogue contains details of all the images that are ran through the VAST pipeline as part of VAST Extragalactic DR1. Table \ref{tab:image_cat} shows the first 10 entries of the image catalogue. The image catalogue contains information for 2945 images (rows), with 16 columns:
\begin{itemize}
    \item \texttt{image\_id} - The unique identifier of the image, which can be used to match to the measurements table.
    \item \texttt{name} - The filename of the image.
    \item \texttt{datetime} - The date and time of the observation, read from the FITS header.
    \item \texttt{jd} [days] - The date and time of the observation in Julian Days.
    \item \texttt{duration} [sec] - The duration of the observation taken from the FITS header.
    \item \texttt{physical\_bmaj}, \texttt{physical\_bmin}/ [deg] - Field of view of the image. For VAST Extragalactic DR1 these are equal, as we crop the image to a square in post-processing (See Section \ref{sec:pp_crop_compress}).
    \item \texttt{radius\_pixels} [deg] - 	Estimated 'diameter' of the useable image area. For VAST Extragalactic DR1 this is the side of the square image.
    \item \texttt{beam\_bmaj}, \texttt{beam\_bmin}, \texttt{beam\_bpa} [deg] - The size of the major/minor axis and position angle of the image restoring beam.
    \item \texttt{rms\_median}, \texttt{rms\_min}, \texttt{rms\_max} [mJy/bm] - The median/minimum/maximum rms value derived from the rms map.
    \item \texttt{centre\_ra},\texttt{centre\_dec} [deg] - The central right ascension/declination coordinate of the image.
    \item \texttt{SBID} - ASKAP Scheduling Block ID 
\end{itemize}
    
\begin{table*} 
\begin{minipage}{\textwidth}
\begin{center}
\centering
\small
\begin{tabular}{lllll} 
\\ \hline
image\_ & name* & datetime & jd & dura-\\
id &  &  &  & tion\\ \hline
 &  &  & days & sec\\ \hline
7251 & \verb|VAST_Extragal_DR1_VAST_0126+00.SB50516.image.fits| & 2023-06-14 01:36:10.1 & 2460109.567 & 716.6\\
7252 & \verb|VAST_Extragal_DR1_VAST_0126+06.SB50515.image.fits| & 2023-06-14 01:22:53.8 & 2460109.558 & 736.5\\
7253 & \verb|VAST_Extragal_DR1_VAST_0151+00.SB50517.image.fits| & 2023-06-14 01:49:56.2 & 2460109.576 & 726.6\\
7254 & \verb|VAST_Extragal_DR1_VAST_0216+00.SB50518.image.fits| & 2023-06-14 02:03:02.5 & 2460109.585 & 716.6\\
7255 & \verb|VAST_Extragal_DR1_VAST_0216-06.SB50519.image.fits| & 2023-06-14 02:16:48.6 & 2460109.595 & 726.6\\
7256 & \verb|VAST_Extragal_DR1_VAST_0241+00.SB50520.image.fits| & 2023-06-14 02:30:44.7 & 2460109.605 & 726.6\\
7257 & \verb|VAST_Extragal_DR1_VAST_0241+06.SB50521.image.fits| & 2023-06-14 02:43:51 & 2460109.614 & 716.6\\
7258 & \verb|VAST_Extragal_DR1_VAST_0306+00.SB50523.image.fits| & 2023-06-14 03:09:53.7 & 2460109.632 & 716.6\\
7259 & \verb|VAST_Extragal_DR1_VAST_0306+06.SB50522.image.fits| & 2023-06-14 02:56:47.4 & 2460109.623 & 726.6\\
7260 & \verb|VAST_Extragal_DR1_VAST_0331+00.SB50525.image.fits| & 2023-06-14 03:35:56.3 & 2460109.650 & 716.6\\ \hline 
 \end{tabular}
 \end{center}
\end{minipage} 

\begin{minipage}{\textwidth}
\begin{center}
\centering
\small
\begin{tabular}{llllllllllll} 
 \\ \hline
 physical\_ & physical\_ & radius\_ & beam\_ & beam\_ & beam\_ & rms\_ & rms\_ & rms\_ & centre\_ & centre\_ & SBID\\
bmaj & bmin & pixels & bmaj & bmin & bpa & median & min & max & ra & dec & \\ \hline
deg & deg & deg & deg & deg & deg & mJy/bm & mJy/bm & mJy/bm & deg & deg & \\ \hline
6.67 & 6.67 & 6791.8 & 4.31e-03 & 3.44e-03 & 8.75e+01 & 0.23 & 0.15 & 7.26 & 21.724 & -0.001 & 50516\\
6.67 & 6.67 & 6791.8 & 4.08e-03 & 3.69e-03 & -6.29e+01 & 0.24 & 0.15 & 1.22 & 21.724 & 6.294 & 50515\\
6.67 & 6.67 & 6791.8 & 4.28e-03 & 3.39e-03 & 8.39e+01 & 0.22 & 0.15 & 0.86 & 27.931 & -0.000 & 50517\\
6.67 & 6.67 & 6791.8 & 4.22e-03 & 3.36e-03 & 7.98e+01 & 0.25 & 0.16 & 1.57 & 34.138 & -0.001 & 50518\\
6.67 & 6.67 & 6791.8 & 4.28e-03 & 3.11e-03 & 7.92e+01 & 0.21 & 0.14 & 0.86 & 34.138 & -6.300 & 50519\\
6.67 & 6.67 & 6791.8 & 4.22e-03 & 3.36e-03 & 8.12e+01 & 0.24 & 0.16 & 2.00 & 40.345 & -0.001 & 50520\\
6.67 & 6.67 & 6791.8 & 4.14e-03 & 3.67e-03 & -5.98e+01 & 0.22 & 0.15 & 1.97 & 40.345 & 6.295 & 50521\\
6.67 & 6.67 & 6791.8 & 4.28e-03 & 3.42e-03 & 8.55e+01 & 0.22 & 0.15 & 0.92 & 46.552 & -0.001 & 50523\\
6.67 & 6.67 & 6791.8 & 4.03e-03 & 3.72e-03 & -6.95e+01 & 0.24 & 0.15 & 10.96 & 46.552 & 6.294 & 50522\\
6.67 & 6.67 & 6791.8 & 4.31e-03 & 3.42e-03 & 8.62e+01 & 0.23 & 0.15 & 6.36 & 52.759 & -0.001 & 50525\\ \hline 
  \end{tabular} 
\end{center} 
\caption{The first 10 lines from the image catalogue. The columns are described in \ref{app:lcdb_images}. *: Image names of post-processed images that are ingested in the pipeline, which can be found on \url{https://data.csiro.au/collection/csiro:72199}.}
 \label{tab:image_cat}
 \end{minipage}
\end{table*}

\section{Dynamic spectra probable stars} \label{app:dynamic_spectra}
\begin{figure}
    \centering
    \includegraphics[width=\linewidth]{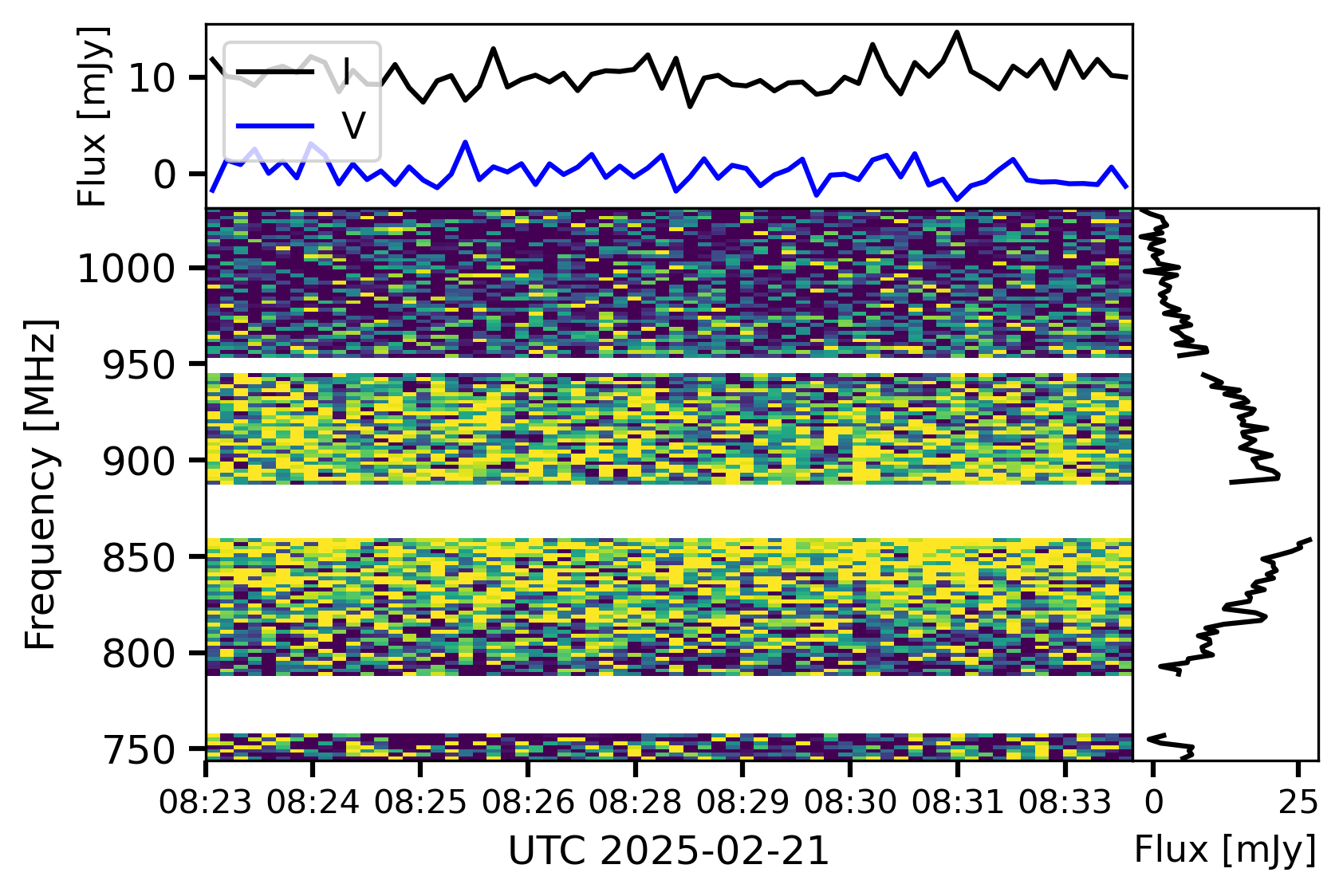}
    \caption{Stokes I dynamic spectrum of the brightest detection of ASKAP J014700.3$+$075128, a probable radio star. The light curves in the top panel show the total (I, black) and circular (V, blue) intensity, respectively. The right panel shows the spectrum.}
    \label{fig:13805206_DS}
\end{figure}

\begin{figure}
    \centering
    \includegraphics[width=\linewidth]{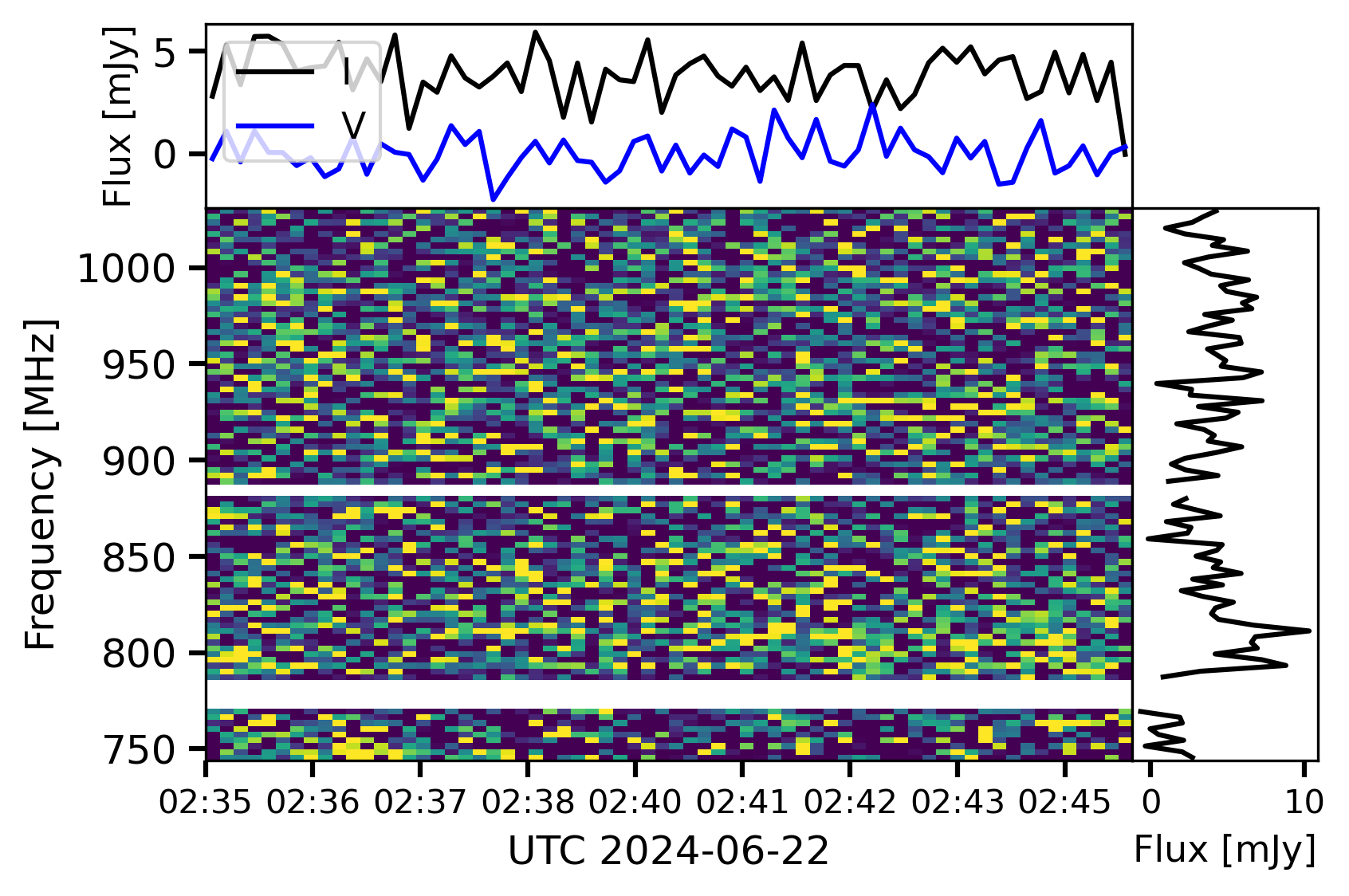}
    \caption{Stokes I dynamic spectrum of the single detection of ASKAP J053954.1-051119, a probable radio star. The light curves in the top panel show the total (I, black) and circular (V, blue) intensity, respectively. The right panel shows the spectrum.}
    \label{fig:13829699_DS}
\end{figure}

This appendix shows the dynamic spectra of the probable stars discussed in Section \ref{sec:untargeted_variability_search}.
Figure \ref{fig:13805206_DS} shows the dynamic spectrum of the brightest detection of ASKAP J014700.3$+$075128, the radio emissions is limited to part of the band between 800 and 950 MHz. Figure \ref{fig:13829699_DS} shows the dynamic spectrum of the single flare of ASKAP J053954.1-051119. This dynamic spectrum has been averaged in frequency compared to the dynamic spectrum in Figure \ref{fig:13805206_DS} to try and detect any frequency structure despite the faintness of the detection (6.6 mJy). For both sources, there is no indication of Stokes V emission; the upper limits are $\lesssim 2\%$ and $\lesssim 4\%$ for ASKAP J014700.3$+$075128 and ASKAP J053954.1-051119, respectively.

\bibliography{bibtemplate}


\end{document}